\documentclass[final,5p,times,twocolumn,authoryear]{elsarticle}
\usepackage{amsmath}
\usepackage{nccmath}
\usepackage{amssymb}
\usepackage[binary-units=true,detect-all,mode=text]{siunitx}

\usepackage{graphicx}
\usepackage{epsfig}

\usepackage[utf8]{inputenc}

\usepackage{xcolor}
\definecolor{codegreen}{rgb}{0,0.6,0}
\definecolor{codegray}{rgb}{0.5,0.5,0.5}
\definecolor{codepurple}{rgb}{0.58,0,0.82}
\definecolor{backcolour}{rgb}{0.95,0.95,0.92}

\usepackage{listings}
\lstdefinestyle{mystyle}{
    backgroundcolor=\color{backcolour},   
    commentstyle=\color{codegreen},
    keywordstyle=\color{magenta},
    numberstyle=\tiny\color{codegray},
    stringstyle=\color{codepurple},
    basicstyle=\ttfamily\footnotesize,
    breakatwhitespace=false,         
    breaklines=true,                 
    captionpos=b,                    
    keepspaces=true,                 
    numbers=left,                    
    numbersep=5pt,                  
    showspaces=false,                
    showstringspaces=false,
    showtabs=false,                  
    tabsize=2
}
\usepackage{hyperref}
\hypersetup{
    colorlinks=true,
    linkcolor=blue,
    filecolor=blue,      
    urlcolor=blue,
    }
\usepackage{lineno}
\usepackage{ulem}

\journal{Nuclear Physics B}

\begin{document}

\begin{frontmatter}

%% Title, authors and addresses
%% use the tnoteref command within \title for footnotes;
%% use the tnotetext command for theassociated footnote;
%% use the fnref command within \author or \affiliation for footnotes;
%% use the fntext command for theassociated footnote;
%% use the corref command within \author for corresponding author footnotes;
%% use the cortext command for theassociated footnote;
%% use the ead command for the email address,
%% and the form \ead[url] for the home page:
%% \title{Title\tnoteref{label1}}
%% \tnotetext[label1]{}
%% \author{Name\corref{cor1}\fnref{label2}}
%% \ead{email address}
%% \ead[url]{home page}
%% \fntext[label2]{}
%% \cortext[cor1]{}
%% \affiliation{organization={},
%%            addressline={}, 
%%            city={},
%%            postcode={}, 
%%            state={},
%%            country={}}
%% \fntext[label3]{}

\title{Toward the End-To-End Optimization of the SWGO Array Layout}

%% use optional labels to link authors explicitly to addresses:
\author[1,2,3,4]{Tommaso~Dorigo}
\author[2,8]{Max~Aehle}
\author[3]{Cornelia~Arcaro}
\author[1,3,4,5]{Muhammad~Awais}
\author[13]{Fabiola~Bergamaschi}
\author[2,4,6]{Julien~Donini}
\author[5,3]{Michele~Doro}
\author[2,8]{Nicolas~R.~Gauger}
\author[9]{Rafael~Izbicki}
\author[2,10]{Jan~Kieseler}
\author[2,7]{Ann~Lee}
\author[2,7]{Luca~Masserano}
\author[2,3,5,6]{Federico~Nardi}
\author[11]{Raaghav~Rajesh}
\author[12,3,5]{Luis~Recabarren~Vergara}
\author[7]{Alexander~Shen}

\affiliation[1]{organization={Lule\aa \, University of Technology},
%            addressline={}, 
            city={Lule\aa},
%            postcode={}, 
%            state={},
            country={Sweden}}
\affiliation[2]{organization={MODE Collaboration},
            %addressline={}, 
            %city={Earth},
            %postcode={}, 
            %state={},
            %country={}
            }
\affiliation[3]{organization={Istituto Nazionale di Fisica Nucleare - Sezione di Padova},
            addressline={via Marzolo 8}, 
            city={Padova},
            postcode={35131}, 
            %state={},
            country={Italy}
            }
\affiliation[4]{organization={Universal Scientific Education and Research Network (USERN)},
            %addressline={}, 
            %city={Earth},
            %postcode={}, 
            %state={},
            country={Italy}
            }
\affiliation[5]{organization={Dipartimento di Fisica e Astronomia G.Galilei, University of Padova},
            addressline={via Marzolo 8}, 
            city={Padova},
            postcode={35131}, 
            %state={},
            country={Italy}
            }
\affiliation[6]{organization={University Clermont Auvergne},
            %addressline={}, 
            %city={Earth},
            %postcode={}, 
            %state={},
            country={France}
            }
\affiliation[7]{organization={Carnegie Mellon University},
            %addressline={}, 
            %city={Earth},
            %postcode={}, 
            %state={},
            country={USA}
            }
\affiliation[8]{organization={University of Kaiserslautern-Landau (RPTU)},
            %addressline={}, 
            %city={Earth},
            %postcode={}, 
            %state={},
            country={Germany}
            }
\affiliation[9]{organization={Universidade Federal de Sao Carlos},
            %addressline={}, 
            %city={Earth},
            %postcode={}, 
            %state={},
            country={Brazil}
            }
\affiliation[10]{organization={Karlsruhe Institute of Technology (KIT)},
            %addressline={}, 
            %city={Earth},
            %postcode={}, 
            %state={},
            country={Germany}
            }
\affiliation[11]{organization={Birla Institute of Technology and Science},
            %addressline={}, 
            city={Pilani},
            %postcode={}, 
            state={Rajasthan},
            country={India}
            }
\affiliation[12]{organization={Centro di Ateneo di Studi e Attivita Spaziali "Giuseppe Colombo", Via Venezia 15, I-35131},
            %addressline={}, 
            city={Padova},
            %postcode={}, 
            %state={},
            country={Italy}
            }
\affiliation[13]{organization={Dipartimento di Scienze Statistiche, University of Padova},
            %addressline={}, 
            city={Padova},
            %postcode={}, 
            %state={},
            country={Italy}
            }

\begin{abstract}
In this document we consider the problem of finding the optimal layout for the array of water Cherenkov detectors proposed by the SWGO collaboration to study very-high-energy gamma rays in the southern hemisphere.  We develop a continuous model of the secondary particles produced by atmospheric showers initiated by high-energy gamma rays and protons, and build an optimization pipeline capable of identifying the most promising configuration of the detector elements. The pipeline employs stochastic gradient descent to maximize a utility function aligned with the scientific goals of the experiment. We demonstrate how the software is capable of finding the global maximum in the high-dimensional parameter space, and discuss its performance and limitations.
\end{abstract}

%%Graphical abstract
%\begin{graphicalabstract}
%\includegraphics{grabs}
%\end{graphicalabstract}

%%Research highlights
%\begin{highlights}
%\item Research highlight 1
%\item Research highlight 2
%\end{highlights}

\begin{keyword}
%% keywords here, in the form: keyword \sep keyword, up to a maximum of 6 keywords
Gamma rays \sep Optimization \sep Cosmic Rays \sep Cherenkov detectors

%% PACS codes here, in the form: \PACS code \sep code

%% MSC codes here, in the form: \MSC code \sep code
%% or \MSC[2008] code \sep code (2000 is the default)

\end{keyword}

\end{frontmatter}

\tableofcontents

%
%%%%%%%%%%%%%%%%%%%\linenumbers

%% main text

\section{Introduction}

The optimal choice of layout, characteristics, materials, and information-extraction procedures of a measuring instrument for research in fundamental science constitutes a loosely constrained problem, featuring a very large number of free parameters related by non-obvious correlations and constrained by a collection of diverse external factors, such as budget, physical hindrances, environmental impact, and science policy. Although typically quite complex, similar problems may sometimes still be tractable by standard means, when an analytical expression of the behavior of the system allows the definition of a likelihood function $\mathcal{L}(\theta) = p(x|\theta)$ given simulated data $x$, and  a solution by minimization of $-\ln \mathcal{L}$ with respect to the modeling parameters $\theta$. When instead the instrument bases its functioning on quantum phenomena such as those governing the interaction of radiation with matter, the optimization problem is typically intractable: the probability $p(x|\theta)$ of observing data $x$ given underlying parameters $\theta$ may not be written explicitly. In the latter circumstance, one has access at best to the generating function of the observed data through forward simulation, a setting commonly referred to as likelihood-free or simulation-based inference~\citep{cranmer2020frontier}. 

 In the latter case discussed above, which is the most common in fundamental physics applications, instrument design is normally informed by extensive studies that sample the space of possible configurations in a sparse, discrete manner. These approaches often rely on heuristic reasoning, parameter scans, or evolutionary algorithms, focusing on optimizing individual sub-components rather than the full detector layout. However, the computational cost of accurately simulating particle interactions and detector responses severely limits the number of configurations that can be explored, making such methods infeasible for problems with thousands of free parameters. The combination of extremely high dimensionality and simulation complexity not only renders traditional approaches computationally prohibitive but also prevents them from systematically identifying and exploiting hidden correlations between design parameters.  

 A promising solution to this challenge comes from recent advances in deep learning and differentiable programming. These techniques enable a fully continuous representation of the detector layout and its response, allowing gradient-based optimization to explore the entire parameter space systematically. Unlike discrete methods, differentiable optimization not only facilitates a more efficient and comprehensive search for globally optimal solutions but also directly constructs a continuous objective function that encapsulates the full end-to-end performance of the detector, ensuring a principled and scalable approach to design optimization. 

\medskip
 Motivated by the above considerations,  in this work we propose an end-to-end optimization strategy that exploits differentiable programming and a fully continuous model of the physical processes of relevance, to address the issue of what is the best layout for the detection elements comprising the planned Southern Wide-Field Gamma-ray Observatory (SWGO). SWGO's main scientific goal is the study of gamma-ray fluxes from the southern sky in the \si{\tera\eV}-\si{\peta\eV}\footnote{\si{\tera\eV}= Tera Electron-Volt = $10^{12}$\,eV. \si{\peta\eV}=Peta Electron-Volt = $10^{15}$\,eV.} range. Cosmic gamma rays entering Earth's atmosphere generate a shower of secondary particles (mainly electrons, positrons, and photons) whose number grows until it is sustained by enough energy for multiplication to continue. By placing a detector of the atmospheric shower particles at high altitude (above 4,000\,m a.\,s.\,l.), it becomes feasible to reconstruct energy, direction of incidence, and time of arrival of primary gamma rays. For PeV showers, the pattern of secondaries on the ground spans from several hundred thousands to millions of square meters. Such a large area can be sampled with extended detector arrays. A ``fill factor'' --the fraction of the ground instrumented by sensitive detection elements-- of \SI{5}{\percent} to \SI{80}{\percent}  can be achieved in areas of one or a few square kilometers by deploying several thousands individual detector units. These are particle detectors such as Cherenkov water tanks, scintillation planes, or resistive plate counters. The number, spacing, and layout of these units, along with their performance, determine the instrument's standard figures of merit: sensitivity, angular and energy resolution, and signal to noise ratio. Noise is mostly represented by hadronic showers--{\it e.g.}, atmospheric showers generated by the much more abundant ($10^3$--$10^5$ times more so) charged cosmic rays. Their effective discrimination from gamma rays is one of the main challenges of such an experiment. 

%A maximization of the performance of the SWGO array can be pursued through the search of the most advantageous geometrical layout of a given number of detection units. The exact definition of a performance metric must reflect the collaboration's appraisal of different achievable scientific goals, and is by definition determined by scientific as well as political arguments. Its ingredients are however rather simple to define in terms of attainable sensitivity, precision of estimated gamma-ray flux, energy, and direction, and robustness to modeling systematics. We will address some of the above elements in the definition of a few possible utility metrics, which for this first work, in which we simplify the detector modeling to make the problem more tractable, are only provided as an example. The proposed metrics however constitute a good starting point as they contain the mathematical ingredients that play a role in determining the scientific output of the instrument. This enables us to address in detail the issue of how the end-to-end optimization problem can be solved in a differentiable manner. 

 A maximization of the performance of the SWGO array can be pursued through the search of the most advantageous geometrical layout of a given number of detection units. The exact definition of a performance metric must reflect the collaboration’s appraisal of different achievable scientific goals, and is by definition determined by scientific as well as political arguments. However, to ensure a quantitative assessment of the proposed approach, we define key performance indicators that guide the optimization process. These include improvements in the array’s sensitivity relative to existing benchmark configurations, the precision of reconstructed gamma-ray fluxes, and robustness to systematic modeling uncertainties. By explicitly incorporating these criteria, we aim to assess the feasibility of this approach in situations which go beyond what can be approached with heuristic-based methods, and to evaluate the scientific merit of the optimized configurations. We will address some of the mentioned ingredients in the definition of a few possible utility metrics, which for this first work, in which we simplify the detector modeling to make the problem more tractable, are only provided as an example. The proposed metrics, however, constitute a good starting point as they contain the mathematical ingredients that play a role in determining the scientific output of the instrument. This enables us to address in detail the issue of how the end-to-end optimization problem can be solved in a differentiable manner. 

 The study we present here lays within the realm of problems of interest of the MODE Collaboration\footnote{Web site: \url{https://mode-collaboration.github.io}}, a group of physicists and computer scientists who try to solve optimization problems with differentiable programming. A snapshot of the problems that have so far been attacked by the MODE collaboration is provided in a few recent publications \citep{modewhitepaper,elbaproceedings,modeprogresspaper}.

The present document is organized as follows. 
In \autoref{sec:swgo} we briefly introduce the SWGO detector and its proposed components, focusing on those characteristics which most influence the choices made in designing the optimization procedure. 
\autoref{s:methodology} introduces the structure of the optimization pipeline, which is composed of modular blocks performing the loosely connected tasks that convert the physical processes into statistical inference. 
\autoref{s:reconstruction} describes the continuous model of atmospheric showers we developed based on full simulation, and the reconstruction of shower parameters by a maximum likelihood technique. 
In \autoref{s:sfuncertainty} we discuss the test statistic we designed to discriminate gamma rays from hadron-induced showers. 
\autoref{s:utility} introduces two possible choices for a utility function. These allow us to produce in \autoref{sec:performance} a rough assessment of the relative performances of different geometries proposed by the SWGO collaboration as benchmarks for the array layout, as estimated by our model and reconstruction technique.  
In \autoref{sec:end-to-end} we describe the full pipeline and the end-to-end optimization loop, which iteratively updates the detector geometry based on the gradient of the utility. Preliminary results are shown and discussed in \autoref{s:results}. 
We offer some concluding remarks in \autoref{sec:conclusion}.
\section{The SWGO observatory}
\label{sec:swgo}

%\mdc{I would like to review the information in this section because I believe we do not explain sufficiently the distribution of particles in the ground. I will discuss this with Luis.}
The Southern Wide-Field Gamma-ray Observatory\footnote{\url{https://www.swgo.org}} (SWGO) is a proposed detector dedicated to the study of cosmic gamma rays in the TeV to PeV range from astrophysical sources located at southern sky declinations. 

  These instruments  detect   charged particles  and radiation  of extended atmospheric showers (EAS) generated in the upper atmosphere by the interactions of high-energy gamma rays and protons or light nuclei.  
Their density and distribution within an EAS depend on the primary particle type, energy and direction. 

\begin{figure} [h!]
    \centering
    \includegraphics[width=0.99\linewidth]{ 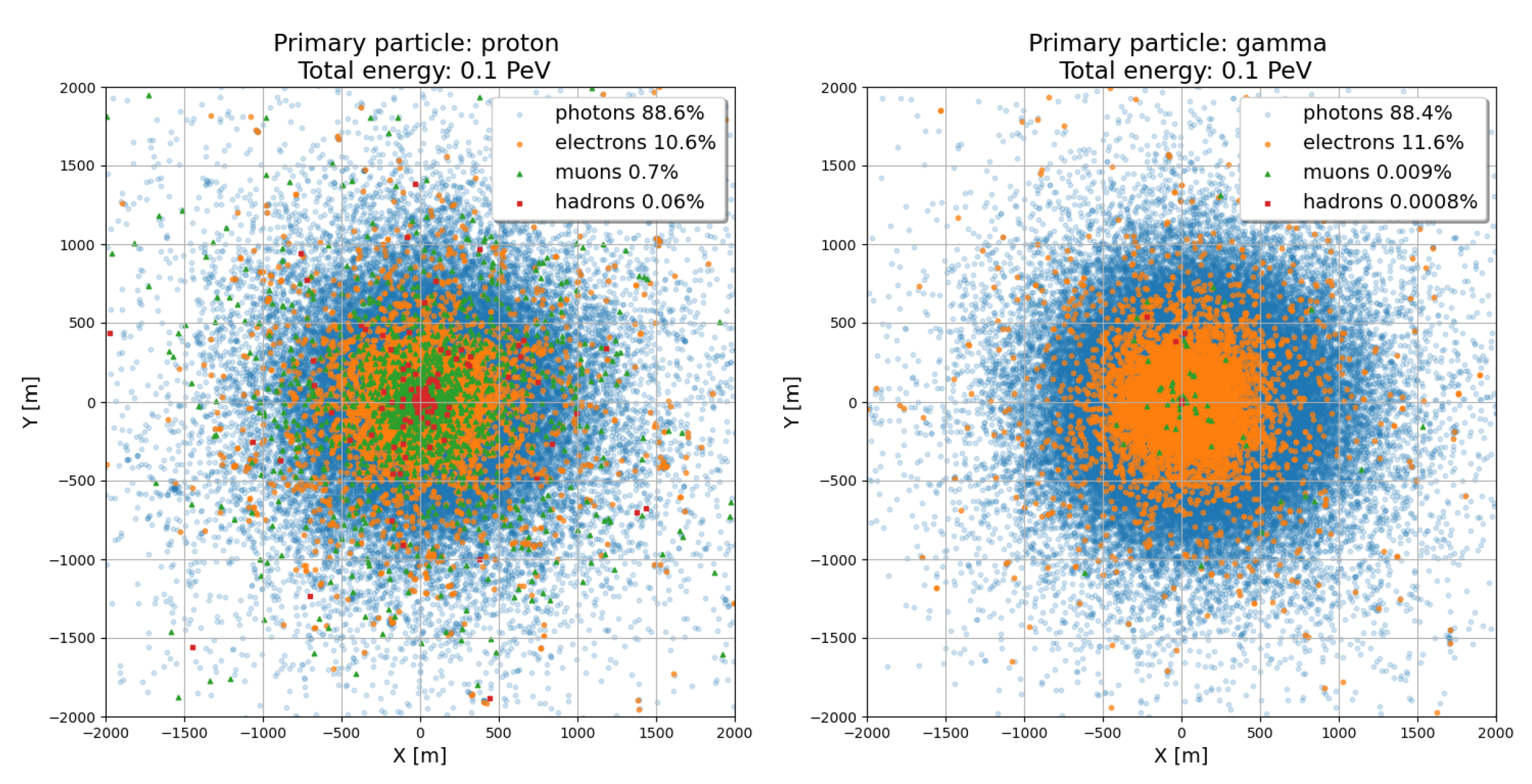}
    \caption{Radial distribution of secondary particles on the ground, originated at \SI{4.8}{\kilo\meter} altitude by a \SI{100}{\tera\eV} proton (left) or photon shower (right). The shower particle footprint at ground strongly depends on the primary particle kind, energy and direction. }
    \label{f:radial}
\end{figure}

An EAS develops laterally, 
so that a ground area ranging from a few hundred meters (for shower energy of hundreds of \si{\giga\eV}), to more than a \si{\kilo\meter} (for \si{\peta\eV} showers) is hit by particles,  for the duration of a few tens of nanoseconds, as shown in \autoref{f:radial}. Electromagnetic showers generate mainly electrons and positrons (collectively electrons hereafter) and secondary gamma rays, plus a very small amount of muons at very high energy. Hadronic showers, besides containing electromagnetic sub-showers, generate in addition a significant flux of muons and heavier particles. The size of the EAS footprint on the ground largely depends on the proximity of the shower maximum development, which in turns depends on the energy of the primary (the larger the primary energy, the closer the shower maximum to the ground) and its angle of incidence from the vertical (hereafter, the polar angle $\theta$). 

To detect this footprint of secondary particles, SWGO will consist of several thousand detector units spread over a large area.  At the time of writing, a number of different configurations   are being considered. 
The detector units share the need to separately count the collective number of electrons and photons reaching the ground\footnote{ In alternative to their number, a measurement of the total energy of the secondaries by species would equivalently allow to extract the needed information. In the absence of a finalized detector design, we base our initial study on the hypothesis that the number of particles be estimated with good precision by the detector units.}, as well as the number of muons. Water tanks may be able to distinguish muons from electrons and photons, {\it e.g.}, if they are segmented  vertically in two parts, where the upper part absorbs the light generated by the electromagnetic component; due the limited range of electrons in water, only muons get through to the lower part of the tank and produce Cherenkov light there. Muons are very rare in gamma ray-originated showers, especially if these primaries have energy in the lower range considered by the instrument; they thus constitute a powerful discriminator of hadron backgrounds. 

A loosely connected issue to the one of the most performing tank design concerns the arrangement of tanks on the ground. The  capability of the array to appropriately measure the energy of the primary particle in the low-energy range of interest to some specific measurements (tens to hundreds of GeV) requires the collection of as large a fraction of the collective signal of particles on the ground as possible. This mandates that tanks be laid out as tightly packed as possible --{\it i.e.}, with a fill factor (FF) of \SI{50}{\percent} or more. On the other hand, in order to be sensitive to the very low flux of photons of extremely high energy (\SI{1}{\peta\eV} and above), the total footprint of the array needs to be maximized with detectors distributed with a low FF and well spaced from one another, meeting cost constraints. However, a too low FF will hinder both a precise energy reconstruction and a successful separation of  backgrounds, whose rate at those high energies exceeds that of gamma rays by a factor of $10^4$ and above.

\subsection{Detector unit design}
\label{sec:tanks}

 For the scope of this work we are not interested in the design details of the different tanks, whose inclusion in the optimization procedure would be premature. Therefore we only consider a reference tank A, with a diameter of \SI{3.82}{\meter} and a depth of \SI{3}{\meter}. We further assume full efficiency for particle detection and identification: that is, a probability $p=1$ that units detect and correctly reveal and identify secondary particles. These assumptions are certainly not valid in a strict sense, but we believe that to first order they factor out of the layout optimization task, as they have limited impact in the variation of the utility with respect to a variation of the layout of tanks on the ground. 
%\mdc{The previous sentence is pretty strong. I think we should validate it further, or soften it. We should discuss this Tommaso.}

%\begin{figure}
%    \centering
%    \includegraphics[width=0.98\linewidth]{ tanks.png}
%    \caption{Schematic design of different SWGO tank designs. Tank design $A$ is called the reference one; it consists in a double layer unit. Each layer holds a photomultiplier tube, both of equal size. With same dimensions tank $B$ replaces a photomultiplier tube by a larger one. Tank design $C$ ($D$) is a smaller (larger) version of tank $A$ ($B$). Tank designs $E$ and $F$ are single-layer units with three (one) photomultiplier tubes of equal size, respectively.}
%    \label{fig:tanks}
%\end{figure}

\subsection{Benchmark arrays}
\label{sec:array}

\begin{table*}[h!t]
\begin{center}
\begin{footnotesize}
\begin{tabular}{cc|ccc|ccc|ccc}
\hline
ID & $N_\text{det}$ & \multicolumn{3}{c}{Zone 1} &  \multicolumn{3}{c}{Zone 2} & \multicolumn{3}{c}{Zone 3} \\
& & FF(\si{\percent}) & Radius (\si{\meter}) &  Units &  FF(\si{\percent}) &  Radius (\si{\meter}) &  Units &  FF(\si{\percent}) &  Radius (\si{\meter}) &  Units \\
\hline
A1 & 6589 & 80 & 160 & 5731 & 5 & 300 & 858 & & \\
A2 & 6631 & 80 & 138 & 4303 & 2.5 & 600 & 2328  & & \\
A3 & 6823 & 80 & 138 & 4303 & 2.5
%* 
& 600 & 2520  & & \\
A4 & 6625 & 80 & 140 & 4429 & 4.0 & 400 & 1518 & 1.25 & 600 & 678  \\
A5 & 6541 & 40 & 234 & 6109 & 5.0 & 300 & 432  & & \\ 
A6 & 6637 & 88 & 162 & 6469 & 1.0 & 300 & 168  & & \\
A7 & 6571 & 80 & 101 & 2335 & 2.5 & 600 & 2394 & 0.63 & 1200 & 1842 \\
B1 & 4849 & 80 & 131 & 3865 & 5 & 300 & 984  & & \\
C1 & 8371 & 80 & 137 & 6829 & 5 & 300 & 1542  & & \\
D1 & 3805 & 80 & 166 & 3367 & 5 & 300 & 438  & & \\
E1 & 5461 & 80 & 150 & 4639 & 5 & 300 & 822  & & \\
E4 & 5455 & 80 & 140 & 3403 & 4.0 & 400 & 1428&  1.25 & 600 & 624  \\
F1 & 4681 & 80 & 188 & 4303 & 5 & 300 & 378  & & \\
\hline
\end{tabular}
\caption{\label{t:layouts_sec2} Main features of the benchmark layouts considered for the SWGO Cherenkov tanks. The zones refers to concentric rings with different fill factors (FF). See \autoref{f:layouts} for a bird's eye view of the layouts. }
\end{footnotesize}
\end{center}
\end{table*}

\begin{figure}[h!t]
\centering
\includegraphics[width=0.98\linewidth]{ 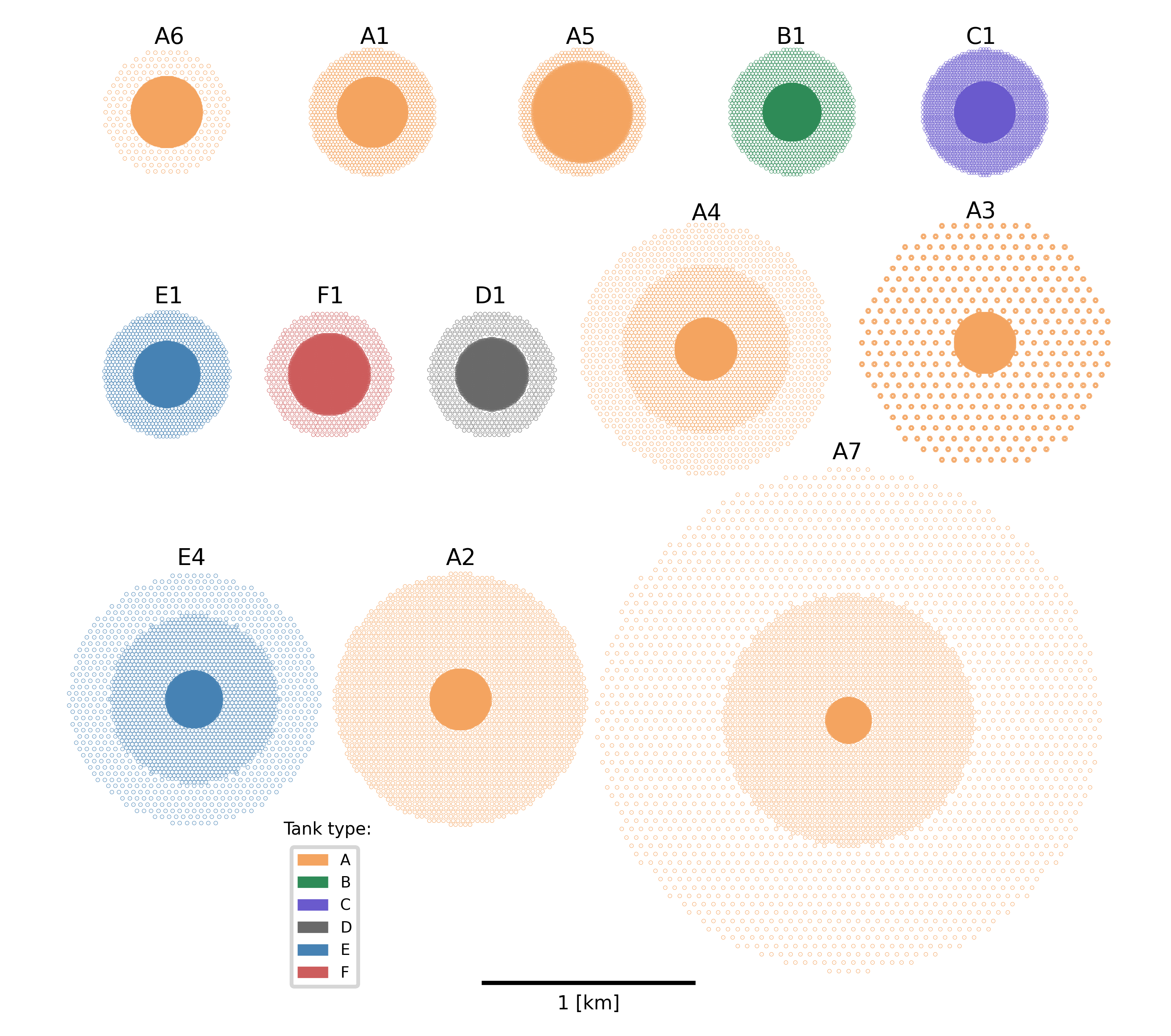}
\caption{Preliminary arrays considered by the collaboration ordered by size according to \autoref{t:layouts_sec2}. Each array is composed by one type of tank labeled by the color, and explores different zones with a decreasing FF at outer radii. }
\label{f:layouts}
\end{figure}

A number of possible arrangements of the units on the ground have been considered by the SWGO collaboration as benchmarks for initial appraisal studies. \autoref{t:layouts_sec2} describes the overall features of those layouts, and \autoref{f:layouts} shows their graphical representations to give an idea of their relative sizes and fill factors. They all share a radial symmetry and a higher FF in the core, with a smaller FF at larger radii. The number of units of these designs ranges between 3805 and 8371, with a median of about 6000 units. The different arrays also include different tank designs in order to keep the total cost of the array the same. $A1$ is the so-called reference configuration. The rationale of a dense core is to allow for detection of lower-energy (multi-\si{\giga\eV} to \si{\tera\eV}) photon showers. At these energies the flux of primaries is high, but their footprint on the ground has a limited extension; 
%\mdc{I am not sure the extension is that much smaller than for higher energy shower. What certainly is smaller is the density of particle at the ground, but is it the hit area?} 
the precise reconstruction of shower parameters is limited by the number of detectable particles, which is much smaller than in PeV-energy showers; for effective discrimination one should then be able to sample a fraction as high as possible of the shower footprint.  A tightly packed layout of detection tanks produces an effective shield from laterally entering electrons and photons: each detector, whose height from the ground is of the order of 1.5-3 meters according to the design, can effectively prevent the soft secondaries component from penetrating sideways and reaching into the bottom of the traversed tank --a phenomenon producing a confounding background to the hard component of the showers. On the other hand, the rationale of a further extended area instrumented at lower FF is to allow for measurements over a much larger area, becoming sensitive to ultra-high-energy (PeV) photons whose rate per unit area is extremely low. For the sparse outer array, a rejection of lateral through-going particles can be still obtained with different solutions all connected to shielding the lower chamber of the tank (in case of double-chamber tanks) either by water or ground. That issue does not exist in the case of single-chamber tanks.

The two requirements of precision at low energy and flux maximization at high-energy are in tension with one another, and they already indicate that an optimization of the layout requires a precise definition of the relative scientific importance of the diverse objectives of the SWGO scientific program. We are not deterred by this complication, as we are familiar with similar situations in other branches of experimental physics, where compromises are routinely struck between a large number of conflicting demands. 
The classical example of this situation is the trigger menu of a hadron collider experiment, where the limited bandwidth available for event storage calls for a very high rejection factor (typically of the order of $10^4$); the acquisition of events of interest for different physics measurements and searches is thus subjected to restrictions, based on a continuous reassessment and difficult decisions on the rate budget of each channel of interest.

 The conflicting requirements mentioned above have led the SWGO collaboration to define a set of different benchmark layouts that share as a common feature a dense core and an extended region more sparsely populated, as shown in \autoref{f:layouts}. Investigations are ongoing by the collaboration to evaluate in quantitative terms the relative benefit of the proposed layouts for the various astrophysics cases of interest to the collaboration~\citep{BarresdeAlmeida:2021xgv}. 

\subsection{Full simulation datasets}
\label{sec:swgo_sim}

 Atmospheric showers may be accurately generated via the publicly available \texttt{CORSIKA} software~\citep{corsika}.  \texttt{CORSIKA} simulates the full development of particle showers, and, once the detector elevation above sea level is fixed, stores all ground particle information including particle position, type, momentum, direction and direction. This step of the simulation is quite CPU- and storage-expensive, with increasing time and storage required as the primary energy increases. The simulations we employed to obtain the parametric model discussed {\it infra} (\autoref{s:methodology}, \hyperref[sec:appendix]{Appendix}) are detailed in \autoref{t:corsika} below. In all cases listed there, the detector altitude has been fixed at 4,800\,m above sea level, which roughly adapts to the sites under consideration. We make use of a large simulation of showers with energy between \SI{0.1}{\peta\eV} and \SI{10}{\peta\eV} we produced at the Italian CNAF data center in Bologna. The choice of the energy range reflects our interest to study the highest-energy gamma rays, yet the purpose of this initial study is to prove an end-to-end approach to the problem rather than to produce usable results --consistently with the fact we noted above, {\it i.e.}, that a complete answer to the optimization question may only come after external constraints and scientific goals are defined and agreed upon.

\begin{table}[h!]
\begin{center}    
\begin{tabular}{lcccc}
 & ID & Energy & Angle & Events \\ 
\hline
1-8 & $\gamma$  & 0.1/0.3/1/3 PeV & \SI{20}{\degree}/\SI{40}{\degree} & 10,000 \\
9 & $\gamma$  & 0.1-10 PeV & \SI{0}{\degree}-\SI{65}{\degree} & 100,000 \\
10-11 & $p$ & 0.1-10 PeV & \SI{20}{\degree}/\SI{40}{\degree} & 200,000 \\
12 & $p$ & 1.0-10 PeV & \SI{0}{\degree}-\SI{65}{\degree} & 200,000 \\
\end{tabular}
\caption{\texttt{CORSIKA}--produced datasets of proton and photon showers used in the present work.}
\label{t:corsika}
\end{center}
\end{table}

%Once ground particles are obtained, the next step is to consider their interaction in the detectors constituting the array. Within the SWGO collaboration, this is achieved with a proprietary software adapted from the HAWC collaboration. This software extract from the CORSIKA sample those particles that actually hit the tank, computes the Cherenkov light within each tank and save the energy, time and flux of Cherenkov photons at the tank photosensor(s). As mentioned above, we assume a total efficiency for Cherenkov light detection within the tank unit. For this reason, we limit ourselves to use the tanks position and extension in order to compute the event rate per tank. This is done with custom Python codes.

\section{The modeling pipeline } 
\label{s:methodology}

 In this section we describe in a block-like fashion the software pipeline that allows for the end-to-end modeling of the SWGO experiment; this includes functional blocks that model the detector and the physical processes responsible for data generation, modules that enable the reconstruction of meaningful low-dimensional summaries of the observed data, and ones that evaluate the experiment sensitivity and utility.

\subsection{SWGO tank pattern on the ground}

For the purpose of this study, we consider a set of $N_\text{det}$ water Cherenkov detectors of type~A (see \autoref{sec:tanks}), all of the same characteristics: three-meter-tall cylinders of radius $R_{tank}=1.91$ m with axes oriented along the vertical $z$ direction, filled with purified water and instrumented with light-sensitive photo-detectors in an arrangement capable of effectively discriminating light signals produced by the soft component (photons, electrons, and positrons) from signals produced by muons. 

In order to simplify the geometry optimization problem, we further decouple it from the detector optimization issue by operating a number of assumptions. We assume that each tank is capable of detecting with \SI{100}{\percent} efficiency particles hitting their top cross-sectional area with kinetic energy in excess of a pre-set energy threshold, thus freeing ourselves from the complex task of suitably modeling the direction of secondaries on the ground as well as modeling their detection efficiency. We also assume that each detector can measure the number of electrons or positrons, and separately the number of muons, with an uncertainty of $5\%$.
%\mdc{Is this in contradiction with p=1 sentence in the previous section?}. 
Finally, a time resolution of $10$~ns is assumed on the signal of particles detected by each unit. With the above assumptions, the set of detecting units is specified solely by the position of the center of each unit on the ground, ${(x_i,y_i)},\;i=1...N_\text{det}$, and by their radius $R$. 

By simulating a large number of showers and reconstructing their parameters, it is possible to mimic a full-fledged measurement under each of the hypotheses for the SWGO layouts of \autoref{t:layouts_sec2}, obtaining a relative appraisal of the different configurations. This task, which entails considering up to 13 different designs, is a straightforward exercise to carry out once a reconstruction method and a utility function are specified; we provide results based on our model in \autoref{s:benchmarks}. The more principled question, however, concerns the investigation of the much higher-dimensional space of {\it all possible ground configurations of $N_\text{det}$ detectors}. Since a detector position is specified by its ground coordinates ${(x_i,y_i)},\;i=1...N_\text{det}$, the space of configurations lives in ${\mathcal{R}}^{2N_\text{det}-3}$. So, {\it e.g.}, the layout of three units involves the choice of three real parameters: by setting the first unit in $(0,0)$, the second can be set at $(x_2,0)$ without loss of generality; the third can then be specified by $(x_3, y_3)$, when $y_3$ can be chosen in the positive semi-axis without loss of generality. With over 6000 units to be deployed, the space of possibilities grows prohibitively large and becomes impossible to probe with discrete sampling methods.

\subsection{Parametrization of the particle flux on the ground}

The task of scanning the large parameter space in a continuous way with gradient descent techniques can be attempted if we endow ourselves with a parametrization of the flux of the different secondary particles on the ground 
$dN_{i}/dR(E,\theta, \phi)$ (with $i=e, \mu, \gamma$ for electrons, muons, or photons)
%$dN_{e,\mu,\gamma}/dR$ 
as a function of the distance from the shower axis, resulting from a cosmic ray shower of given energy $E$, polar angle $\theta$, and azimuthal angle $\phi$. Ideally such parametrization should be available in closed form for both gamma-- and proton--originated showers (the dominant background to discriminate against), and for the particle types to which the detectors are differently sensitive: electrons, positrons\footnote{Electrons and positrons can be to first approximation considered together, due to their similar behavior in interacting with water at energies much above the critical energy $E_\text{crit}$, which is of $\SI{76.2}{\mega\eV}$ for positrons and $\SI{78.3}{\mega\eV}$ for electrons \citep{PDG}.}, photons, and muons. An analytical description of these dependencies is reviewed by \citet{Greisen:1960wc}. As EAS arising from high-energy primaries involve a number of complex stochastic processes, precise parametrizations are difficult to produce; yet even a mildly inaccurate choice may be able to successfully inform a continuous scan of the parameter space, if a full simulation is then used to validate the results in the vicinity of configurations identified as optimal by the approximate model.

We use \texttt{CORSIKA}
%~\citep{corsika} 
simulations of gamma and proton showers at very high energies (from \SI{100}{\tera eV} to \SI{10}{\peta\eV}) described in \autoref{sec:swgo_sim} to extract a model of the density of secondary particles on the ground as a function of their distance from shower axis \footnote{Our choice of initially focusing on the highest-energy range of detectable gamma rays reflects both our scientific goals and our interest in probing the least-understood part of the radiation spectrum; in addition, at high energy the discrimination problem becomes tougher, and the showers footprint is large; these two factors make the choice of the array layout more complex and rich.}. The resulting model is the basis of the optimization pipeline which is described in what follows. 

\begin{figure*}[h!t]
\begin{center}
    \includegraphics[width=0.95\linewidth]
    {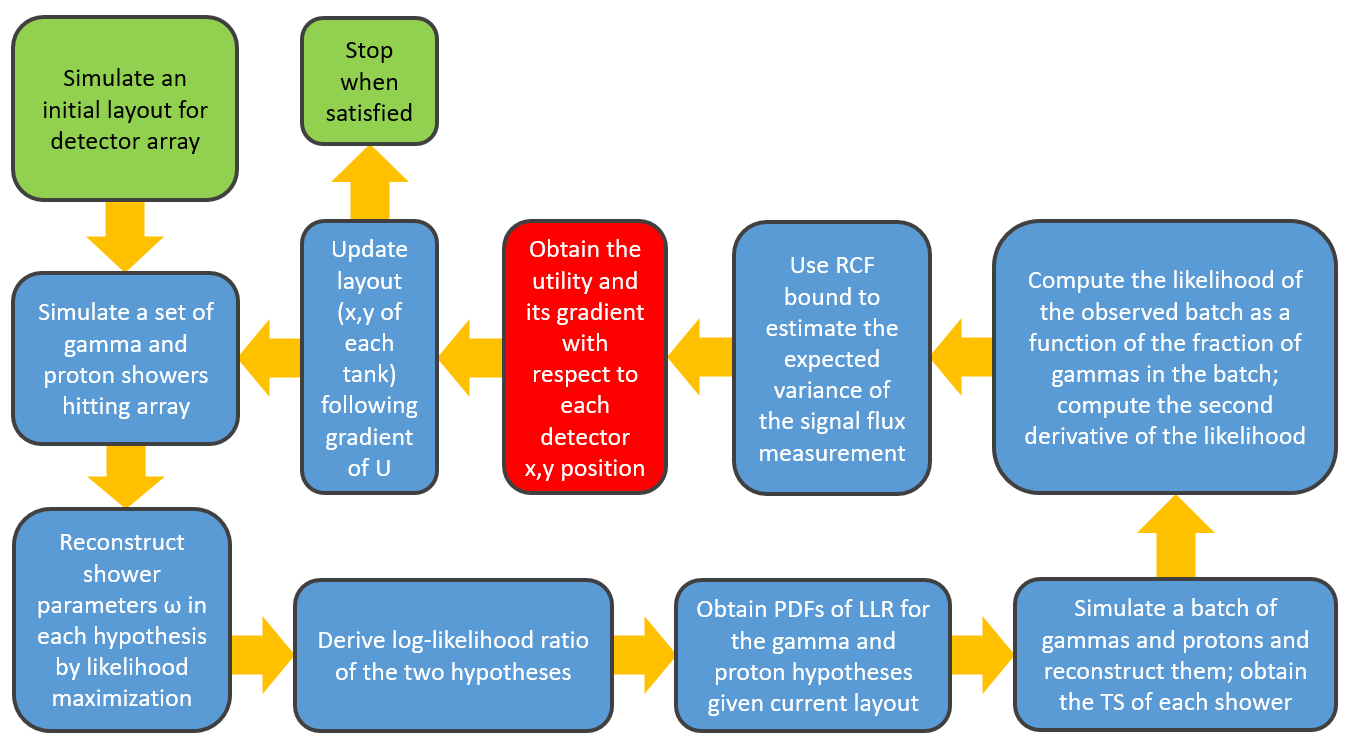}
    \caption {Block diagram for the optimization of the SWGO detector array. }.
    \label{f:pipeline}
\end{center}
\end{figure*}

\subsection {The optimization pipeline}

\begin{enumerate}
    \item The starting point is the generation of an initial ground configuration of $N_\text{det}$ detector units,  
    ${(x_i,y_i)}_{i=1,...N}$. These $2N$ parameters can be defined by choosing one of a number of pre-definite layouts, which depend on a few shape parameters (spacing in $x,y$ or radius) that may be specified by the user.
    \item A set of gamma and proton showers are simulated by considering an intersection of the shower axis with the ground at a position $(X_0,Y_0)$ randomly chosen within a region extending wide away on all sides from the initial footprint assumed for the detector tanks, such that showers at the edge of this region have a negligible probability of being detected by the array. We find that if meaningful triggering requirements are enforced ({\it e.g.}, that at least 30-50 tanks are hit by one or more particles), \SI{2.0} km of radial extension away from the detectors for the generated shower centers is more than sufficient (see \autoref{f:pvsdande}) even for the highest-energy showers we considered. Given shower energy $E$ and polar angle $\theta$, the position of the shower center on the ground, together with the azimuthal angle $\phi$ defined as the angle between the horizontal component of the shower axis and the positive $x$ direction, determines through our parametrization (\autoref{s:model}) the expectation value of the number of particles of different kinds ($\lambda_\mu, \lambda_e$) that will hit the sensitive area of each of the $N_\text{det}$ detector units. 
\item The number of particles seen by each detector can be sampled from Poisson distributions of means equal to the expectation values of the fluxes at the detector position on the ground, optionally after the application of a further Gaussian smearing that mimics detection resolution effects
%\footnote{For our initial tests we set the default value of the smearing $\sigma_R$ at $5\%$.}
: {\it e.g.}, for the number of detected muons we have $N_\mu = Poisson[G(\lambda_\mu,\sigma_R \lambda_\mu)]$, where as stated {\it supra} we set $\sigma_R=5\%$. More detail on the procedure is offered in \autoref{s:reconstruction}.
    \item Through a likelihood maximization, estimates of the shower parameters are obtained by exploiting an assumed full knowledge of the flux model parameters, and under the assumption that tanks offer perfect discrimination power between the various particle species. The likelihood is maximized in turn under each primary hypotheses (gamma and proton) for each shower, regardless of the true primary particle species.
    \item The obtained likelihood values, maximized over shower parameters E, $\theta$, $\phi$, $X_0$, and $Y_0$, are used to compute a likelihood ratio test statistic $\mathcal{T}$  for each generated shower, \par
    
    \begin{equation}
        \mathcal{T} = \frac{\log L_\gamma (\hat E_\gamma, \hat \theta_\gamma, \hat \phi_\gamma, \hat {X_{0,\gamma}}, \hat{Y_{0,\gamma}})}{\log L_{p}(\hat E_p, \hat \theta_p, \hat \phi_p, \hat {X_{0,p}}, \hat{Y_{0,p})}}
    \end{equation}

    \noindent
    where the hat symbol on the parameters indicates that the likelihoods are evaluated at the set of parameters that maximize the considered hypothesis. The sampling of a large number of showers of different parameters allows to construct a probability density function (PDF) of $\mathcal{T}$  for both hypotheses, by substituting each value with a Gaussian kernel of sigma equal to the uncertainty on $\mathcal{T}$ estimated with the Rao-Cramer-Frechet (RCF) bound~\citep{rao}.
    \item A new batch of gamma and proton showers is generated, and a distribution of $\mathcal{T}$  values obtained.
    %\footnote {For the determination of the utility and its gradients it is necessary to generate a different set of showers from that used for the PDF evaluation, for reasons explained in more detail {\it infra}. }. 
%% CHECK WHERE THIS IS DONE!    
    The distribution of $\mathcal{T}$ is then fit with a two-component mixture model, to extract the fraction of gamma-rays in the batch as well as its uncertainty --the latter again estimated using the RCF bound.
    \item Using the uncertainty on the gamma fraction, a flux-related utility function term is computed to mimic the goals of the experiment  (see \autoref{s:utility}). A simple proxy to sensitivity of the detector array to signals at high energy may be a properly weighted sum of the inverse of the relative uncertainty on the flux of gamma-rays in a batch of showers, where the batch represents the data collectible in a given time interval; pointing and energy resolution terms can also be added to define a global utility function, or directly combined in an ``integration-time-for-5-sigma" estimate of the discovery sensitivity for hypothetical monochromatic point sources of photons in the sky (see \autoref{s:utilityps}). 
    \item A propagation of derivatives then allows to extract the derivative of the utility function over the set of $2N_{det}$ displacements $(\delta x_i$, $\delta y_i)\;i=1,...,N_\text{det}$. These are used to update the detector positions.
    \item The cycle can be repeated by generating a new set of gamma and proton showers and deriving a new PDF of $\mathcal{T}$ (a necessary step as the position updates affect the distribution of $\mathcal{T}$), then fitting a new batch of showers, recomputing the utility and its gradients, and iteratively updating detector positions.
\end{enumerate}

The software pipeline described above is summarized in \autoref{f:pipeline}. The algorithm converges to layouts that optimize the utility function. Of course, in its present form the program still suffers from approximations that may compromise the value of its results, awaiting the definition of more complex utility functions than the ones mentioned above, to model the real objectives of the  core scientific program of the SWGO collaboration.  Then, after a final decision will be taken on the design of the detector units, a parametrization of the detection of particles of different species, a non-perfect detection efficiency, and more realistic resolution terms could be included in the pipeline, improving the precision of the model. In any case, the true relative gains of optimized layouts over baseline ones may be appraised with precision at any stage, by exploiting a full simulation of atmospheric showers and of the detection of secondaries by the detector units. 

 In the following sections we describe the modeling of each functional block in the pipeline we summarized above.

\section{ Definition  of the utility function}
%and \textcolor{red}{calculation of} its derivatives}
\label{s:utility}

 As discussed above, a true end-to-end optimization may only result from the consideration of every ingredient turning the physical processes of interest into the final result of the measurement. One crucial part of this program is played by the precise definition of the quantity that measures the success of the measurement --the utility function. In this section we  consider the problem of writing an expression for the global utility, {\it i.e.} a function which might serve as a proxy to the scientific goals of the SWGO experiment. It would be naive 
to assume we can encode into a single number the vast range of measurements and searches that the collaboration is planning without first staging a thorough discussion within the collaboration. However, the goal of this document is to show how such a number may successfully guide the design of the array. With that in mind, we discuss a couple of possible choices, which may be used in alternative as a crucial ingredient of the optimization pipeline we aim to test.  At the end of this section we also add a discussion of our modeling of two possible additional terms connected to the geometrical arrangement of units on the ground and the total area they cover, which that may take part in determining the monetary cost and the construction effort of the experiment. 

\subsection{A three-component utility definition}

 A considerable scientific interest in astroparticle physics today is offered by the recent discovery of 12 PeV emitters in the northern hemisphere by the LHAASO collaboration~\citep{LHAASO}, where few or none were expected. Therefore,  strong expectations are placed for an instrument such as SWGO to detect further PeV emitters in the southern hemisphere. The measurement of the gamma flux in the \SI{100}{\tera\eV} to \SI{10}{\peta\eV} region is therefore taken as a good initial bid for one of the main goals of the SWGO experiment in this document. We proceed to define the following utility function: \par

\begin{equation}
    U_{GF} = \frac{\Phi_\gamma}{\sigma_{\Phi} \sqrt{\rho}},
    %\int_{0.1}^{10} \left[ \frac{1}{\sigma_{\Phi_{\gamma}}} %\frac{d\Phi_{\gamma}}{dE} \right] dE 
\label{eq:U1}
\end{equation}

\noindent
where $\Phi_\gamma$ is the gamma-ray flux, $\sigma_{\Phi}$ its uncertainty, and  $\rho$ a density factor that will be discussed below. The function above is an integrated measure of the precision in the determination of the gamma-ray flux $\Phi_{\gamma}$ in the (0.1 PeV, 10 PeV) range; it could optionally be subdivided into pieces that describe the relative value of a precise flux measurement within smaller energy intervals. Such a fine-tuning is straightforward, and has been neglected at present.
We anyway notice how the choice of the coefficient in the power law that describes the energy distribution of the generated showers effectively plays the same role of a weighting factor: a large negative value of the power corresponds to giving all the importance to the low-energy end of the sample, and a large positive value instead privileges the contribution of high-energy showers. For most of our studies we kept the energy distribution of generated showers flat in energy, %\mdc{flat you mean independent on the energy? most targets can be roughly modeled as a power law with spectral index -2/-3}, 
in order to explicitly avoid any bias of that kind in the generated solutions; this choice must be kept in mind at all times when interpreting the results. 

The density factor $\rho$ in the expression above keeps track of the number of showers that were generated at each step during the optimization loop: following a move of the detectors, the area on the ground that must be illuminated with shower centers must vary; if a fixed number of showers is instead used, the density factor accounts for the effect. We note that while a scaling of the flux uncertainty at the denominator of $U_{GF}$ with $1/\sqrt{N_\text{batch}}$ is a reasonable assumption, it would be better to avoid it altogether, and to generate at every epoch during optimization a sufficiently large flux of showers in a fixed region surrounding the array wide enough to allow detector units to move around without seeing a decreased flux. In the future we do plan for longer runs of the pipeline that may accommodate this safer but more CPU heavy mode of operation; results presented in this document are instead making use of the above assumed scaling. The utility definition thus contains a factor $1/\sqrt{\rho}$, where the density is computed per batch. 

Correspondingly, at each epoch the area where we generate showers is determined by the following procedure:\par
\begin{enumerate}
    \item The average radial extension $\overline{R}$ and RMS $\sigma_R$ of the array are evaluated\footnote{Although a cylindrical symmetry is present in the problem, we do not assume it in our work; regardless, having set the initial array in a region centered at $(0,0)$ coordinates in the $(x,y)$ plane the radial extension from that point remains a good measure of the overall scale of the array during any optimization loop.};
    \item a constant $R_\text{slack}$ (set to \SI{2.0}{\kilo\meter} by default\footnote{The use of the average radial extension of the array and its RMS may cause, in later stages of an optimization run, units that have drifted to high radius to ``see'' a smaller than \SI{2.0}{\kilo\meter} radius of generated showers around them. This is a marginal effect and an acceptable one for the exploratory runs we made, and can also be considered an implicit limit in the radial extension we want to consider for the array. Using the average array radius instead of the radius of the outermost unit avoids issues with the differentiability of the utility.}; see \autoref{s:triggering}) is added to $\overline{R}$;
    \item shower cores are generated at a random position (with constant density in $x,y$) within a circle of radius\par 
    
    \begin{equation*}
        R_\text{tot}=\overline{R}+2\times \sigma_R +R_\text{slack}
    \end{equation*}
    
    from the origin of the axes on the detector plane.

\end{enumerate}

\noindent
Core positions generated at a distance exceeding $R_\text{slack}$ from any of the array detectors are discarded and new ones are generated, such that the illuminated region on the ground only includes core positions with a distance no larger than $R_\text{slack}$; 
    %\mdc{I think newly generated is confusing};
The number of total generated core positions (including discarded ones) $N_\text{trials}$ is then stored; the procedure stops when $N_\text{batch}$ core positions are obtained. Energy, polar and azimuthal angle of incidence, and identity of each shower are then assigned at random following their assumed prior.

At the end of the above procedure, which is repeated at every epoch during optimization, the area uniformly covered by the $N_\text{batch}$ showers is computed as:
\begin{equation}
    A = \pi R_\text{tot}^2 \frac{N_\text{batch}}{N_\text{trials}}.
\end{equation}

\noindent
The density of showers per unit area on the ground in a batch consequently scales (omitting the $\pi$ term) with:
\begin{equation}
    \rho \propto \frac{N_\text{trials}}{R_\text{tot}^2}.
    \label{eq:rho}
\end{equation}

An important additional ingredient in a global utility function is the precision of the estimate of relevant parameters of gamma-originated showers: energy, as well as polar and azimuthal angle of incidence of the primary particle on the upper Earth atmosphere. On the contrary, the two parameters describing the core position on the ground ($X_0,Y_0$) should be considered as nuisance parameters in the estimation problem: they contain no physical information on the primary flux. 

A suitable function sensitive to the overall energy resolution ($IR$, for integrated resolution) achievable for gamma-ray showers with a given detector geometry is the following:\par

\begin{equation}
U_{IR} = \frac {\sum_{k=1}^{N_\text{batch}} P_\text{tr}(k) w(k)} {\sum_{k=1}^{N_\text{batch}} P_\text{tr}(k) w(k) \frac{\hat{\sigma}_{E_k}}{E_{k,t}}}.
\label{eq:uir}
\end{equation}

\noindent
where the $P_\text{tr}(k)$ terms are non-zero only for gamma-ray showers. Above, the $\hat{\sigma}_{E_k}/E_{k,t}$ term in the sum at the denominator of the $U_{IR}$ function has the effect of letting $U_{IR}$ grow when overall the energy of showers is measured more precisely. The choice of averaging the relative energy uncertainty rather than the absolute uncertainty implies that as a baseline the $U_{IR}$ function defined above gives equal value to low- and high-energy shower reconstruction --a choice which has of course a significant impact in the optimization task. While keeping that factor fixed, the weighting factor $w(k)$ may instead be used to give more importance to the precise energy reconstruction of higher- or lower-energy showers, as already noted above. The weight $w(k)$ has the following form %\mdc{WhY this form?}: 
\begin{equation}
    w(k) = 1 + \omega \ln(E(k)/E_\text{min})
\end{equation}

\noindent
with $E_\text{min}=\SI{0.1}{\peta\eV}$, the lower range of the considered shower energies. For simplicity, in the model used for the results presented in this work we fixed $\omega = 0$, removing any dependence on shower energy of the contributions to the integrated resolution utility. 

A further comment on the definition of $U_{IR}$ is in order. In \autoref{eq:uir} we use an estimate of the uncertainty of the energy measurement, but divide it by the true shower energy in the ratio $\hat{\sigma}_{E_k}/E_{k,t}$. The reason for this choice is to focus on the magnitude of the former (which depends on the particular layout that measures the showers) and avoid the smearing of random fluctuations in the estimate of the latter (which carries no information on the layout): this improves the value of the gradient of the $U_{IR}$ function in pointing toward more advantageous configurations of the detector units. Finally, we mention that the choice of averaging $\sigma_E / E $ at the denominator of $U_{IR}$ rather than averaging $E/\sigma_E$ at the numerator is driven by the need to avoid large occasional contributions arising from abnormally small estimates of the energy uncertainty through the RCF bound.

A third factor can be considered in the global utility if we wish to endow it with sensitivity to the precision of the pointing resolution (PR) that the detector obtains on gamma-ray sources. This third factor, $U_{PR}$, can be constructed with the residuals of measurements $\theta_k$, $\phi_k$ of polar and azimuthal angle of incidence of the primary gamma ray. We define it as\par

\begin{equation}
   U_{PR} = \frac 
{\sum_{k=1}^{N_\text{batch}} P_\text{tr}(k) w(k) \frac{\delta_{min}}{\Delta R}}{\sum_{k=1}^{N_\text{batch}} P_\text{tr}(k) w(k)} 
\end{equation}

\noindent
where we compute the angular deviation\footnote{Angle differences are duly corrected to avoid periodicity issues.} $\Delta R$ as \par

\begin{equation}
\Delta R = [(\theta_t-\theta_m)^2 + (\phi_t-\phi_m)^2 + {\delta_{min}}^2]^{0.5}
\end{equation}

\noindent
and let $\delta_{min}$ be a small regularization parameter (set to $\delta_{min}=0.001$) that avoids instabilities in the denominator of the expression. We found this definition of a pointing resolution easier to use and more stable than one based on estimated uncertainties on the angular shower parameters in analogy with \autoref{eq:uir}.

The three factors $U_{GF}$, $U_{IR}$, and $U_{PR}$ can be used in alternative, or combined in a multi-objective utility, once suitable multipliers assessing their relative value are chosen. We have experimented with different combinations, but for our preliminary investigations --whose purpose is mainly to verify the functioning of the software-- and the demonstrative results presented in this work we have used the following definition:
\begin{equation}
    U_1 = \eta_{GF} U_{GF} + \eta_{IR} U_{IR} + \eta_{PR} U_{PR}.
    \label{eq:U}
\end{equation}

\noindent
Without loss of generality we may set the first of the three multipliers to unity. We observe that a choice of $\eta_{IR} = 0.2$ and $\eta_{PR}=0.0008$ provides in the performed tests an approximate balancing of the values of the three components of the utility, considering the excursion that they may withstand due to the optimization of $U$. This choice is totally arbitrary at this point, but it has the merit of allowing us to study the interplay of the three components in the optimization loops without the variation of any one of them dominating the others.

%While the $U_{IR}$ and $U_{PR}$ factors are to first order independent on the number of showers that are generated on the ground at each epoch, and are only sensitive to the maximum distance from the detectors that is illuminated by the generated flux of primaries (the larger this distance, the worse the reconstruction of showers), the $U_{GF}$ factor clearly depends on the density of generated showers per unit area. When we wish to retain the possibility of comparing results that employ different numbers of showers per batch, and configurations that have arbitrary areas covered with detectors, a correction is necessary. The correction is provided by the already discussed factor  $1/\sqrt{\rho}$ (see \autoref{eq:U1}), where $\rho$ is the number of generated showers per batch per unit square kilometer. The factor compensates for the variance of the flux estimate, as the latter scales with the Poisson mean. 

\subsection {An alternative utility definition \label{s:utilityps}}

The definition of the $U_1$ function discussed above (\autoref{eq:U}) requires the user to arbitrarily specify the relative importance of the three pieces of the utility. While such a feature gives flexibility to the function, which may adapt to different experimental demands, it seems useful to show of how a more unified definition for an experiment-wide utility can be constructed. We thus produced a formula that summarizes into a single number the overall capability of the array to measure transient, localized phenomena in the sky. We imagine the performance of the instrument as directly connected with its capability to discover a point source of gamma rays in the sky, when the gamma energy is fixed to a specific value. While being an idealization, this setup is simple to define and it allows to study how the optimality condition depends on the energy of the source.

We consider the following situation: a point source emitting gamma rays of energy $E_{PS}$ from angular position $\theta_{PS}$, $\phi_{PS}$ is sought by considering a signal window extending in angle in the range$(-1.4\sigma_{\theta_{PS}},+1.4 \sigma_{\theta_{PS}})\times(-1.4\sigma_{\phi_{PS}},+1.4 \sigma_{\phi_{PS}})$ and with energy in the range $(-1.4 \sigma_{E_{PS}},+1.4 \sigma_{E_{PS}})$\footnote{The choice of windows of width equal to 2.8 times the resolution in each measured parameter optimizes the significance of a Gaussian signal on top of a flat background.}, where the angular resolutions $\sigma_{\theta_{PS}}$, $\sigma_{\phi_{PS}}$ and the energy resolution $\sigma_{E_{PS}}$ are computed on a shower-per-shower basis. A ``sideband'' background-estimation region instead extends to an angular region of the sky spanning an area $10\times 10$ times larger around the signal box; background gamma rays are considered if they have energy compatible to that of the sought-for source.

Under the above conditions, we have a classic on/off problem where the statistical significance of the observed signal can be expressed in a simplified way by considering the sideband region of the sky as a source of prediction of the expected background in the signal region. A five-sigma pseudosignificance measure to such a counting experiment can then be simply expressed by the approximate formula \par

\begin{equation}
    Z = 2[\sqrt{B+S} - \sqrt{B}]
\end{equation}

\noindent
where $S$ and $B$ represent the expected number of event counts due to the signal and background sources inside the signal region, respectively.
By setting $Z=5$ in the above expression we obtain a requirement for the minimum number of gamma rays observed in the signal region $N_{PS}=S$: \par

\begin{equation}
    N_{PS,5\sigma} \geq \frac{25+20\sqrt{B}}{4}.
\end{equation}

\noindent
While approximated, the above formulation has the advantage of being easily differentiable and to offer a valid proxy to the sensitivity of the instrument. We may thus define a "point-source" utility function as \par

\begin{equation}
U_{PS} = \frac{k} {N_{PS,5\sigma}}
\end{equation}

\noindent
where the factor $k$ tracks the exposure time provided by the considered number of showers generated for the batch, and is equivalent to $k=1000/\sqrt{\rho}$ with the density factor $\rho$ defined in \autoref{eq:rho}.

\subsubsection {Data generation for the point source utility case}

Given the above discussed restrictions on the energy range of interest for the idealized search of a monochromatic point source, the generation of $N_{ev}$ showers used for the test statistic determination and of $N_{batch}$ events for the utility and gradients evaluation is modified to only consider showers with primary energy suitably covering the interval of interest. We reduce the range of generated energies to the interval $(0.5E_{PS}, 1.5 E_{PS})$, which corresponds to allowing for consideration of showers whose true energy is under- or overestimated by at most 50\%. The user may then specify a desired point source energy (in the range $(0.2, 6.66)$ PeV, such that the generated range mentioned above does not exceed the (0.1 PeV, 10 PeV) bound) to investigate different experimental situations. 

As far as the polar angle of showers is concerned, when running with the point source utility we do not need to modify its distribution --consistently with an assumption that a source in the sky may have any observed location, so that the same solid-angle-driven sampling distribution in polar angle used for the standard utility can be used. The code also allows the user to optionally switch from generating showers with a $\sin \theta$ distribution to generating them with a uniform distribution, which corresponds to assuming that the source crosses the zenith of the observatory in its daily motion in the sky, {\it i.e.} to assuming that the source is located at the same latitude of the observatory. This can be used to check how strongly does source location affect optimality conditions on the layout.

\subsection {Penalty terms for area and length  \label{s:penalty}}

A larger area instrumented with detection units corresponds to a higher construction cost, because of a number of factors. Even assuming the absence of physical obstructions (mountains, rivers, lakes, inhabited areas, civil roads, {\it etcetera}) one must consider the cost and availability of land, the cost of leveling the area out, the cost of removing rocks or other hindrances, and last but not least the cost of creating necessary infrastructure (transportation roads for construction and maintenance). In addition, the units must be connected to a counting room by cabling to transfer there the electronic signals, and (at least, in the simplest electrification scheme) to receive power. A fence delimiting the perimeter of the instrumented area can also be envisioned. Because of all the above, a true optimization of the layout of the SWGO array must carefully include a dependence on the instrumented area in the utility, by adding a cost penalization to designs that extend the area of the array to larger values, or that foresee a large distance of tanks from a common center.

We implemented two factors that can be used in conjunction with any combination of the above utility pieces. They subtract a penalty to the value of the utility, thereby discouraging the system from expanding into arbitrarily large areas.

The first factor is called ``area cost", and it uses an estimate of the area instrumented with detector units. This is not the physical area occupied by tanks, which is always equal to $\pi R_{tank}^2 N_{det}$, but rather the area of the convex hull that includes all the units. While the convex hull is potentially an overestimate of the actual area required to construct a given array (as, {\it e.g.}, it ignores the possible absence of detectors in a central region, which many of our optimized layouts prefer), it is a good representation of the monetary cost of an extended layout.

We compute the convex hull area with a standard algorithm \citep{convexhullalgo}, and use it to determine the area cost as \par

\begin{equation}
    U_{TA} = \max \left[-\eta_{TA} (e^{(A-A_0)/A_0}-1), 0. \right]
\label{eq:areacost}
\end{equation}

\noindent
where $A$ is the current hull area, and the default value of $A_0$ is set by default to be the area of the A7 array (which has a radius of 1.2 km). This term is therefore null until the hull area exceeds about 4.52 square kilometers. The inclusion of the $U_{TA}$ penalization is sufficient to dampen layouts that exceed the area of the largest among the 13 SWGO benchmark configurations, when $\eta_{TA}$ is chosen to have its default value.

The second factor, which may be used in combination with the first or by itself, uses as a metric of cost the total distance of the $N_{det}$ units from the center of the array,\par

\begin{equation}
    L_{tot} = \sum_{i=1}^{N_{det}} (x_i^2 + y_i^2)^{0.5}
\end{equation}

\noindent
In analogy to $U_{TA}$, a cost related to the distance from the array center can be defined as\par

\begin{equation}
    U_{TL} = \max \left[ -\eta_{TL} (e^{(L-L_0)/L_0}-1),0. \right]
\end{equation}

\noindent
where this time the default value of $L_0$ is chosen to be the total length of a uniform arrangement of $N_{det}$ units in a circle of 1.2 km radius (again, the radius of A7)\footnote{Note, however, that this is not the total distance of the A7 units from the center of that array, because A7 has a non-uniform distribution of units in the instrumented area.}.

The terms $U_{TA}$ and $U_{TL}$ are differentiable and their gradients with respect to the units coordinates are straightforward to obtain. Their inclusion in optimization loops increases the realism of the utility model.
\section{Shower parametrization and reconstruction}
\label{s:reconstruction}

 In this section we describe the ingredients of the modeling pipeline described in \autoref{s:methodology}. In \autoref{s:model} we provide details of the surrogate model we developed for gamma and proton showers; we next discuss how shower parameters can be extracted from the data collected by the detectors on the ground in \autoref{s:reconstruction}. This is followed by a description of the likelihood function which is used to derive a test statistic discriminating gamma-ray from proton showers in \autoref{s:likratio}. The section concludes with an appraisal of the reconstruction model, and a discussion of how the triggering probability is modeled by the software. All these ingredients require a precise formulation as they will be used in later sections to construct an analytical description of the derivatives of the utility function. 

\subsection{Continuous parametrization of secondaries from gamma and proton showers \label{s:model}}

\noindent
 We use the full simulation datasets of EAS mentioned in \autoref{s:methodology} to extract a simple parametric model that determines, as a function of the radius from the shower axis and at an altitude of \SI{4800}{\meter}, the density of secondaries of different species possessing momenta above a suitable threshold, given a primary particle identity (proton or photon), energy $E$, and polar angle $\theta$. For the time being we do not focus on a precise modeling of the individual detection of secondary particles in the tanks, and we therefore choose a low energy threshold (\SI{10}{\mega\eV}).
This  conforms  to the  novel  strategy we pursue in this work: the one of initially setting near-perfect detection conditions when studying geometric effects; those  conditions can then be revisited after an initial exploration of the layout has been performed. The rationale of such a {\it modus operandi} is to not allow assumptions on the limitations of  any given  choice for the hardware solutions to affect the search for the best configuration: any initial limitation in the hardware performance constitutes a reduction of the phase space of theoretically achievable solutions; the reduction cannot then be recovered by subsequent refinements. 
%\mdc{I don't understand how to reason for the previous sentence.}

%\mdc{I am not sure this paragraph is really needed.} 

We further simplify our modeling task by  lumping together 
 the fluxes of electrons, positrons and photons, by reasoning that their signal in the Cherenkov tanks is not distinguishable with the considered detector designs and their radial distribution is also anyway very similar. We thus produce two sets of distributions for electromagnetic particles overall, depending on primary identity ($p$ / $\gamma$), energy and polar angle; we sum together positive and negative muon fluxes into a corresponding second set of distributions. 

For modeling purposes we initially divide the primary energy into 20 bins equally spaced in the logarithm of energy, in the \SI{0.1}{\peta\eV} - \SI{10}{\peta\eV} energy range, and primary polar angles into 4 bins in the 0 - 65 degrees angle. There are thus 320 distributions: 80 for electromagnetic particles from gamma primaries, 80 more for muons from gamma primaries, and a total of 160 more of the same kinds from proton primaries. We omit to consider showers with a polar angle larger than 65 degrees, which are anyway not liable to be precisely reconstructed by ground-based detectors, due to extinction of the particle flux in the increased atmospheric thickness they encounter. %\md{I thinkg figs 6,7,8 are not readable in such small format. I propose to either overlap each pad into a single pad for each figure or b) select a pad among those or c) select a pad and add full figure into the appendix}

We obtain fluxes of particles on the ground as a function of the radial distance from the shower axis by projecting on the plane transversal to the shower axis their ground position, and normalizing their flux to particles per square meter. As mentioned above, this is done in 20 energy bins and 4 polar angle bins for each primary ($p$, $\gamma$) and for both classes of modeled secondaries (($e^+, e^-, \gamma$), ($\mu^+,\mu^-$)). The flux histograms are fit with the following functional forms: \par
\small
\begin{align}
\label{eq:flux}
    \frac{dN^{\mu,\gamma}(E_{\gamma} , \theta_{\gamma})}{dR} &= p_0^{\mu,\gamma}(E_{\gamma} , \theta_{\gamma}) \cdot \exp\left(-p_1^{\mu,\gamma}(E_{\gamma} , \theta_{\gamma}) R^{p_2^{\mu,\gamma}(E_{\gamma} , \theta{\gamma})}\right)
\\
    \frac{dN^{e, \gamma}(E_{\gamma} ,\theta_{\gamma})}{dR} &= p_0^{e, \gamma}(E_{\gamma} ,\theta_{\gamma}) \,  \exp\left(-p_1^{e,\gamma}(E_{\gamma} ,\theta_{\gamma}) R^{p_2^{e, \gamma}(E_{\gamma} ,\theta_{\gamma})}\right)
\\
    \frac{dN^{\mu,p}(E_{p} ,\theta_{p})}{dR} &= p_0^{\mu,p}(E_p ,\theta_p) \,  \exp\left(-p_1^{\mu,p}(E_p ,\theta_p) R^{p_2^{\mu,p}(E_p ,\theta_p)}\right)
\\
    \frac{dN^{e,p}(E_{p} ,\theta_{p})}{dR} &= p_0^{e,p}(E_p ,\theta_p) \,  \exp\left(-p_1^{e,p}(E_p ,\theta_p) R^{p_2^{e,p}(E_{p} ,\theta_p)}\right)
\end{align}
\normalsize

\noindent
The dependence of the sets of parameters $p_0$, $p_1$, $p_2$, and $p_3$ on $E, \theta$ for each combination of primary and secondary particle species are in turn obtained from a parametric model, such that all fluxes are continuous and differentiable with respect to radius, energy, and angle.
Details of the result of these fits and the functional forms modeling the variation of the four sets of parameters ${p_0, p_1, p_2}$ are provided in the \hyperref[sec:appendix]{Appendix}.
%\autoref{sec:appendix}. 
%Appendix. 
%\autoref{sec:appendix}. 
The typical precision in the flux estimate produced by the parametrization is better than \SI{10}{\percent} on the shower axis, and better than \SI{20}{\percent} at radii of \SI{1.5}{\kilo\meter} away from the shower axis, where particle fluxes have typically decreased down to a part in 100,000 or less of the central flux. For gamma-originated showers, fluxes of electrons and gammas (resp. muons) overall have a root mean square deviation from the model equal to  \SI{15.6}{\percent} (resp. \SI{18.9}{\percent}). For proton showers, standard deviations are of \SI{3.8}{\percent} for electron plus gamma fluxes (resp. \SI{8.4}{\percent} for muon fluxes). We deem this precision more than sufficient for the scope of the present work, which focuses on the optimization procedure and on the {\it derivatives} of the utility function rather than on the actual value of the utility of any configuration. 

\subsubsection{  Modeling of accidental coincidences} 

Although the very large flux of secondary particles from a primary gamma or hadron in the TeV or PeV energy range cannot be easily modified by accidental coincidences from the flux of other unrelated cosmic ray interactions, we need to account for the latter as it has the potential to affect the discrimination of gamma rays from hadrons, which in our surrogate model is mostly reliant on the dearth of muons from the former of the two sources.

We model that contribution by considering a spatially uniform, continuous flux of muons and electrons on the ground. The integration window within which particles are counted by the considered detectors has a width of 128~ns. SWGO plans to continuously acquire data from all detector units, reconstruct the most probable energy and direction of showers, and refine the reconstruction by matching the integration window to the expected arrival time of particles on the ground, assuming a flat shower front of secondaries. In our simplified modeling task, we are therefore interested in modeling the effect of that integration window. The flux of muons and electrons described in the preceding subsection is consequently modified to account for an additional rate of muons (electrons) equal to $18.3 \times 10^{-7}$ (resp., $2.0 \times 10^{-8}$) particles per square meter per nanosecond. Considering the $128$~ns integration window and the $A1$ tank specifications, this additional rate corresponds to $2.7 \times 10^{-3}$ particles (mostly muons) per unit per triggering window. Such a number seems at first of little relevance, until one realizes that with 6000 detectors the average number of units observing at least one muon from these accidental coincidences is 16.2. Given the very small signal to noise ratio of primary gammas to hadrons at high energy, and the reliance on the absence of muons in gamma-originated showers, this rate of accidentals is not irrelevant and it must therefore be considered in a modeling effort, as we do here.

\subsubsection{ Area of generated showers} 

\begin{figure*}[h!t]
\begin{center}
%\includegraphics[width=0.85
%\linewidth]%{Figures/Ptrigger_gt10.png}
\includegraphics[width=0.9\linewidth]{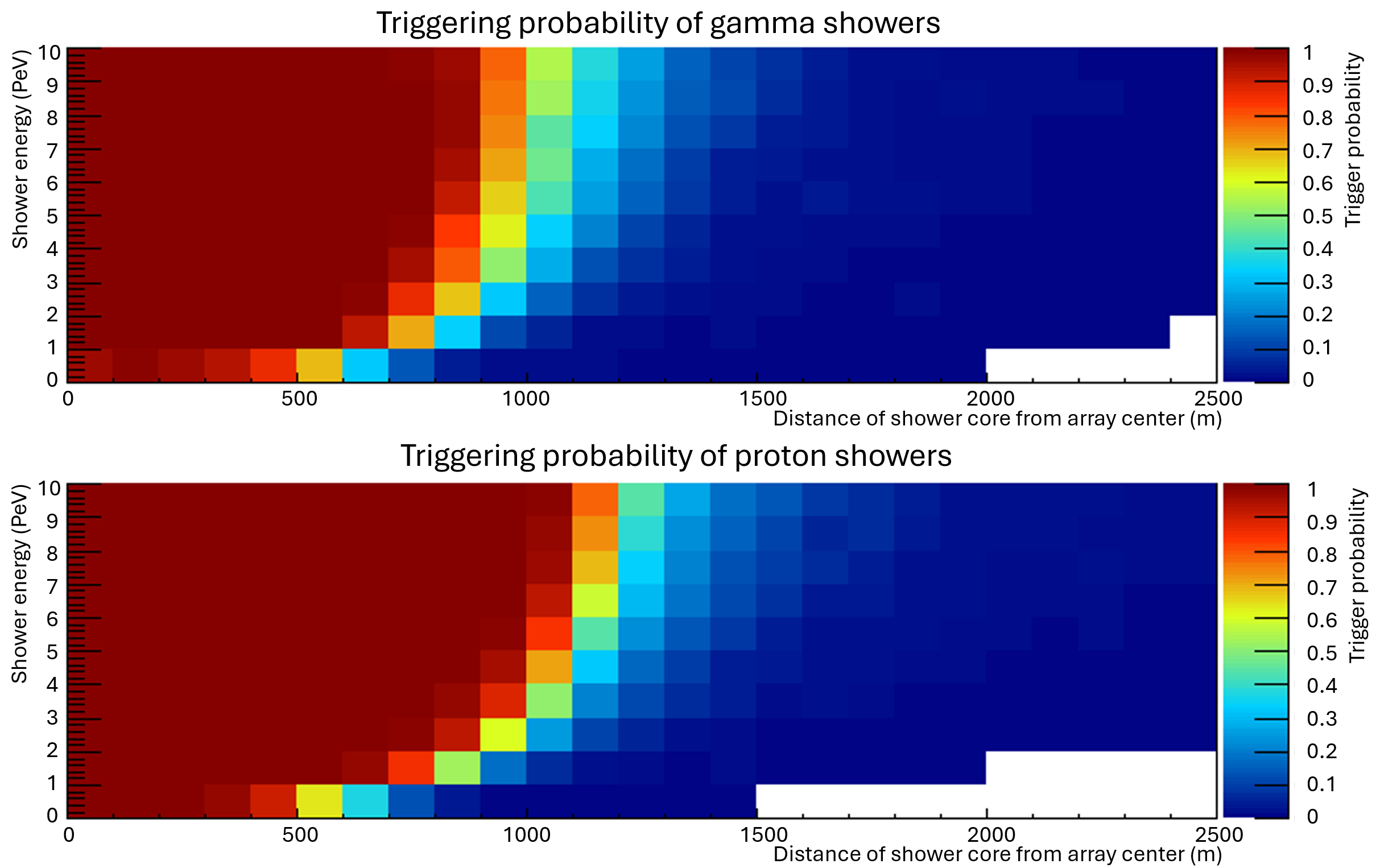}
    \caption{Triggering probability for primaries of different energy (vertical axis, in \si{\peta\eV}) and distance from the closest detector of a circular array (horizontal axis, in meters). Top: photon primaries; bottom: proton primaries.  To create the figure, we consider an array consisting in 361 units arranged in a tightly packed hexagonal arrangement, such that the array extension on the ground is small enough to give a well-defined meaning to the distance of the array from the core of generated showers.}  
    \label{f:pvsdande} 
\end{center}
\end{figure*}

The probability that a shower leaves a signal in a sufficient number of detector units is a function of energy and distance from the array core. In order to ensure that we fully test the functionality of arrays, we need to carefully choose the area illuminated by showers in the detector plane, such that we extend it past the sensitive area. Shower generation and reconstruction are CPU-expensive tasks in the optimization loop, so the two criteria are conflicting --a wider illuminated area tests more carefully the array performance, but it also spreads a fixed number of generated showers per epoch to smaller density, which is detrimental for the precision of the inference; or, if the density is kept fixed, it requires a larger number of showers per batch.  In \autoref{f:pvsdande} we show the probability that a shower leaves a signal in at least 50 detector units for the two cases of primary identity (gamma or proton) as a function of the distance of the shower core on the ground from the closest detector, and as a function of the energy of the primary particle in \si{\peta\eV}.   We observe that showers can hardly be reconstructed by an array located over 2000 meters away from the instrumented region, so that the initial sampling of shower cores within 2000 meters from the closest detector is a very safe condition in the generation procedure.

\subsubsection{  Derivatives of the fluxes \label{s:modelderiv}} 

 As a preliminary step to the calculation of  the gradient  of the utility function  we defined in \autoref{s:utility},  we compute  below the  derivatives of the fluxes as a function of shower parameters. The calculation is straightforward but tedious. We provide  only  its essential results.

The derivative of the flux equations (\autoref{eq:flux}) with respect to the distance $R$ of a detector unit from the shower center is straightforward to obtain: \par

\begin{equation}
\frac{d \left( dN/dR \right)}{dR} = -p_1 p_2 R^{p_2-1} \frac{dN}{dR}
\end{equation}

The calculation of the flux derivative with respect to shower energy $E$, however, is more contrived. We get it from:
\begin{equation}
    \frac{d \left( dN/dR \right)}{dE} = \left( \frac{1}{p_0} \frac{dp_0}{dE} -R^{p_2} \frac{dp_1}{dE} - p_1R^{p_2} \ln R \frac{dp_2}{dE} \right)  \frac{dN}{dR}
\end{equation}

\noindent
for which we need interpolated values of the derivatives of flux parameters $dp_i/dE$ $(i=1,2,3)$, which are extracted by interpolation from the parametrized model described in the \hyperref[sec:appendix]{Appendix}, and stored in lookup tables. 

Finally, the derivative of the flux with respect to polar angle can be obtained as:
\begin{align}
    \frac{d (dN/dR)}{d \theta} = & \Big( \frac{1}{p_0} \frac{dp_0}{d\theta} -R^{p_2}\frac{dp_1}{d\theta}- p_1 R^{p_2} \ln R \frac{d p_2}{d\theta} 
    \nonumber \\ & -p_1 p_2 R^{p_2-1} \frac{dR}{d\theta} \Big) \frac{dN}{dR}
\end{align}

\noindent
which also makes use of the lookup tables for the derivatives of the flux parameters, $dp_i/d\theta$ $(i=1,2,3)$.

\subsection{Reconstruction of shower parameters \label{s:likratio}}

 Below we concern ourselves with providing an analytical formulation of the procedure that turns detector readouts into estimates of the parameters defining observed showers. 
Given a layout of $N_\text{det}$ Cherenkov tanks, specified by their ${(x_i, y_i)}_{i=1...,N_\text{det}}$ positions on the ground, and the assumptions we have discussed above, the data that the array collect upon being hit by an atmospheric shower can be specified in the form of $4 N_\text{det}$ numbers: the number of detected e.m. (muon) secondaries $N_{e,i}$ ($N_{\mu,i}$, respectively) by detector $i$, and the time of arrival at detector $i$ of the e.m. (muon) component $T_{e,i}$ ($T_{\mu,i}$, respectively), with $i=1... N_\text{det}$. We compute recorded arrival times of each component as an average over all recorded secondaries (including backgrounds), for simplicity.

Once we define the parametrization of the differential fluxes of the two kinds of particles on the ground as a function of distance from the shower axis defined as described above, assuming no uncertainty on the model itself, we may write an expression for the probability of observing the set of $4 N_\text{det}$ variables, under the hypothesis that the shower is originated by a gamma-ray or proton primary:
  \begin{align}
  \label{eq:lik}
  \mathcal{L}^\gamma(\lambda_{e,i},\lambda_{\mu,i}, t_{e,i}^\text{exp}, t_{\mu,i}^\text{exp}) = \nonumber \\ \prod_i P(N_{e,i}|\lambda^\gamma_{e,i}) P(N_{\mu,i} | \lambda^\gamma_{\mu,i}) G(T_{e,i}-t_{e,i}^\text{exp}) G(T_{\mu,i}-t_{\mu,i}^\text{exp}) \nonumber \\
\mathcal{L}^p(\lambda_{e,i},\lambda_{\mu,i}, t_{e,i}^\text{exp}, t_{\mu,i}^\text{exp})= \nonumber \\ \prod_i P(N_{e,i}|\lambda^p_{e,i}) P(N_{\mu,i} | \lambda^p_{\mu,i}) G(T_{e,i}-t_{e,i}^\text{exp}) G(T_{\mu,i}-t_{\mu,i}^\text{exp}) 
\end{align}

\noindent
In the expressions above we have denoted with $P(N|\lambda)$ the Poisson probability of observing in a detector $N$ secondary particles of each species from a shower whose expected number of secondary particles of the specified kind inside the detector area at the detector location is $\lambda$, and with $G(T-t^\text{exp})$ the probability of observing a time difference $T-t^\text{exp}$ of the signal with respect to that expected from the shower development. $G$ is a Gaussian distribution with expected value of 0 and variance $\sigma_T^2$, and it is meant to model the experimental resolution on the arrival time of secondaries on the detection units (we assume speed of light propagation of secondaries). The $\lambda$ and $t^\text{exp}$ values are provided by the surrogate model described in \autoref{s:model}. For the parameter $\sigma_T$ we choose the value of $10$ ns, which is large enough to impose no strict constraints on the specific characteristics of the detection units for the time being.

\subsubsection { Likelihood maximization} 

Once the set of data $N_{e,i}, N_{\mu,i}, T_{e,i}, T_{\mu,i}$ is specified, the logarithm of the expression given in \autoref{eq:lik}  is a function of the hypothesis on the identity of the primary particle, and of five parameters that fully determine the shower: the intersection of the shower axis on the ground ${X_0, Y_0}$ (dubbed `shower core'); the polar and azimuthal angles ${\theta, \phi}$ defining the shower direction; and the energy $E$ of the primary particle. The maximization of the log-likelihood can be performed numerically once derivatives with respect to each of the five parameters are estimated.

The shower model describes particle densities as a function of the distance of the detecting unit location $(x_i, y_i, 0)$ (where we have redefined the elevation $z$ to be 0 m at the detector plane) from the shower axis. When the polar angle $\theta$ is non-null, this distance can be computed by expressing the coordinates of points on the shower axis as a function of a parameter $\xi$, and finding the value of $\xi^*$ corresponding to the closest point to the unit by setting to zero the derivative of the expression of the distance with respect to $\xi$ to find the minimum of $d$:
\begin{align}
    x &= X_0 + \xi \sin{\theta} \cos{\phi} \quad
    y = Y_0 + \xi \sin{\theta} \sin{\phi} \quad
    z = \xi \cos{\theta} \nonumber \\ 
    d &= ((x_i-x)^2 + (y_i-y)^2 + z^2)^{0.5} \nonumber \\
    \xi^* &= (x_i-X_0) \cos{\phi} \sin{\theta} + (y_i-Y_0) \sin{\theta} \sin{\phi}  
\end{align}
\noindent
The distance of detector $i$ from the shower core can then be computed as:
\begin{equation}
    R_i = \sqrt{(x_i-X_0)^2+(y_i-Y_0)^2-{\xi^*}^2}
\end{equation}

We may now express the derivative of the log-likelihood function over shower parameters $X_0, Y_0, \theta, \phi$ as follows: 
\begin{align}
    \frac{d \ln \mathcal{L}}{d X_0} &= \sum_i \left[ \frac{d \ln \mathcal{L}}{dR_i} \frac {dR_i}{dX_0} + \frac{d \ln \mathcal{L}}{dt_i} \frac{dt_i}{dX_0}\right],
  \nonumber \\\frac{d \ln \mathcal{L}}{d Y_0} &= \sum_i \left[ \frac{d \ln \mathcal{L}}{dR_i} \frac {dR_i}{dY_0} + \frac{d \ln \mathcal{L}}{dt_i} \frac{dt_i}{dY_0}\right],
\nonumber \\
    \frac{d \ln \mathcal{L}}{d \theta} &= \sum_i \left[ \frac{d \ln \mathcal{L}}{dR_i} \frac{dR_i}{d \theta} + \frac{d \ln \mathcal{L}}{dt_i} \frac{dt_i}{d \theta} \right],
\nonumber \\
    \frac{d \ln \mathcal{L}}{d \phi} &= \sum_i \left[ \frac{d \ln \mathcal{L}}{dR_i} \frac{dR_i}{d \phi} + \frac{d \ln \mathcal{L}}{dt_i} \frac{dt_i}{d \phi} \right],
\end{align}

\noindent
with
\begin{align}
    \frac {dR_i}{dX_0} &= \frac{-(x_i-X_0) +\xi^*\sin{\theta}\cos{\phi}}{R_i}, \nonumber \\
    \frac {dR_i}{dY_0} &= \frac{-(y_i-Y_0) +\xi^*\sin{\theta}\sin{\phi}}{R_i}.\nonumber \\
    \frac{dt_i}{dX_0} &= - \sin{\theta} \cos{\phi}/c, \nonumber \\
    \frac{dt_i}{dY_0} &= - \sin{\theta} \sin{\phi}/c \nonumber \\
    \frac{dR_i}{d \theta} &= -\xi \cos{\theta} \left[ (x_i-X_0) \cos{\phi} + (y_i-Y_0) \sin{\phi} \right] / R_i, \nonumber \\
    \frac{dR_i}{d \phi} &= -\xi \sin{\theta} \left[ -(x_i-X_0) \sin{\phi} + (y_i-Y_0) \cos{\phi} \right] / R_i. \nonumber \\
    \frac{dt_i}{d \theta} &= \cos{\theta} \left[ (x_i-X_0) \cos{\phi} + (y_i-Y_0) \sin{\phi} \right] / c,\nonumber \\
    \frac{dt_i}{d \phi} &= \sin{\theta} \left[ -(x_i-X_0) \sin{\phi} + (y_i-Y_0) \cos{\phi} \right] / c  
\end{align}

\noindent
where the latter two expressions are derived for simplicity by assuming that the shower front moves at the speed of light $c$ and is flat --that is, orthogonal to the axis.

To complete the calculations we also need explicit derivatives of the log-likelihood function by $E$, which directly depends on the shower model, and by detector distances and expected arrival times $R_i$ and $T_i$; these are computed as follows:
\begin{align}
    \frac{d \ln \mathcal{L}}{dE} &= \sum_i \left[ \frac{d \ln \mathcal{L}}{d \lambda_{e,i}} \frac{d \lambda_{e,i}}{d E} + \frac{d \ln \mathcal{L}}{d \lambda_{\mu,i}} \frac{d \lambda_{\mu,i}}{d E} \right],
\nonumber \\
    \frac{d \ln \mathcal{L}}{dR_i} & = \frac{d \ln \mathcal{L}}{d \lambda_{e,i}} \frac{d \lambda_{e,i}}{dR_i} + \frac{d \ln \mathcal{L}}{d \lambda_{\mu,i}} \frac{d \lambda_{\mu,i}}{dR_i} \nonumber \\
    \frac{d \ln \mathcal{L}}{dt_{e,i}^\text{exp}} &= \frac{T_{e,i} - t_{e,i}^\text{exp}}{\sigma_T^2},   
\nonumber \\
    \frac{d \ln \mathcal{L}}{dt_{\mu,i}^\text{exp}} &= \frac{T_{\mu,i} - t_{\mu,i}^\text{exp}}{\sigma_T^2}.       
\end{align}

\noindent
Finally, the derivatives of the log-likelihood function by the flux expectations $\lambda_{e,i}$, $\lambda_{\mu,i}$ at detector $i$ are simply 
\begin{align}
    \frac{d \ln \mathcal{L}}{d \lambda_{e,i}} &= -1 + \frac{N_{e,i}}{\lambda_{e,i}},
\qquad
    \frac{d \ln \mathcal{L}}{d \lambda_{\mu,i}} = -1 + \frac{N_{\mu,i}}{\lambda_{\mu,i}}.
\end{align}

\noindent
\subsubsection{ The likelihood ratio discriminator } 

 The likelihood definition provided {\it supra} allows us to assemble a performant test statistic with which we may distinguish gamma-ray showers from proton ones. Indeed, 
the biggest challenge of a ground-based gamma-ray observatory is to successfully distinguish the almost purely electromagnetic showers of secondaries originated by gamma rays from the mixed electromagnetic-hadronic showers generated by primary hadrons (protons or light nuclei).   The ratio between the flux of primary gammas and hadrons 
is already much smaller than 1 for GeV-energy showers, and it rapidly falls to values below $10^{-4}$ for the PeV-energy showers whose rate and origin constitute one of the main scientific goals of SWGO.

%\mdc{Shound't we explain more the next sentence} 
There are multiple discriminating features which can be exploited to distinguish gamma-ray primaries from protons; by far the most striking one is  the  number and distribution of the  particles constituting what is called its  ``hard'' component, composed of energetic muons produced by pion and kaon decays in the atmosphere. Leveraging the simplified model of secondaries density on the ground described in \autoref{s:model}, we will concern ourselves only with the radial distribution of these particles in the following\footnote{Inclusion of topological information on the distribution of secondaries in time and position on the ground will be possible with a neural network model, which  will be  the subject of future studies.}. 
%\textcolor{yellow}{A more detailed model that included clumpiness of the secondaries on the ground --especially muons-- would allow for a more precise match to the information exploitation that is possible by a real experiment. We plan to include such effects in a future version of the code. }

The classification problem can be cast as one of hypothesis testing, when Neyman-Pearson's (NP) lemma ensures that the most powerful test statistic to distinguish two simple hypotheses is a $\ln$-likelihood ratio \citep{NPlemma}. Although we are in this case not under the conditions of validity of the NP lemma, given that the hypotheses under test are composite (the $\ln$-likelihood values depend on the five parameters defining a shower under each of the two primary particle hypotheses), the $\ln$-likelihood ratio still retains good properties, once we evaluate it for the value of parameters that maximize the two likelihoods: this is thus a likelihood ratio
\begin{equation}
\mathcal{T} = \ln \frac{\mathcal{L}_{\gamma}(\hat{X}_{0\gamma},\hat{Y}_{0\gamma},\hat{\theta_{\gamma}},\hat{\phi_{\gamma}},\hat{E}_{\gamma})}{\mathcal{L}_{p}(\hat{X}_{0p},\hat{Y}_{0p},\hat{\theta_p},\hat{\phi_p},\hat{E}_{p})} =
\ln (\mathcal{L}^\text{max}_{\gamma}) - \ln (\mathcal{L}^\text{max}_{p}).
\label{eq:Tdef}
\end{equation}

\noindent
Note that in the above expression the estimators $\hat{X}_{0\gamma},\hat{Y}_{0\gamma},\hat{\theta_{\gamma}},\hat{\phi_{\gamma}},\hat{E}_{0\gamma}$ at the numerator are different from those at the denominator, as the two sets refer to different primary hypotheses (gamma at the numerator, proton at the denominator).
We use the likelihood ratio as our primary tool to extract inference on the flux of gamma primaries from a set of detected showers; that inference will be used to compute a utility function that tracks the sensitivity of SWGO to the gamma-ray spectrum at different energies. The optimization loop searching for the most advantageous configuration of detectors on the ground will then require us to obtain derivatives of that utility function with respect to the detectors position, ${x_i,y_i}$. We provide those calculations in \autoref{sec:performance}.

\subsection{Performance of shower reconstruction}

Since secondary densities decrease sharply both as a function of the distance of a detection unit from the shower axis, and as a function of the energy of the primary, the minimization of the above $-\ln\mathcal{L}$ function is not trivial; special care must be taken to prevent the algorithm from getting stuck in local minima. In particular, the correct identification of the shower core (parameters $X_0$, $Y_0$) is problematic when that point on the ground is far from the detector array, and coinsequently only a limited number of detector units are hit by secondaries: {\it e.g.}, a decrease of the distance from the axis or an increase of the primary energy $E$ would have a very similar effect on the observed counts, making a precise axis/energy determination challenging.

In order to ease the identification of the absolute minimum of $-\ln\mathcal{L}$ we initially devised an initialization of the parameters by computing the value of $-\ln\mathcal{L}$ in a grid of points, in two steps. We started with angular parameters $\theta$, $\phi$, whose value is mostly sensitive to the timing information and is relatively decoupled from that of the other three parameters. 
%\mdc{Why is that?} 
We used a grid of 4 by 4 values of $(\theta,\phi)$ in $(\SI{0}{\degree}, \SI{65}{\degree})\times(0, \pi)$, and set $(X_0,Y_0) = (\SI{0}{\meter},\SI{0}{\meter})$ and $E=\SI{1}{\peta\eV}$ as values for the shower core and energy; the grid values of $\theta$ and $\phi$ corresponding to the largest  of the 16 likelihoods were used as initialization values for the following step.

The initialization procedure for the $X_0$, $Y_0$ and $E$ coordinates is CPU-heavy, as we need to scan three dimensions and the residuals from the true value of these parameters are strongly correlated. After some studies we settled to a grid of $1000$ values ($10$ by $10$ values of ${X_0,Y_0}$ and $10$ values of energy). The five values of the parameters corresponding to the highest likelihood could then be taken as initialization for the full likelihood maximization, which is performed by gradient descent using the ADAM optimizer~\citep{kingma2017adam}.

We verified that the procedure described above guarantees that the likelihood maximization converges to the absolute maximum. However, this result is obtained at the cost of very significant CPU expense. We then observed that if we initialized the likelihood with the true value of the five parameters for the true primary hypothesis, we ended up obtaining the same results for shower parameters and likelihood ratio test statistic. Since high-energy cosmic showers are rare, in a real experiment there will be enough time for as fine a scan of initial likelihood parameters of each collected shower as desired. Therefore, as far as our optimization study is concerned, there is no loss of generality in using the {\it a priori} unknown true values of the shower parameters to initialize the likelihood maximization procedure. This simplification comes with a significant benefit in terms of computing speed; likelihood maximization does still require on average about fifty evaluations of the function and its derivatives to converge to the absolute maximum; but the 1,000 ancillary initial evaluations are then avoided. 

\begin{figure}[h!t]
\includegraphics[width=0.99\linewidth]{ 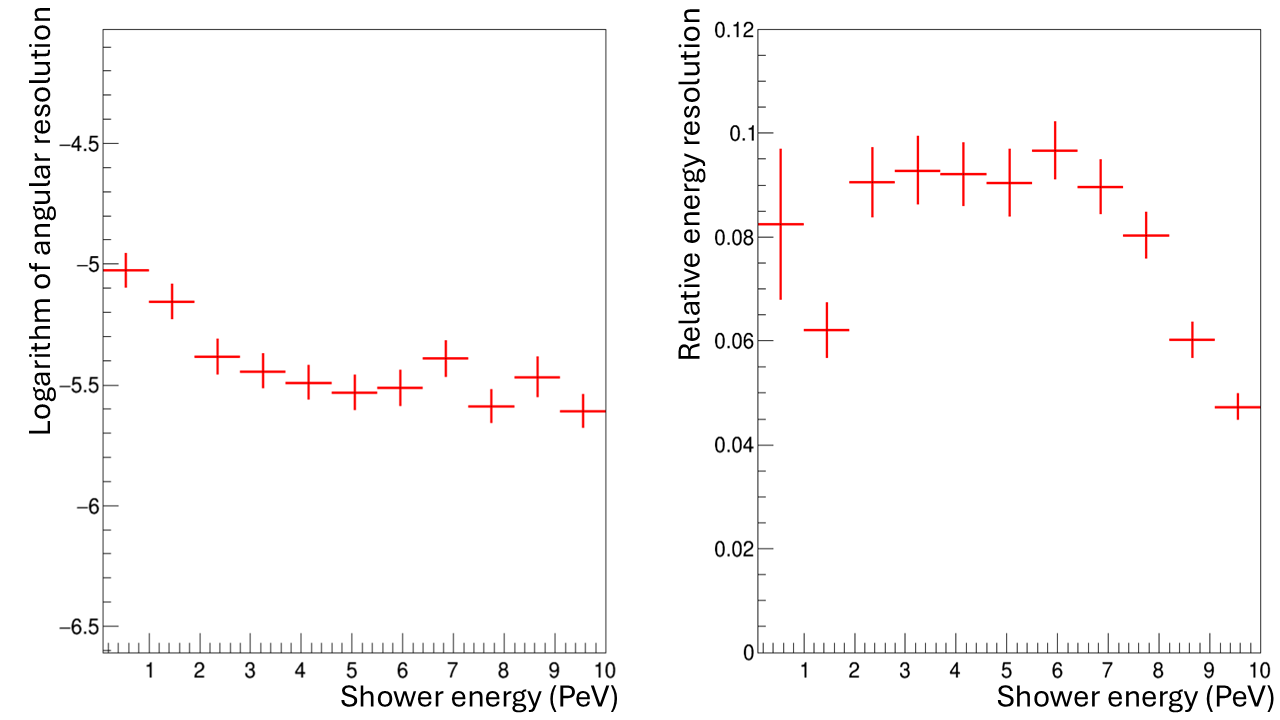}
\includegraphics[width=0.99\linewidth]{ 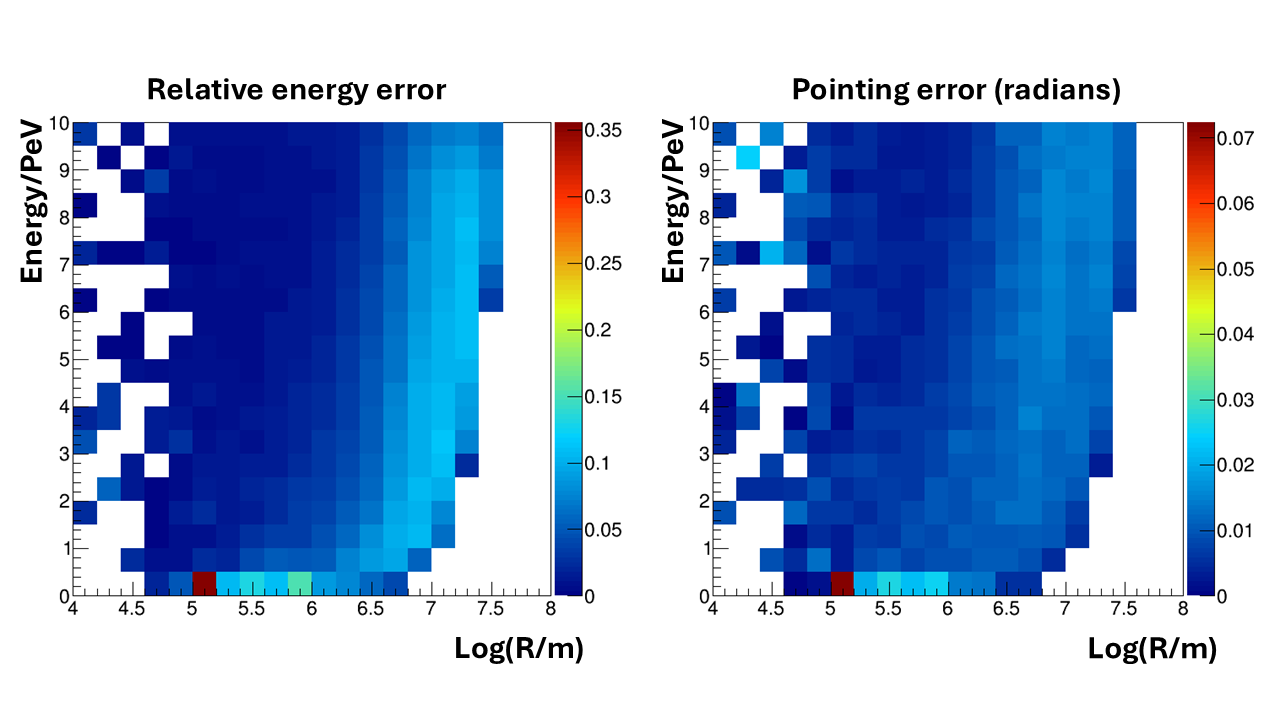}
\caption{\it Top: Logarithm of the mean relative angular error (left) and mean relative energy error (right) for gamma showers reconstructed by an array of 6288 tanks arranged in a compact layout corresponding to the A5 benchmark. Bottom: Mean relative energy and pointing error as a function of energy and distance of the core from the center of a small circular array of 361 tanks with high fill factor, corresponding to the proposed first phase of SWGO (SWGO-A).}
\label{f:resolutions}
\end{figure}

For an appraisal of the reconstruction performance, we study a set of gamma showers detected by an array of 6288 detector units arranged in 331 19-tank macro-tank aggregates (see {\it infra}, \autoref{s:results}) arranged in a uniform circular pattern of fill factor and extension roughly corresponding to those of the core of the A5 benchmark (a compact 300 m wide array with a core of 234 m radius, see \autoref{t:layouts_sec2}). We find that the reconstruction procedure yields an angular resolution in the range of 0.4 to 0.2 degrees for shower energy in the studied 0.1-10 PeV range; the relative energy resolution instead is found to vary from $10\%$ to $5\%$ in the same range (see \autoref{f:resolutions})

%\mdc{I think the next sentence is rather vague and can create some confusion} 
While the reconstruction procedure described above could certainly be improved in a real offline analysis of individual detected showers --{\it e.g.}, by the use of deep-learning-powered supervised regression techniques-- we believe that the added precision on shower parameters would only be very loosely coupled with the signs of gradients on the utility function, and would thus not significantly modify the  optimality conditions we are searching for on the location of detector units; the CPU-limited and potentially sub-optimal reconstruction procedure we are using in this work is thus a useful starting point to fuel a search for optimal layout. 

%In \autoref{f:angresiduals} we show, for the same array and for showers at different distances from the array core, residuals of the polar and azimuthal angle reconstruction as a function of the polar angle. We observe a reasonable behaviour, with long tails to high residuals for showers that are only marginally detected by the array. 

\subsection{Triggering}
\label{s:triggering}

 Before we can proceed with an explicit calculation of the gradient of the utility function, we need to consider the triggering strategy of the SWGO array, and produce a model of the resulting acceptance/rejection effects on the showers hitting the array.  

A parameter which has the potential of affecting the optimal layout of detection units is the number of tanks which record a signal (of muons or e.m. particles) for the EAS to be accepted in our selection. If that number is too small, our inference on the origin of the shower or on the shower parameters is insufficiently precise, so it makes sense to enforce a lower threshold on its value. The SWGO collaboration is also presently using a threshold in its official reconstruction software, with a minimum number of coincident units set to values between 30 and 50~\citep{swgoreco}. For the studies reported in this document we therefore set the threshold for accepting showers in the calculation of the gamma-ray flux and the utility to $N_{trigger} \geq 50$ detection units, as we are considering the highest-energy range of the spectrum of primary particles of interest to SWGO. We then need to also inform the gradients of the utility function of the probability that a shower fails the $N_{trigger}$ condition if we move a detector from its current position. 

For the purpose of updating gradients correctly, we compute the probability that a detector observes at least one secondary particle of any kind (muons or e.m. particles) by evaluating
\begin{equation}
S(N_\text{obs} \geq 1) = \sum_{i=1}^{N_\text{det}} \left( 1 - \exp(-\hat{N}_{\mu,i}-\hat{N}_{e,i})\right)
\end{equation}

as the sum of probabilities over detector units to detect at least one particle of any kind, where $\hat{N}_{\mu,i}, \hat{N}_{e,i}$ are expected fluxes of secondaries at each detector. We use its value to derive:
\begin{equation}
p(N_\text{active}\geq N_\text{trigger}) = 1 - \sum_{k=0}^{N_\text{trigger-1}} \left[ \text{Poisson}(k,S) \right].
\end{equation}

\begin{figure*}[h!]
\includegraphics[width=0.99\linewidth]{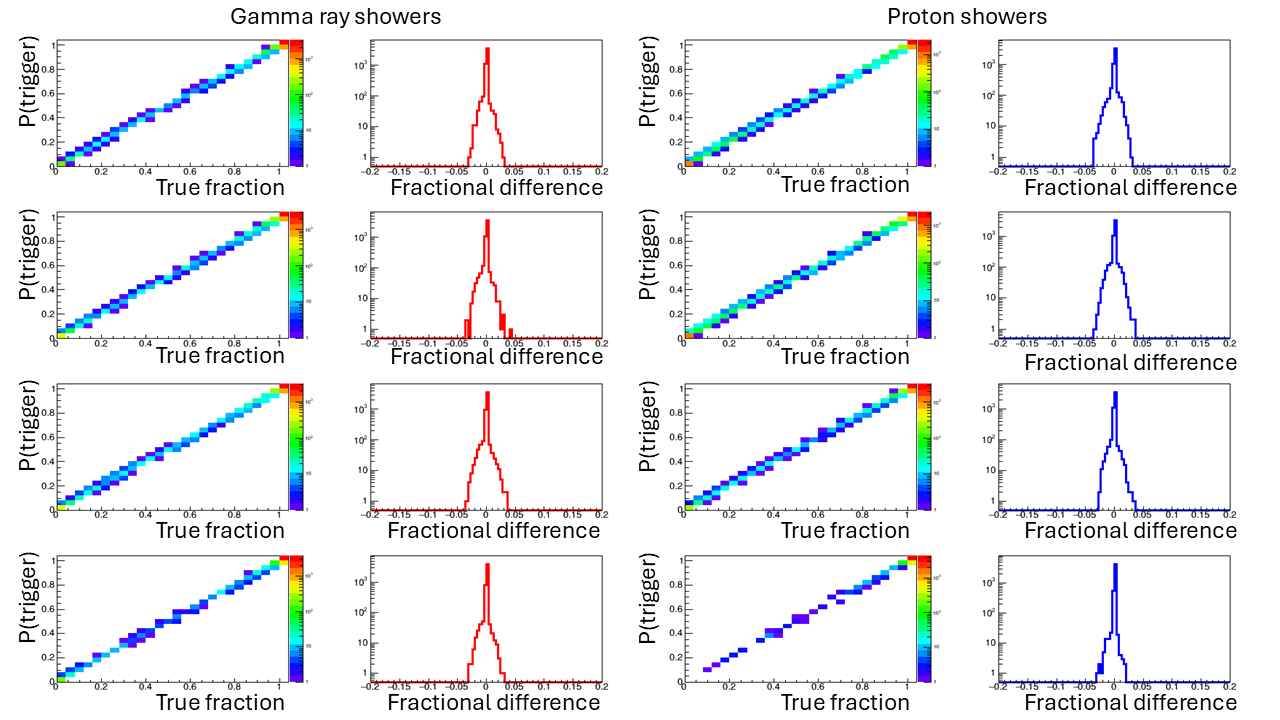}
\caption{Test of the approximated calculation of triggering probability discussed in the text. For four different energies (the four rows of graphs) graphs in the first and third columns show, for gamma- and proton-originated showers respectively, the comparison of approximated $P(N \geq 50)$ triggering probability versus true triggering fraction; graphs in the second and fourth columns show, again for gammas (red) and protons (blue) respectively, the fractional difference between estimated and true probability. The results shown are obtained with an array comprising 91 19-unit macro-tanks arranged in a circular array of 500 meter radius, hit by showers with a core at a distance of less than 1500 meters from the center. For this test the simulated shower energies range from 0.1 to 2.6 PeV (top to bottom row); higher-energy showers almost always pass the trigger criteria.}
\label{f:PTrigger_ge50c}
\end{figure*}

\noindent
In other words, we approximate the combinatorial probability that out of a total of $N_\text{det}$ detectors at least $N_\text{trigger}$ of them have seen a signal, by using the approximated combined probability of small p-values as their sum. The approximation above works quite well for our purposes, as we verify by toy simulations that it estimates with accuracy close to \SI{1.5}{\percent} the true probability in practical situations of relevance to this study, which are restricted to cases when the number of detectors giving a signal is not too different from the threshold $N_\text{trigger}$ (and is thus sensitive to it). Results of one such test are shown in \autoref{f:PTrigger_ge50c}.

The formulas discussed above can be easily differentiated with respect to the position of detectors $x_i$, $y_i$ by using the chain rule and by observing that changes of detector positions affect the expected fluxes $\hat{N}_{\mu,i}$, $\hat{N}_{e,i}$. So, {\it e.g.}, to obtain the derivative of the probability that a shower passes the trigger condition with respect to a $dx_{i^*}$ movement of detector $i^*$, we compute \par

\begin{equation}
\frac{dp}{dx_{i^*}} = - \frac{d}{dx_{i^*}} \left[ \sum_{j=0}^{N_\text{trigger}-1} \left( {\mathrm e}^{-S} S^j / j! \right) \right].
\end{equation}

\noindent
By recalling the definition of $S$, now making explicit the expected fluxes into detector $i^*$, \par

\begin{equation}
    S = \sum_{i=1}^{N_\text{det}} \left[ 1-{\mathrm e}^{(-\lambda_{\mu,i} - \lambda_{e,i})}\right]
\end{equation}

\noindent 
we get \par

\begin{equation}
    \frac{dP_\text{active}}{dx_{i^*}} = 
    %- 
    \sum_{j=0}^{N_\text{trigger}-1} \left[ \frac{1}{j!} \left( {\mathrm e}^{-S} (S^j - jS^{j-1}) \right) \frac{dS}{dx_{i^*}}\right]
\end{equation}
%        // The latter term can be computed as follows (PActive = 1 - sum(1:Ntr-1)):
%        //     dPActive_m/dxi = -d/dxi [sum_{j=0}^{Ntr-1}(e^{-Sm}*Sm^j/j!)]
%        // with
%        //     Sm = Sum_{i=1}^Ndet [1-exp(-lambda_mu^i-%lambda_e^i)_m]
%        // We get
%%        //     dPActive_m/dxi = - Sum_{j=0}^{Ntr-1} 1/j! %[e^{-Sm}*(Sm^j-j*Sm^(j-1))*dSm/dxi]
%        // Now for dSm/dxi we have
%        //     dSm/dxi = d/dxi [ Sum_{i=1}^{Ndet} (1-e^(-%lambda_mu^i-lambda_e^i)_m))]
%        //               = -d/dxi (e^[-lambda_mu^i-%lambda_e^i]_m) =
%        //               = e^[-lambda_mu^i-lambda_e^i]*%(dlambda_mu^i/dxi + dlambda_e^i/dxi)
%        // and the latter are computed in the flux %routines.

\section{ Ingredients of the Utility Derivatives} \label{s:sfuncertainty} 

 In this section we consider the model ingredients defined {\it supra}, working our way to a full definition of the derivatives of the utility function with respect to the position of detection units on the ground, which is the main ingredient in the iterative gradient descent that allows our end-to-end optimization procedure to succeed. 

\subsection { Construction of the PDF of the test statistic} 

A measurement of the flux of gamma rays entails the reconstruction of the showers observed in a given time interval, as well as the estimation of the fraction of observed showers that are genuinely due to gamma primaries. As described in the previous section, we rely on a likelihood ratio test statistic to quantify the separation power of gamma from proton primaries produced by the array. Once we simulate a batch of gamma and proton EAS corresponding to a given data collection time, we reconstruct all showers under each of the two hypotheses in turn, and proceed to compute the test statistic value for every shower. Rather than using the test statistic to select a gamma-rich subset, however, we directly fit for the fraction of gamma primaries in the batch by likelihood maximization. This allows us to avoid the non-differentiable operation of selecting test statistic values above a hard-set cut, as well as the loss of information related to discarding a fraction of the signal failing a hard-set cut.

In a batch of $N_\text{batch}$ showers, for each of which we have a measured value of $\mathcal{T}$, we can estimate the fraction of photons $f_{\gamma}$ by maximizing the following likelihood:
\begin{equation}
    \mathcal{L}(f_{\gamma}) = \prod_{i=1}^{N_\text{batch}} \left[ f_{\gamma} P_{\gamma}(\mathcal{T}_i) + (1-f_{\gamma}) P_p(\mathcal{T}_i) \right],
\end{equation}

\noindent
where with $P_{\gamma}(\mathcal{T}_i)$ ($P_p(\mathcal{T}_i)$) we have expressed the probability density function of the test statistic for gamma (proton) showers. 

We construct a model of the density function of the test statistics $\mathcal{T}_{\gamma}$ ($\mathcal{T}_{p}$) by
substituting a Gaussian kernel to each value of $\mathcal{T}$ for a set
of gamma (respectively, proton) showers. The standard deviation of the Gaussian kernel needs to be a fair estimate of the true uncertainty of the value of $\mathcal{T}$, which depends on the varying conditions under which different showers are measured. The procedure is the following:

\begin{enumerate}
\item We simulate $N_\text{events}$ showers, half of which are from gamma primaries and half from proton primaries. The location of the core (defined by nuisance parameters $X_0, Y_0$) is sampled in an area covering the one instrumented by detectors, with a margin wide enough to guarantee that showers that fell outside of it would never produce a triggering signal in the array (see \autoref{f:pvsdande} in \autoref{s:modelderiv}). An acceptance/rejection procedure is also performed to allow for the calculation of the effective area to which the array is sensitive, as explained in detail {\it infra} (\autoref{s:utility}). Polar and azimuthal angles are sampled by assuming uniform location of the sources in the sky. For the polar angle this results in a  distribution proportional to $\sin \theta$, while the azimuthal angle $\phi$ remains uniform. We constrain the polar angle $\theta$ to be smaller than 65 degrees, a value beyond which SWGO will do limited target investigation because the shower reconstruction for large polar angles becomes problematic; in any case observable rates fall down quickly due to atmospheric absorption. Primary energy is instead sampled from a falling power law distribution with user-defined hardness in the $(\SI{0.1}{\peta\eV},\SI{10}{\peta\eV})$ range.
\item A trigger condition is enforced to select showers that hit a minimum of $N_{trigger}$ detector units with at least one particle of any species. The threshold is set by default to $N_{trigger}=50$ but may be changed for specific studies. 
\item We proceed to reconstruct the triggering showers under both primary hypotheses using the likelihood maximization described in \autoref{s:reconstruction} above, obtaining a test statistic value from each from the log-likelihood difference $\mathcal{T}(i) = \ln{(\mathcal{L}_\text{max}^{\gamma})} - \ln{(\mathcal{L}_\text{max}^\text{p})}$;

\item We compute the uncertainty on the value of $\mathcal{T}$ for each shower $k$ by error propagation;

%, \par
%\begin{align}
%    \sigma_{\mathcal{T}_k}^2  & = \sum_{i=1}^{N_\text{det}} \Big[ \Big( \frac{d\mathcal{T}_k}{d\lambda_{\mu,i}^\gamma} \Big)^2 \sigma^2 \lambda_{\mu,i}^\gamma 
%     + \Big( \frac{d\mathcal{T}_k}{d\lambda_{e,i}^\gamma} \Big)^2 \sigma^2 \lambda_{e,i}^\gamma \\
%    &+ \Big( \frac{d\mathcal{T}_k}{dt_{\mu,i}^p} \Big)^2 \sigma_{t_{\mu,i}^p}^2 
%    +  \Big( \frac{d\mathcal{T}_k}{dt_{e,i}^p} \Big)^2 \sigma_{t_{e,i}^p}^2 
%   + \Big( \frac{d\mathcal{T}_k}{d\lambda_{\mu,i}^p} \Big)^2 \sigma^2 \lambda_{\mu,i}^p \nonumber  \\  
    %&+ \Big( \frac{d\mathcal{T}_k}{d\lambda_{e,i}^p} \Big)^2 %\sigma^2 \lambda_{e,i}^p 
    %+ \Big( \frac{d\mathcal{T}_k}{dt_{\mu,i}^p} \Big)^2 %\sigma_{t_{\mu,i}^p}^2 
    %+  \Big( \frac{d\mathcal{T}_k}{dt_{e,i}^p} \Big)^2 \sigma_{t_{e,i}^p}^2 \Big] \nonumber
%\end{align}

%where the variances are computed using the RCF bound of the relative likelihood terms.

% Remark (by Max): The \begin{strip}...\end{strip} environment seemed to cause trouble. Large parts of section 5.1 "Construction of the PDFs of the Test Statistic" were missing from the pdf output and this was fixed when removing the two commands. The same problem was observed for the two other uses of the strip environment. I removed it everywhere; if it served a purpose, please describe the desired effect and I can try to achieve it in some other way.
%\begin{strip}
%\end{strip}

\item We model the PDF of gamma and proton test statistics by using all reconstructed showers passing the trigger condition, summing a Gaussian distribution for each shower, each with width equal to the corresponding estimated uncertainty and centered at the observed value of $\mathcal{T}_k$ of the set of selected gamma-rays and protons, respectively, and by renormalizing the resulting distributions to unit integral: 
\begin{align}
    P(\mathcal{T}_{\gamma}) &= \frac{1}{N_\gamma} \sum_{k=1}^{N_\text{events}} I_k(\gamma) G(\mathcal{T}_k,\sigma(T_k)) \\
    P(\mathcal{T}_{p}) &= \frac{1}{N_p} \sum_{k=1}^{N_\text{events}} I_k(p) G(\mathcal{T}_k,\sigma(\mathcal{T}_k))
    \label{eq:Tconstr}
\end{align}

where the indicator function $I_k$ equals 1 if the primary particle is of the requested kind and it passes the triggering condition, and zero otherwise.
\end{enumerate}

\noindent
Although the above described sum of Gaussian kernels centered at the discrete values of $\mathcal{T}$ produces continuous models of the PDF of the test statistic for gamma and proton showers, these models must be used with caution. In particular, at each loop in the optimization cycle we need to separately generate and reconstruct batches of showers having different parameters from those of the showers used for the construction of the above density functions. If we did not do that, the calculated gradients described {\it infra} (\autoref{s:utility}) would be strongly biased --any variation of the test statistic $\mathcal{T}$ of a shower, produced by a movement of a detector on the ground, would be likely to reduce the estimated density, because the new value $\mathcal{T}'$ at which the PDF is evaluated would not correspond anymore to the peak of the kernel of the identical shower used in the PDF construction. 

\subsection {Estimation of the variance of the gamma-ray fraction}

In order to construct a meaningful utility function to optimize we do not need an estimate of $f_{\gamma}$, which is fixed at 0.5 in the event generation\footnote{Although this is far from the real experimental conditions (gamma showers are rare in comparison to proton ones), this value does not affect the variation of the uncertainty with the array layout, and in fact it optimizes the determination of the latter quantity.} and is then modified by the triggering criterion, but rather an estimate of its uncertainty given the experimental conditions under which it has been extracted.

To be precise, we are interested in the mean squared error of the gamma-ray fraction in the sample of collected showers, including in our consideration a possible bias in the fraction estimate. A bias in the estimate may arise from the measurement process or other factors affecting data collection. We consider that in most experimental situations a bias can be corrected for {\it a posteriori}, with studies of simulated samples or calibration data sets; in our case this can be done by employing the high-fidelity simulation of EAS, provided by \texttt{CORSIKA}, and of the interaction of secondaries in the apparatus, provided by \texttt{Geant4}. Hence, we concentrate on the variance alone. The variance of ${f_{\gamma}}$ can be estimated by using Rao-Cramer-Frechet's bound~\citep{rao}:
\begin{equation}
    {\sigma_{f_{\gamma}}}^2 = \left[ -\frac{d^2 \ln \mathcal{L}}{d f_{\gamma}^2} \right]^{-1}.
    \label{eq:rcf}
\end{equation}

\noindent
The second derivative of the log-likelihood function can be obtained as:
\begin{equation}
- \frac{d^2 \ln \mathcal{L}}{d f_{\gamma}^2} = \sum_i \left[ \frac{[P_{\gamma}(\mathcal{T}_i)- P_p(\mathcal{T}_i)]^2}{[f_{\gamma}P_{\gamma}(\mathcal{T}_i) + (1-f_{\gamma})P_p(\mathcal{T}_i)]^2} \right]
\end{equation}

\noindent 
and from the above expression we obtain an estimate of the variance on $f_{\gamma}$, using \autoref{eq:rcf}.

\subsection {Derivatives of the utility}

The calculation of the derivatives of the utility function is a CPU-heavy task, given that it entails the need of iterating, for every detector, on the batch of showers generated at each epoch, tracking the value of partial derivatives that inform the utility about the effect of its movement in $x$ or $y$. We provide below the essential ingredients of the tedious calculation, which has been checked with the {\texttt{Wolfram Mathematica}} 
program.

\subsubsection{Derivatives of $U_{GF}$ term}

The expression of $U_{GF}$ contains an estimate of the flux of photons, given a certain integration time, and an estimate of the uncertainty on that flux estimate. As discussed in \autoref{s:sfuncertainty}, both quantities are obtained from the derivatives of the log-likelihood function as a function of the fraction of gamma-originated showers, given the set of observed values of test statistic in a batch of showers. 

To compute the derivative with respect to the $x_i$ position of detector $i$, we need to evaluate:\par

\begin{align}
    \frac{dU_{GF}}{dx_i} & = U_{GF} \left(\frac{1}{f_{\gamma}} \frac{df_\gamma}{dx_i} - \frac{1}{\sigma_{f_{\gamma}}} \frac{d \sigma_{f_\gamma}}{dx_i} -\frac{0.5}{\rho}\frac{d\rho}{dx_i}
    \right) \\ \nonumber
& = U_{GF} \biggl[ \frac{1}{f_{\gamma}} \frac{df_\gamma}{dx_i} - \frac{1}{\sigma_{f_{\gamma}}} \frac{d \sigma_{f_\gamma}}{dx_i} +  
\frac{1}{R_\text{tot}} \frac{dR_\text{tot}}{dx_i} \\ \nonumber & - \frac{0.5}{N_\text{trials}}\frac{dN_\text{trials}}{dx_i} \biggr]
\end{align}

\noindent

 Let us consider each term in the expression above in turn. The estimate $f_\gamma$ is the result of setting to zero the derivative over $x_i$ of the two-component likelihood function, \autoref{eq:lik}.
We may compute this by implicit differentiation. We treat $f_\gamma$ as the dependent variable and the probabilities $p_\gamma(k)$, $p_p(k)$ as independent variables, and use the chain rule to relate the derivation by $p_\gamma$ and $p_p$ to derivation by $x_i$. Let us first define \par

\begin{equation}
    H = \frac{d \ln \mathcal{L}}{d f_\gamma} = \sum_k \left[ \frac{p_\gamma(k) -p_p(k)} {f_\gamma p_\gamma(k) + (1-f_\gamma) p_p(k) } \right] = 0
\end{equation}

\noindent
as the equation that defines $f_\gamma$. We now differentiate with respect to $x_i$, obtaining:
\begin{equation}
    \sum_k \left[ \frac{dH}{dp_\gamma(k)} \frac{dp_\gamma(k)}{dx_i} + 
    \frac{dH}{dp_p(k)} \frac{dp_p(k)}{dx_i} \right] +
    \frac{dH}{df_\gamma} \frac{df_\gamma}{dx_i}   = 0
\end{equation}

\noindent
whence 
\begin{equation}
    \frac{df_\gamma}{dx_i} = - \frac{1}{\frac{dH}{df_\gamma}} \sum_k \left[ \frac{dH}{dp_\gamma(k)} \frac{dp_\gamma(k)}{dx_i} +
    \frac{dH}{dp_p(k)} \frac{dp_p(k)}{dx_i}\right].
\end{equation}

\noindent
Now we have:
\begin{equation}
    \frac{dH}{df_\gamma} = - \sum_m \left[ \frac{(p_\gamma(m)-p_p(m))^2}{(f_\gamma p_\gamma(m) + (1-f_\gamma) p_p(m))^2} \right] 
\end{equation}

\noindent
(Note that the last expression is minus the inverse of the estimated variance on $f_\gamma$). If we further define, for brevity:
\begin{equation}
    \Delta(m) = p_\gamma(m)-p_p(m),
\qquad
    D(m) = f_\gamma p_\gamma(m) + (1-f_\gamma) p_p(m),
\end{equation}

\noindent we can write:
\begin{equation}
    \frac{dH}{dp_\gamma(k)} = \frac{\left[ D(k) - \Delta(k) f_\gamma \right]}{D(k)^2} = \frac{p_p(k)}{D(k)^2} 
\end{equation}

\noindent
and

\begin{equation}
    \frac{dH}{dp_p(k)} = \frac{- \left[ D(k) + \Delta(k) (1-f_\gamma) \right]}{D(k)^2} = -\frac{p_\gamma(k)}{D(k)^2} .
\end{equation}

\noindent
Finally, we obtain
\begin{equation}
    \frac{df_\gamma}{dx_i} = \frac{\left[ p_p(k) \frac{dp_\gamma(k)}{dx_i} - p_\gamma(k) \frac{dp_p(k)}{dx_i} \right]}{D(k)^2 \sum_m (\frac{\Delta(m)^2}{D(m)^2})}.
\end{equation}

\noindent
As far as the $x_i$ derivatives of the probabilities of the observed test statistic under the two hypotheses are concerned, they are obtained by considering that the latter are obtained from sums of Gaussian terms, {\it e.g.}, 
\begin{equation}
    p(\mathcal{T}_\gamma)(k) = \sum_{m=1}^{N_\text{ev}} \left[ P_\text{tr}(m) \frac{\exp[-\frac{(\mathcal{T}(m)-\mathcal{T}(k))^2}{2 \sigma^2_T(m)}]}{(2 \pi \sigma^2_\mathcal{T}(m))^{0.5}} \right]. 
\end{equation}

\noindent
We can differentiate over $dx_i$ directly, obtaining
% the formula and some of the preceding text were not visible in the pdf output, I fixed this by removing the strip environment, see comment in sec4.tex.
%\begin{strip}
\begin{multline}
    \frac{dp_\gamma(k)}{dx_i} = 
    %\Biggl[  
    \sum_m \biggl[ P_\text{tr}(m) G(m,k) \\ \nonumber 
    \Bigl( \frac{\mathcal{T}(m)-T(k)}{\sigma_\mathcal{T}^2(m)} \Bigl(\frac{d\mathcal{T}(k)}{dx_i} - \frac{d\mathcal{T}(m)}{dx_i}\Bigr) \\
    + \frac{[\mathcal{T}(m)-\mathcal{T}(k)]^2}{2\sigma^4_\mathcal{T}(m)} \frac{d\sigma^2_\mathcal{T}(m)}{dx_i}  
    -\frac{1}{(2 \pi \sigma^2_\mathcal{T}(m))^{1.5}} \frac{d\sigma^2_\mathcal{T}(m)}{dx_i} \\ + \frac{1}{P_\text{tr}(m)} \frac{dP_\text{tr}(m)}{dx_i} \Bigr) \biggr] 
%    \times \sum_m P_\text{tr}(m) \\ - \sum_m P_\text{tr}(m)G(m,k) \sum_m \frac{dP_\text{tr}(m)}{dx_i}   
    %\Biggr] 
    %\frac{1}{\sum_m P_\text{tr}(m)}
\end{multline}

% in the code:
% the relevant lines are
% dnum_dxg += pi * pow(Gden,2.) * dsigma2_dx[id][m];
% so the pi factor and the correct power of the norm factor of the gaussian are ok. Instead the description a few tens of lines earlier was incorrect, now fixed it.

%\end{strip}

\noindent
where we have used the concise expression\par 

\begin{equation}
    G(m,k) = \exp[-\frac{(\mathcal{T}(m)-\mathcal{T}(k))^2}{2 \sigma^2_T(m)}]/(2 \pi \sigma^2_\mathcal{T}(m))^{0.5}.
    \end{equation}

\noindent
Similar expressions to those above are found for derivatives versus $y_i$ and for derivatives of $p_p(k)$. 

Finally, we may consider the contributions to $dU_{GF}/dx_i$ of the derivatives of $R_\text{tot}$ and $N_\text{trials}$ factors. As far as the former is concerned we have simply \par

\begin{equation}
    \frac{dR_\text{tot}}{dx_i} = \frac{1}{N_\text{batch}} \frac{x_i}{R_i}
\end{equation}

\noindent
where $R_i$ is the distance of detector $i$ from the origin of axes. For $N_\text{trials}$, which is not differentiable, we need instead to resort to an approximate calculation. We consider the distance $R_{it}$ ($t=1...N_\text{trials}$) of each shower core generated in the trial procedure from detector $i$, and then:\par

\begin{enumerate}
    \item if the trial failed to be included in the set of $N_\text{batch}$ showers, we compute varied distances from shower $t$ $R'_{it,\Delta x}$, $R'_{it,\Delta y}$ corresponding to detector $i$ moves by arbitrary increments $\Delta x$, $\Delta y$ in turn (setting them to the typical movement of detectors at each epoch); if the shower meets the $R<R_\text{slack}$ criterion after the move, counters $\Delta N_x$, $\Delta N_y$ are correspondingly increased;
    \item if the trial shower is part of the batch, we test if its distance from detector $i$ is the only one below the threshold $R_\text{slack}$; in that case, we determine if a movement $\Delta x$ or $\Delta y$ would cause the criterion to fail, and if so we decrement counters $\Delta N_x$ and/or $\Delta N_y$ accordingly.
\end{enumerate}

\noindent
At the end of the procedure, the needed derivatives of $N_\text{trials}$ are simply found as \par

\begin{equation}
    \left( \frac{dN_\text{trials}}{dx_i}, \, \frac{dN_\text{trials}}{dy_i} \right) = \left( \frac{ \Delta N_x}{\Delta x}, \, \frac{\Delta N_y}{\Delta y} \right).
\end{equation}

In our optimization runs we have observed that the inclusion of derivatives of the density term $\rho$ in \autoref{eq:U1} usually dominates the total gradient, causing an expansion to very large radii of the array. When we study the interplay of the gamma flux term \autoref{eq:U1} with the other terms, we have found useful to remove the density-related derivatives. They are instead used in the default mode of calculation, when we include in the utility a penalty factor that accounts for the cost of preparing and instrumenting larger areas (see \autoref{s:penalty}).

\subsubsection{Derivatives of $U_{IR}$ term}

The second term in the utility function defined {\it supra}, $U_{IR}$, contains an explicit dependence on the estimated uncertainty on the energy of gamma-induced showers, as well as terms depending on the triggering probability and weight of each gamma shower. The derivative of the term by the displacements $dx_i$, $dy_i$ is straightforward to obtain. Here we summarize the calculation of the derivative with respect to $dx_i$ for detector unit $i$: \par

\begin{equation}
    \frac{dU_{IR}}{dx_i} = \frac{N'D - ND'}{D^2}
\end{equation}

\noindent
with

\begin{align}
    N &= \sum_k \left[ P_\text{tr}(k) W(k) \right] \notag\\
    D &= \sum_k P_\text{tr}(k) W(k) \frac{\sigma_{E_k}}{E_{k,t}}\notag\\
    N'&= \sum_k  \frac{dP_\text{tr}(k)}{dx_i} W(k) \notag\\
    D' &= \sum_k \biggl[ \frac{dP_\text{tr}(k)}{dx_i} W(k) \frac{\sigma_{E_k}}{E_{k,t}} + \\
    P_\text{tr}(k) W(k) \frac{d \sigma_{E_k}}{dR} \frac{dR}{dx_i} \frac{1}{E_{k,t}} \biggl].
\end{align}

\noindent
where $k$ sums run on gamma showers in the batch. 
%We wrote the expression above using the derivative of the inverse uncertainty $\frac{1}{\sigma_{E_k}}$ because it is more straightforward to obtain. 
We computed and verified the expression with \texttt{Mathematica}; the derivation is tedious and we omit it from this text. However, the derivative of the uncertainty depends both directly and implicitly (through the dependence of $E_{k}$ on it) on the distance $R_{ik}$ of each detector $i$ from the core of shower $k$. 
The extraction of the implicit dependence $d E_k/dR_{ik}$
is a part of the calculation that requires some workaround; we describe the essential steps below.

As the measured energy $E_k$ is the result of the maximization of a likelihood function composed of a large number of terms, computing analytically its dependence on the distance from one detector unit $R_{ik}$ is unfeasible. We resort to observing that if the likelihood reaches a maximum when evaluated for fitted shower parameter $E_k$, and for a given value of $R_{ik}$, then a variation $R_{ik} \to R'_{ik} = R_{ik}+\delta$ will move the solution to $E'_k$ when \par

\begin{equation*}
\frac{d \ln {\cal{L}}(E_k,R_{ik})}{dE_k} = 0 
\end{equation*}
\begin{equation*}
\frac{d \ln {\cal{L}}(E_k+\delta E,R_{ik}+\delta R)}{dE_k}=0
\end{equation*}

\noindent
so we have \par

\begin{equation}
\delta R = - \delta E \frac{\frac{d \ln\cal{L}}{dE_k}}{\frac{d \ln \cal{L}}{dR_{ik}}} 
\end{equation}

\noindent
from which we solve for the derivative $dE_k/dR_{ik} \simeq \delta E / \delta R$.

\subsubsection {Derivatives of the $U_{PR}$ and $U_{PS}$ expressions} 

The calculation of derivatives of the pointing resolution utility term  $U_{PR}$  with respect to detector positions $x_i$, $y_i$ is straightforward. The essential ingredients are the derivatives of the triggering probability of each gamma shower
 $dP_\text{tr}(k)/dx_i, \, dP_\text{tr}(k)/dy_i$  and the derivatives of the measured angular parameters $d \theta_k /dx_i$, $d \phi_k / dy_i$. Again, the complete solution is quite long and tedious, and we do not report it here; it has been computed using the \texttt{Wolfram Mathematica} program.

 Similarly, the derivative of $U_{PS}$ with respect to a variation of the position $x_i,y_i$ of each of the detectors is straightforward to compute, once we accept that in analogy to what happens for the $U_{IR}$ function discussed {\it supra} it requires the indirect evaluation of a variation of the energy estimate as a function of the distance of shower core to detector position, $dE_k/dR_{ik}$.

\section{Performance comparison of the SWGO benchmark arrays}
\label{sec:performance}
\label{s:benchmarks}

 Before exploring the performance of the optimization technique we described in the previous sections, we provide in this section a relative appraisal of the 13 SWGO benchmarks according to the definitions of an experiment-wide utility function we discussed in \autoref{s:utility}.

As mentioned in \autoref{sec:swgo}, our shower model is approximate --it does not, {\it e.g.}, describe disuniformities in the EAS and sub-shower features, nor the angle of incidence of secondaries on the ground or the different distribution of arrival times of different particle species. In addition, the detector units are considered in the simplistic hypothesis that they detect with \SI{100}{\percent} efficiency the relevant particles (electrons and positrons, photons, and muons) that intersect their top surface, provided that they have energy in excess of \SI{10}{\mega\eV} (see \autoref{s:methodology}); only a 5\% uncertainty in the number of detected particles is introduced by the model. Some of the above shortcomings can be addressed relatively easily; they will be studied in future extensions of this analysis. Their impact in the relative value of different configurations is not the focus of the work presented here, and it will be considered only after a verification of the shower model with full simulations of the optimized layouts. Regardless of the above shortcomings, the model we developed can still be used for a fast appraisal of the 13 layouts that have been proposed for the SWGO array. This comparison is useful as a point of comparison with what other independent studies may be finding, and as a preliminary step to an appraisal of the configurations produced by runs of the optimization pipeline, as discussed in \autoref{s:results}. 

In \autoref{t:layouts} we detail the result of runs with the 13 SWGO array configurations (\autoref{t:layouts_sec2} and \autoref{f:layouts}), which can be used to compare the performance of the configurations based on the utility definitions of \autoref{s:utility}. 
By examining the presented results one notices that, unsurprisingly, for the sake of maximizing both the flux utility $U_{GF}$ and the point-source utility $U_{PS}$, the most performing arrays are those with a larger number of detection units spread out to higher radii; other interesting differences are also evident in the pointing resolution and integrated resolution results.

Using the definition of the utility function we have adopted for these exploratory studies, we observe a significant spread in the scientific value of different arrangements. However, by considering tanks of standard \SI{1.91}{\meter} radius in all cases, we have here neglected the different specifications of detector units that ought to be used in some of the above benchmark configurations. The use of larger tanks in any of the designs would cause a corresponding increase of the tabulated utility.

Taken at face value, the results show a mild dependence of both utilities with the number of units $N_\text{det}$, and a much higher dependence with $R_\text{max}$. Indeed, the most performing design among the 13 considered is $A7$ (6571 tanks, max radius \SI{1200}{\meter}). These results are however only demonstrative and should not be used for a relative appraisal of the benchmarks.

\begin{table*}[h!]
\begin{center}
\begin{footnotesize}
\begin{tabular}{l|c|c|c|c|c|c|c|c}
ID & $N_{det}$ & $R_{max} (m)$ & $U_{GF}$ & $U_{IR}$ & $U_{PR}$ & $U_1$ & $U_{PS}$ (2 PeV) & $U_{PS}$ (6 PeV) \\
\hline
A1 & 6589 & 300  & $657 \pm 15$ & $846 \pm 27$ & $781 \pm 4$ & $2284 \pm 20$ & $2445 \pm 83$ & $2501 \pm 44$ \\
A2 & 6631 & 600  & $776 \pm 12$ & $959 \pm 20$ & $1024 \pm 3$ & $2759 \pm 21$ & $3232 \pm 79$ & $3600 \pm 76$ \\
A3 & 6823 & 600  & $755 \pm 13$ & $969 \pm 22$ & $1021 \pm 4$ & $2745 \pm 21$ & $3248 \pm 85$ & $3559 \pm 95$ \\
A4 & 6625 & 600  & $737 \pm 12$ & $924 \pm 23$ & $966 \pm 4$ & $2626 \pm 23$ & $3008 \pm 95$ & $3269 \pm 55$ \\
A5 & 6541 & 300  & $643 \pm 11$ & $914 \pm 22$ & $850 \pm 3$ & $2407 \pm 18$ & $2739 \pm 80$ & $2808 \pm 29$ \\
A6 & 6637 & 300  & $632 \pm 8$ & $914 \pm 15$ & $731 \pm 2$ & $2277 \pm 13$ & $2372 \pm 45$ & $2437 \pm 22$ \\
A7 & 6571 & 1200  & $1174 \pm 10$ & $973 \pm 16$ & $1262 \pm 3$ & $3409 \pm 14$ & $4454 \pm 83$ & $5159 \pm 67$ \\
B1 & 4849 & 300  & $609 \pm 8$ & $884 \pm 14$ & $784 \pm 2$ & $2276 \pm 11$ & $2639 \pm 78$ & $2579 \pm 27$ \\
C1 & 8371 & 300  & $727 \pm 13$ & $827 \pm 19$ & $752 \pm 3$ & $2306 \pm 17$ & $2350 \pm 99$ & $2445 \pm 26$ \\
D1 & 3805 & 300  & $540 \pm 14$ & $1027 \pm 28$ & $795 \pm 4$ & $2361 \pm 26$ & $2603 \pm 90$ & $2846 \pm 47$ \\
E1 & 5461 & 300  & $644 \pm 7$ & $847 \pm 15$ & $786 \pm 2$ & $2276 \pm 14$ & $2505 \pm 76$ & $2553 \pm 22$ \\
E4 & 5455 & 600  & $755 \pm 8$ & $866 \pm 14$ & $1002 \pm 2$ & $2622 \pm 15$ & $3228 \pm 83$ & $3387 \pm 34$ \\
F1 & 4681 & 300  & $606 \pm 7$ & $911 \pm 14$ & $815 \pm 2$ & $2331 \pm 13$ & $2676 \pm 57$ & $2751 \pm 25$ \\
% above, 679 entries out of 764
\end{tabular}
\caption{\label{t:layouts} Utility values for the 13 SWGO benchmark layouts, computed with multiple runs of 2000 showers.   These values are computed with layouts  all employing the same tank of $R=1.91$ m radius, while the original configurations  are meant to have  tanks of different radii. The comparison is thus useful at the level of the relative worth of different geometries and number of units, but it does not  translate into an assessment of the relative performance of the original benchmarks.  %\mdc{I would remove at least one significant digit if not two in the uncertainty and adapt the value of utility. Possibly, I woudl use the scientific notation for better reading}
}
\end{footnotesize}
\end{center}
\end{table*}
\section{The end-to-end optimization loop}
\label{sec:end-to-end}

 In this section we provide a detailed description of the optimization loop, which requires a number of parameters to be defined by the user in order to precisely reflect the specific question to which an answer is sought. 

The optimization loop runs on the pipeline described in blocks in \autoref{s:methodology}. Following an initialization of the optimization run parameters, a definition of the starting layout of the detector array, and of the shower model, the loop aims at computing the utility of the current configuration and the derivatives $d U/dx_i$, $dU/dy_i$ of the utility with respect to the $x_i$ and $y_i$ positions of each detector unit $i$; the derivatives are used to update all detector positions, when the loop can be repeated.

The computing-heavy parts of the optimization cycle are two: a first cycle on the requested gamma-ray and proton showers, which is needed to generate and reconstruct showers by likelihood maximization under the two hypotheses in turn; and a cycle on detector units $i=1...N_{det}$, when derivatives of the utility are computed with respect to positions $x_i$ and $y_i$. These two cycles constitute the bulk of the computation, and are performed by multi-threading loops, when each of the $N_\text{CPU}$ available CPUs is tasked with reconstructing $1/N_\text{CPU}$ of the showers in the first cycle, and then with computing derivatives for $1/N_\text{CPU}$ of the detectors in the second cycle. This arrangement speeds up the execution of the program by a factor that approaches $N_\text{CPU}$, if the number of showers and detectors are both integer multiples of that value.

The code is available in both a standalone {\ttfamily c++} version and as a \texttt{ROOT} macro \citep{Brun:1997pa}. Running under \texttt{ROOT} allows the user to monitor the execution of the program through some service histograms that show current configuration, utility, and a wealth of other metrics, and is useful for debugging purposes; no multi-threading has been implemented in the \texttt{ROOT} macro version at the moment.

\subsection{ Initialization and run parameters}

A starting configuration for the array must be specified during the initialization of the program. This involves deciding a value of the following parameters:\par 

\begin{enumerate}
    \item the number of tanks comprising the array, $N_\text{det}$;
    \item the initial location of each tank on the ground, $x_i,y_i$;
    \item whether some of the tanks are defined as fixed on the ground (in case one wishes to explore the optimization of only a part of the array layout);
    \item the size of each tank, assumed circular, $R_\text{tank}$; all tanks are assumed to be of the same size in the runs we performed so far.
\end{enumerate}

\noindent
The successful execution of an optimization loop also requires the user to specify a number of other parameters that define the behavior of the program and the calculations it performs. We provide here a list to give a view to the functionality and options that are currently implemented in the code:\par

\begin{itemize}
    \item the number of showers generated at each epoch for the calculation of the PDF of the test statistic of proton and gamma-ray showers, $N_\text{events}$;
    \item the number of showers generated at each epoch to test the performance of the array, $N_\text{batch}$; this number is usually chosen to be equal to $N_\text{events}$ for simplicity;
    \item the number of epochs (updating loops) of the optimization task;
    \item the starting epoch. This is useful in case an optimization run is performed on a configuration that has already been worked at previously, so that all the program outputs retain the same numbering;
    \item whether starting detector positions are to be read in from an input file;
    \item an integer parameter that defines the shape of the initial generated array. There are several options available, from a circular uniformly filled array, to sets of detectors placed in a circle at fixed radius, to two or three such sets, or to a set of random values within a square. Special values of this parameter in the range $101-113$ are reserved to specify the shape of the array among the 13 benchmarks described in \autoref{sec:swgo};
    \item the starting learning rate for detector movements;
    \item the coefficients of the utility related to the flux precision, $\eta_\text{GF}$ (this has by default the value of 1.0);
    \item the coefficients of the utility related to the energy resolution part, $\eta_\text{IR}$, and to the pointing resolution $\eta_\text{PR}$;
    \item the coefficients of the components of the utility related to the cost of the spatial extension and sum of lenghts from the center of the array;
    \item whether the point-source utility $U_{PS}$ is elected as the objective of the optimization instead;
    \item if $U_{PS}$ is being optimized, the user must then choose the value of the energy of photons from the source;
    \item the minimum number of tanks that are required to record at least one secondary particle, $N_\text{trigger}$, in order for the shower to be considered and reconstruction to be performed; in case macro-tanks composed of multiple units are used, the number refers to individual tanks;
    \item the slope of the power law determining the energy distribution of generated showers, if the optimization objective is $U_{GF}$;
    \item a multiplier of the utility derivatives that modulates how large are the movements of the detector positions during an update.
\end{itemize}

\noindent
Additional options include the turning off of reconstruction of specific parameters for showers ({\it e.g.}, angles, core position, or energy), the option of generating showers orthogonal to the ground (to simplify reconstruction for sanity checks and debugging purposes), and the number of grid points scanned in the initialization of shower parameters for their reconstruction, if the associated parameter controlling this functionality is turned on.

The specification of some of the above parameters is liable to affect the correct working of the algorithm and therefore its success in improving the utility function. The program includes default values for all parameters, as well as warning printouts and even user-independent hard-resets when parameters are initialized to values too far off their intended range of operation; but it is useful to mention here a couple of points of relevance anyway. 

1) The user must be careful to choose a not too small number of showers, $N_\text{events}$ in the generation of the PDF of the likelihood ratio test statistic $\mathcal{T}$ (see \autoref{eq:Tdef}). This is because $\mathcal{T}$ is estimated at each epoch by summing together $N_\text{events}$ Gaussian distributions centered at the location of each reconstructed shower (and of sigma equal to each estimated uncertainty), as in \autoref{eq:Tconstr}, and the procedure may yield non-smooth PDF distributions otherwise. A ``bumpy'' PDF is not only in general a poor estimate, but it produces erratic results when its gradient is used for the maximization of the $U_{GF}$ utility term.

2) The coefficients of the area and length cost parts of the utility must be properly sized in order to produce the intended effect. For example, a too high relative value of the area cost (as driven by parameter $\eta_{TA}$ in \autoref{eq:areacost}) has the unwanted effect that the utility gradients may ``freeze'' any outward motion of detectors placed on the outer rim of the current array, preventing the exploration of a common change of configuration which may be overall advantageous to the full utility. This happens for large $\eta_{TA}$ when during the optimization loop a configuration has expanded to reach the area $A_0$ (see again \autoref{eq:areacost}). The array will then not be able to modify its outer shape to a more effective one spanning the same area, because no detector on the outer rim will be able to move to a position that causes a temporary increase of the area (which could be reabsorbed by a successive update of another detector's position). 

One additional parameter called \texttt{CommonMode} must be discussed separately. This variable is used to decide whether detector positions are updated independently at each iteration of the gradient descent loop (\texttt{CommonMode}=0), \par

\begin{align}
    x_i &\to x'_i = x_i + \eta(i) \frac{dU}{dx_i} \nonumber \\
    y_i &\to y'_i = y_i + \eta(i) \frac{dU}{dy_i}
\end{align} 

\begin{figure*}[h!]
\includegraphics[width=0.99\linewidth]{ 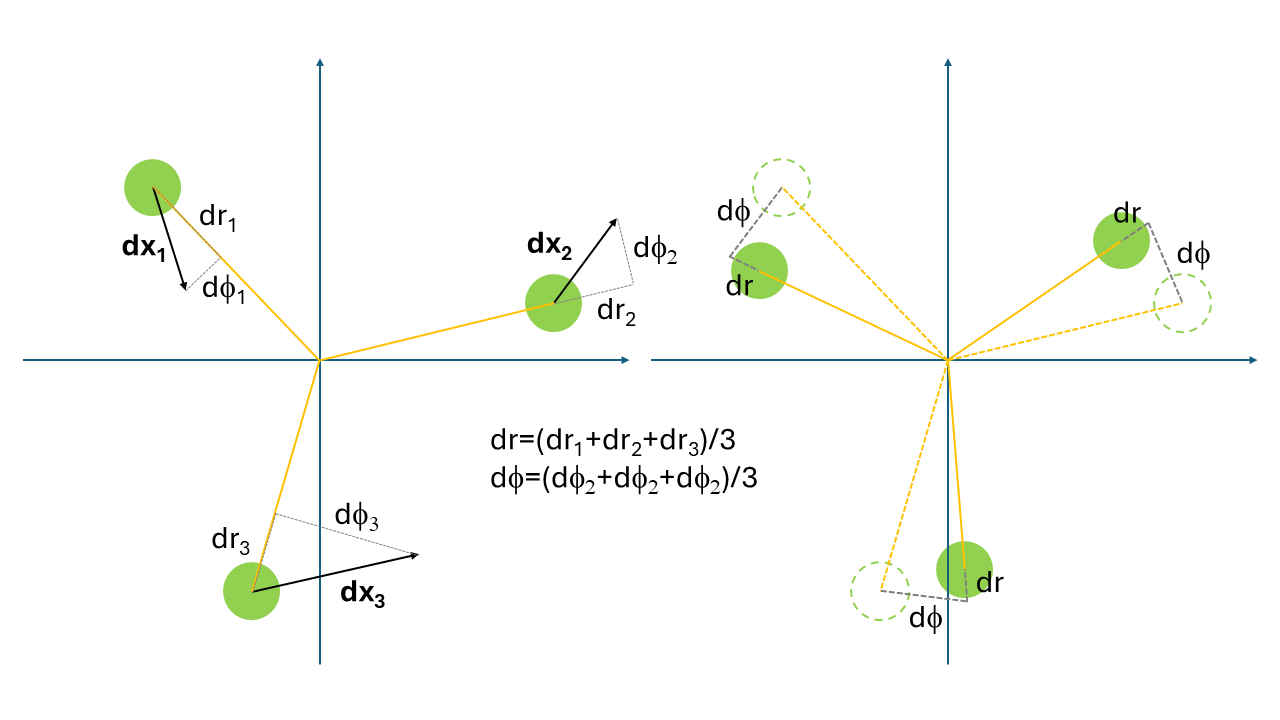}
\caption{Illustration of the gradient update of a triplet of detectors when CommonMode=3. The gradient of each unit is projected along radial and azimuthal components (left), and their averages are used for the update (right). }
\label{commonmode3}
\end{figure*}

\noindent
(where $\eta(i)$ is the total learning rate of unit $i$, see {\it infra}), or whether instead a common movement is computed for classes of detectors that share the same radius (\texttt{CommonMode}=1), or the same symmetrical position around the origin, with a multiplicity for each class decided by the user through the value of \texttt{CommonMode}. For example, if \texttt{CommonMode} is set to three, the initial layout of the detector units is set up such that triplets of detectors are set at the same radius and at azimuthal positions around the origin offset by 0, 120, and 240 degrees, such that they form the pattern of an equilateral triangle. After individual gradients are computed, the new position of each element of the triplet is determined by computing the average radial displacement of the three units, and the average azimuthal rotation (see \autoref{commonmode3}). The three units positions are then updated according to those averages. The procedure reduces the stochasticity of the utility derivatives, and at the same time ensures that the overall layout of the detector retains symmetry according to the multiplicity specified by the parameter. Since this strategy has proven very effective in allowing faster convergence of the optimization loops, most of the results shown in this document are produced with \texttt{CommonMode}=3.

When including in the problem physical constraints that detector must obey to (see {\it infra}, \autoref{s:constraints}), the symmetry of multiplets of detectors around the original center of the array may have to be broken --{\it e.g.}, only one unit out of three in a triplet may have to be displaced to the boundary of a physically allowed region after a gradient update brought it in a forbidden area, while the other two are free to move to the new positions. When that happens, the detector forcibly displaced from the symmetric position gets decoupled from the others, and future gradient updates will move it independently of the others in the multiplet, which will instead continue to move following the combination of their gradients as if the missing element were still in place (but without its contribution to the common gradient update in radius and azimuthal rotation). This arrangement allows to retain approximate symmetry of the array and to exploit the beneficial averaging of noisy gradients, while fulfilling the necessary external constraints on the ground.

\subsection {Learning rate scheduling}

In order to facilitate convergence to a meaningful configuration corresponding to a stable maximum of the utility, a learning rate must be specified and carefully adapted to the evolving task. Each unit (or set of units, if the optimization is run with \texttt{CommonMode} equal to 2 or larger) needs to move by an amount that is on average smaller than the distance of the unit to its neighbors: this improves the effectiveness of the procedure of independently moving each unit in series, as it minimizes the correlations of the effect of movements of neighboring units.

Another condition that the learning rate must fulfil is to slowly decrease as a function of the epoch number, to let the system slowly drift to an optimal configuration. This trend may be however modulated by including oscillatory increases and decreases: such a schedule often helps the search in complex multi-dimensional spaces, enabling the system to jump out of local minima. We have experimented with a few different options and finally settled on an exponential decrease of the learning rate, which gets damped by a factor of $\exp(\alpha_0)$ from the start to the end of the optimization, while oscillating with a squared cosine shape: \par

\begin{equation}
    \eta_{c} = \exp(- \alpha_0 x/N_\text{epochs}) \left[ \alpha_1+(1-\alpha_1) \cos^2(\alpha_2 x/N_\text{epochs}) \right]
\end{equation}

\noindent
where $x$ is the epoch, ranging from 0 to $N_\text{epochs}$. Suitable values for $\alpha_{0,1,2}$ are, {\it e.g.}, $(5,0.3,20)$, whereby the damping over the full cycle amounts to a factor of $\exp(5)$ with about six oscillations. The above factor is a common learning rate, and it applies to all units. On top of that, it is useful to track the individual movements of the units, and correct the learning rate of each unit differently, increasing the mobility of units that consistently move in some direction and damping the mobility of units that oscillate around some particular position. To achieve that goal, the position of each unit is recorded during gradient descent, and the following quantity is evaluated at each epoch:\par

% see comment in sec4.tex
%\begin{strip}
\begin{align}
\resizebox{0.99\hsize}{!}{$
    \cos \theta_{\text{eff},i} = \frac{ (x_i^n - x_i^{n-1})(x_i^{n-1}-x_i^{n-2})+(y_i^n-y_i^{n-1})(y_i^{n-1}-y_i^{n-2})}{\sqrt{\left[(x_i^n-x_i^{n-1})^2+(y_i^n-y_i^{n-1})^2\right]\left[(x_i^{n-1}-x_i^{n-2})^2+(y_i^{n-1}-y_i^{n-2})^2\right]}}
    $}
\end{align}
%\end{strip}

\noindent
The factor above is used together with the global scheduling in determining the overall multiplier of each unit's utility gradient: \par

\begin{equation}
    \eta(i) = \eta_c \exp(k \cos \theta_{\text{eff},i}) 
\end{equation}

\noindent
with $k$ set to producing a small correction per epoch, $k=0.05$; the build up of the correction may however have a significant effect over several epochs (typically, a damping one). Finally, the displacement of each unit in $x$ and $y$ is capped at a maximum allowed value, to prevent abnormally high gradients from disrupting the smooth learning of configurations. This maximum value is usually set to be equal to the initial detector spacing.

\subsection {Miscellanea}

A number of other details need to be specified in order to produce a working model of reconstruction, inference, and meaningful utility gradients. We list here some of them.

\begin{itemize}
    \item In the optimization procedure we are implicitly assuming that stochastic gradient descent will allow the progressive drift of an initial layout to a more advantageous configuration. This is reasonable inasmuch as it is akin to the independent update of each of the weights and biases of a neural network layer. Detector positions are modified in series, one by one, without a guarantee that the combination of the movements produces an overall improvement in the utility. The guarantee of continuous local utility increases would only exist for infinitesimal displacements, but the large CPU demand of the task mandates that the learning rate multiplier of each unit's utility gradients be finite and not too small. We have observed that indeed the learning rates are a critical parameter in the success of an optimization run. We have so far handled this issue by trial and error.
    \item While tanks (or macro-tank aggregates, such as those we employ in our initial investigations) have a physical size (a $R_{tank}=$ 1.91 m radius), the calculation of utility gradients and of consequent required movements are not informed by the position of nearby tanks.  This creates the possibility that the calculated ending position of multiple tanks, following a gradient update, is incompatible with their physical size. An iterative procedure is run every 10 epochs at the end of the position update cycle, to resolve physical overlaps. The procedure identifies all pairs of tanks placed by the gradient update at a distance of less than $2 R_{tank}$ from each other, starting with the closest pair, and moves the two units away by an equal amount along the line connecting their centers until they are at a distance of $2 R_{tank}+S$, with $S=0.6$ m a minimum allowed spacing that ensures that tanks can be physically assembled in that position. The procedure converges quickly to a practically achievable configuration.\footnote{ Since this procedure is operated after the check of physical boundaries on the ground corrects for the presence of detector units in a forbidden region (see below, \autoref{s:constraints}), it may happen that a detector gets temporarily displaced into the forbidden region by the overlap removal. This is a small effect and we currently ignore it, especially since it is only relevant if it happens at the last epoch. }
    
\end{itemize}

 \subsection{Macro-tank aggregates \label{s:macrotanks}}
 
\begin{figure}[h!]
 \begin{center}
     \includegraphics[width=0.9\linewidth]{ 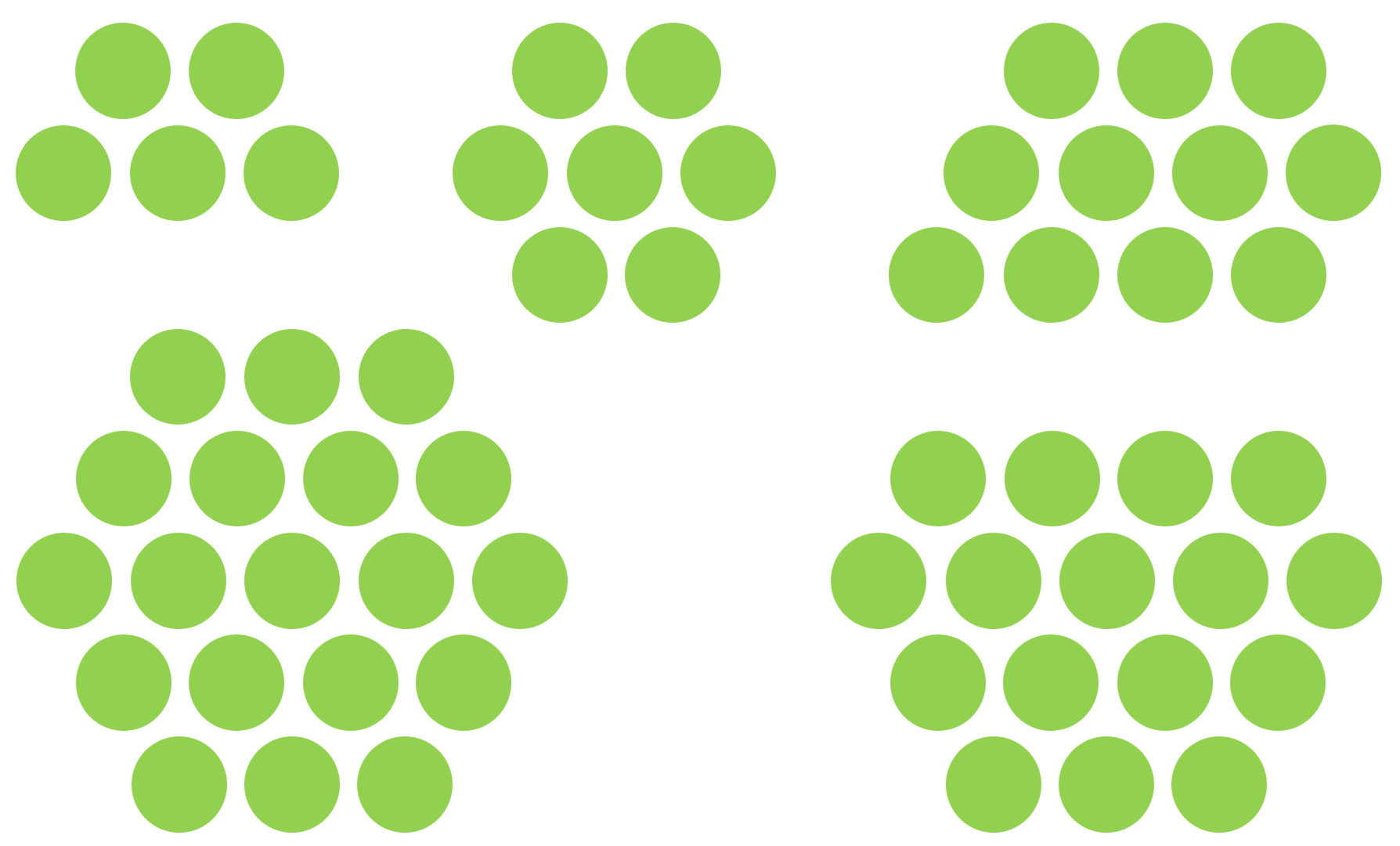}
 \end{center}
\caption{ Sample layout of macro-units comprising 5, 7, 11, 16, and 19 tanks. See the text for detail.} 
\label{f:macrotank}
 \end{figure}

 In order to reduce the computing load and carbon footprint of our optimization runs, for the present work we consider (as we have done in the tests of physical constraints of the previous section) ``macro-tank'' aggregates of 19 or 37 units of 1.91 m radius, tightly packed in a hexagonal setup (See \autoref{f:macrotank}). 
 
More in general, in any conceivable setup, individual units must be separated by a minimum distance of $S=0.6$ m in order to allow for construction and servicing. We have implemented this constraint in our macro-tank templates. Some of the configurations that correspond to the above recipe are shown in \autoref{f:macrotank}. Similarly to individual units, macro-tanks should also be separated by, at a minimum, the same $S$ distance on the ground. The minimum spacing $D$ between a pair of macro-tanks is determined by the number of units in the aggregate $N$, their radius $R$, and the spacing $S$ as follows: we first define the number $k$ of rings for $N$ units set in a hexagonal pattern as \par

\begin{equation}
k = \min \left\{ k \mid 1 + 3(k^2 + k) \geq N \right\}
\end{equation}

\noindent
(where $k=0$ indicates a single tank with no rings around), and then we determine a minimum distance between pairs of macro-tanks that guarantees the needed spacing for any orientation of the detectors placement around the center, \par

\begin{equation}
    D = [(2k+1)(2R+S)].    
\end{equation}

Below is a summary of the relation of $k$, $N$, and $D$ for $N\leq 61$.

\begin{center}
\begin{tabular}{c|c|c}
   $k$  & $N$ & D \\
   \hline
   0  & 1 &  4.42 m   \\
   1  & 7 &  13.26 m \\
   2  & 19 &  22.1 m \\
   3  & 37 &  30.94 m\\
   4  & 61 &  39.78 m\\
\end{tabular}
\end{center}

\subsection{Other physical constraints \label{s:constraints}}

Although the utility we discussed in \autoref{s:utility} includes optional terms that relate to the cost of the spatial extension of the array and the distance of units from the center of the array, those terms cannot guide the movement of units within specific physical constraints that may be present on the ground. In a real-world application, the ground space available to the experiment will not extend indefinitely in all directions, and there may be specific hindrances that must be taken into account in the search of optimal layouts. 

\begin{figure}[h!]
\includegraphics[width=0.99\linewidth]{ 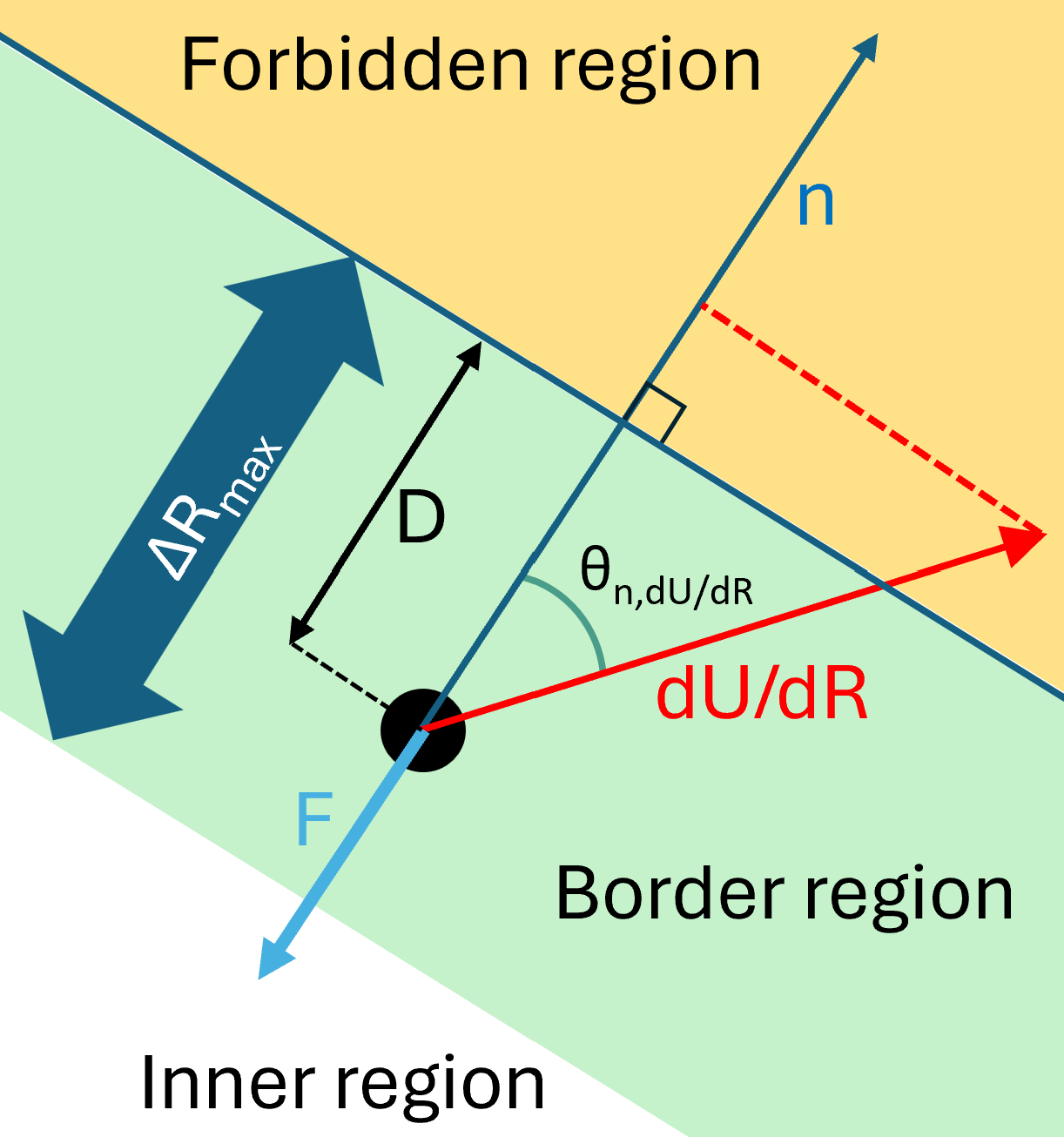}
\caption{Explanation of the repulsive force expressed by the
boundary of the allowed region. See the text for detail.}
\label{f:repulsivepotential}
\end{figure}

Indeed, the presence of non-symmetric constraints makes the problem much richer than what it is in a unconstrained scenario, as they break the cylindrical symmetry of the problem in non trivial ways. In the code we have implemented the possibility of specifying a set of linear or circular boundaries on the ground, which the detectors cannot overtake during optimization updates. The procedure consists in verifying whether the update of a detector position, $(x,y) \to (x',y')$ would place it in the forbidden region of the plane, in which case the position $(x',y')$ is further updated by following the direction normal to the boundary line back to the point where it intersects the boundary. Before the update, a repulsive gradient is also generated whenever a detector is sitting within a distance to the boundary smaller than the maximum length of the possible gradient update step; the direction of the generated repulsive gradient is again orthogonal to the boundary line, directed inward, and its magnitude is equal to a variable fraction of the component of the original gradient of the utility projected along the outward direction of the same orthogonal versor (see \autoref{f:repulsivepotential}):\par

\begin{equation}
F_i = - (1-\frac{D_i}{\Delta R_{max}}) \eta_i \left( \frac{dU}{dR} \right)_i \cos{\theta_{n,(dU/dR)_i}}
\end{equation}

\noindent 
where $D_i$ is the distance from the boundary of detector $i$, $\Delta R_{max}$ is the maximum distance allowed for a gradient update step of the detector, and $\theta_{n,(dU/dR)_i}$ is the angle between the normal to the boundary and the utility gradient of detector $i$. The addition of a repulsive force $F$ to the gradients of the utility effectively dampens the component of motion of tanks toward the boundary, if they are already within a close range from it.

\begin{figure}[h!]
\includegraphics[width=0.99\linewidth]{  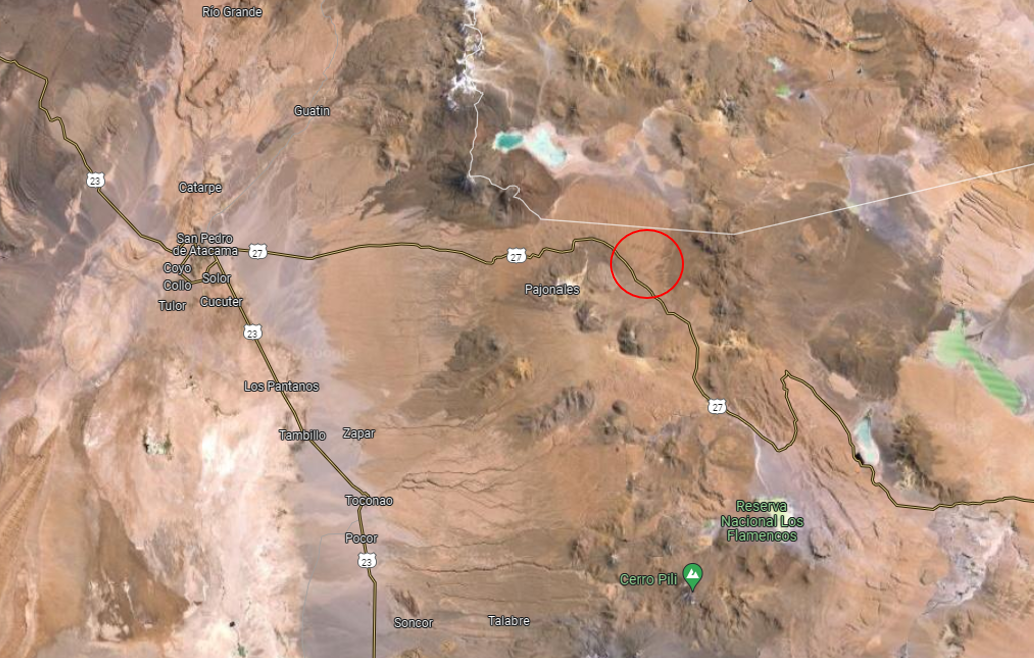}
\includegraphics[width=0.99\linewidth]{ 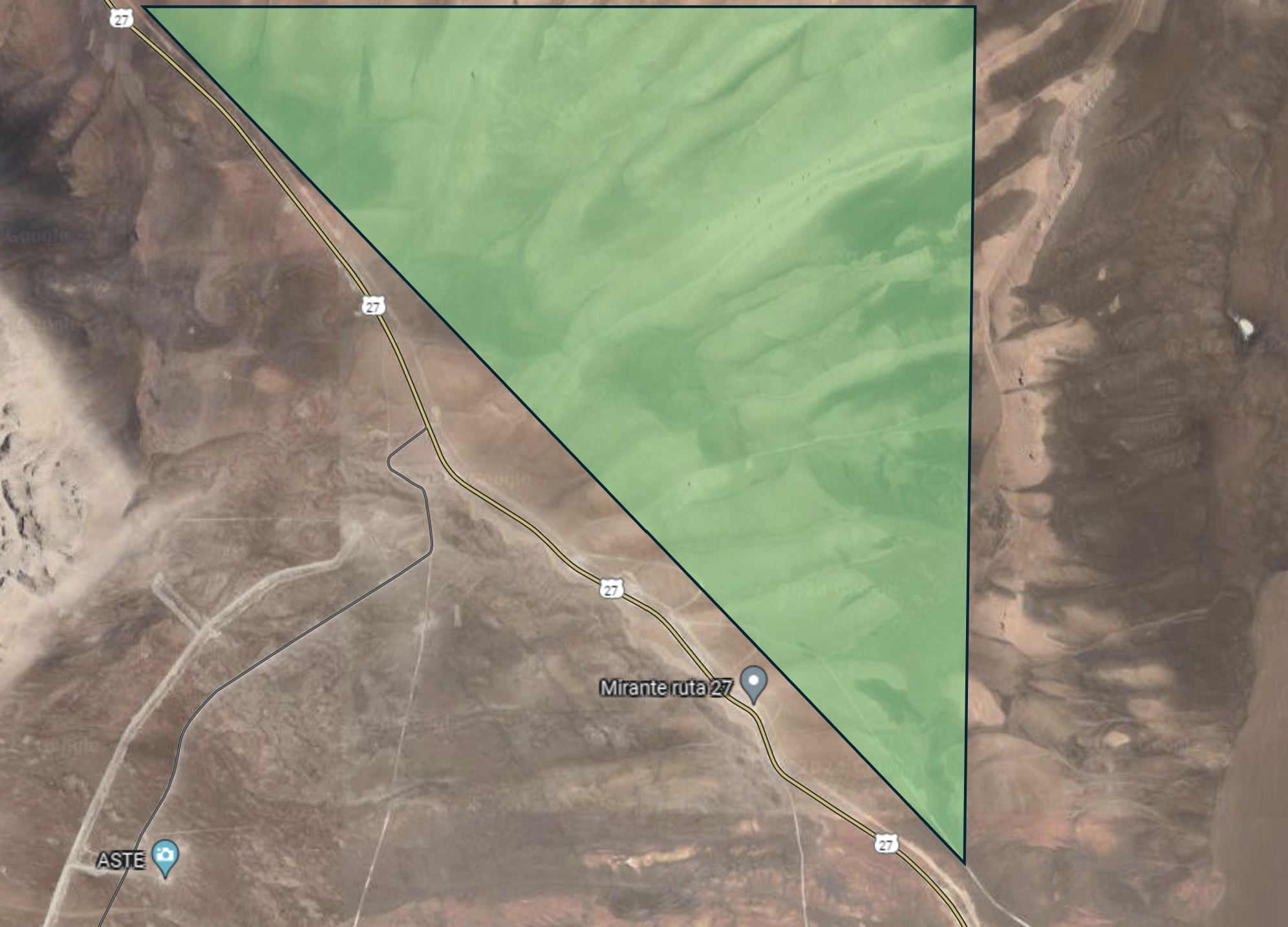}
\caption{ Aerial view of the Pampa La Bola site in the Chilean Andes. The triangle in the zoomed view at the bottom shows the flat area considered for checks of the boundary constraints in the code.} 
\label{f:Pampalabola}
\end{figure}

We tested the procedure by considering the main site proposed for the Chilean option of SWGO, in the Pampa La Bola area of the "Parque Astron\'omico" near the Atacama desert. The region is an excellent choice for the construction of SWGO, as it is located at very high altitude (4800m above sea level), and it has an extended flat region comprised within the area allocated by the Chilean government to the Parque Astron\'omico (see \autoref{f:Pampalabola}). By studying the topography and by direct recoinnassance visits on the ground, we identified a triangle of flat ground extending by 5.3 by 5.5 km where the experiment could deploy its tanks; the area lays entirely within the boundaries of the Parque Astron\'omico, and does not intersect public roads. The identified area is the default region for optimization runs aiming for tests of this functionality.

\section{Results}
\label{s:results}
\label{sec:results}

 In this section we present a few sample results obtained with the algorithm developed as discussed {\it supra},  and comment on the improvements the software may achieve. Due to the large CPU demand of the optimization loop, we have not yet studied systematically the full problem of the layout of O(6000) detector units; such a task should only be performed once the program is modified to include a detailed and close-to-final modeling of tank efficiencies and response versus angle, number of particles, and energies, as well as a shower model capable of reproducing all features of secondary particle flux generated by \texttt{CORSIKA}. The purpose of the results presented in this section is thus  mainly 
  demonstrative: we wish to  show  how the optimization pipeline is capable of identifying non-trivial solutions that maximize the given utility, and how such solutions share common features that, {\it a posteriori}, make mathematical and physical sense.  Regardless of our focus on the working details of the algorithm rather than on its ability to improve the performance of the detector, a comparison of the utility values and of the pointing and energy resolution graphs that are offered in the tables and figures of this section (see \autoref{t:u_330_values}, \autoref{t:330_A7}, \autoref{f:u123_61_500}, \autoref{f:ups6_61_500}, and \autoref{f:330_3200_3} {\it infra}), should allow for a clear assessment of the size of the improvements that are offered by the optimization. 

\subsection {A proof of convergence}

For a proof of the convergence of different initial layouts to similar optimized configurations, we simplified the optimization task by only considering the $U_{GF}$ part of the utility function (\autoref{eq:U1}), and by generating showers orthogonally to the ground\footnote{The advantage of fixing $\theta = 0^\circ$ does not lay in the resulting reduction of the dimensionality of the space of shower parameters in the reconstruction step, but rather in the reduction of diversity of patterns within one batch of given size that the gradient descent procedure must account for, effectively corresponding to a data augmentation.}. We further reduced CPU consumption by assuming that the energy of the showers be measured precisely, only tasking the reconstruction algorithm with estimating the shower core positions. We fix the showers energy to 1.0 PeV, so that all showers have the same footprint on the ground; this further removes stochasticity in the gradient descent loop. We employed small initial layouts of 36 macro-tanks comprising 19 detector units each, arranged in (a) a random distribution of units in a wide circle, (b) a tightly-packed ball (with spacing of \SI{50}{\meter} between units), and (c) a two-annuli configuration. We set the \texttt{CommonMode} parameter to 3, so that each of the 12 triplets of units can move freely but respecting the triangular symmetry of their initial configuration. Using 3000 events for the PDF estimation and batches of 3000 showers at every epoch, each job takes about one day of running for 400 epochs on a single CPU. The problem of finding optimal values of the $x,y$ position of 36 macro-units, with the three-fold symmetry, is thus reduced to a merely 21-dimensional one.

\begin{figure*}
    \centering
    \includegraphics[width=0.95\linewidth]{ 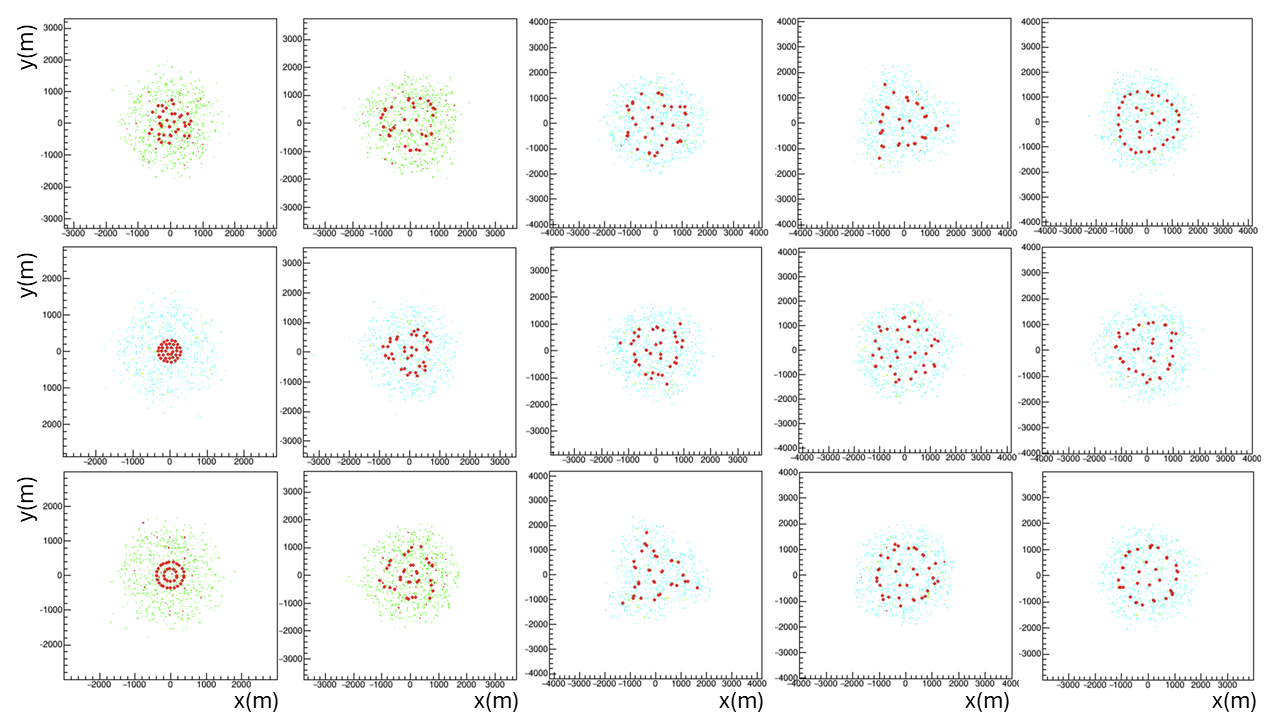}
    \caption{Convergence of three initial layouts (top to bottom: wide random layout, packed circle, two annuli) during a 400-epochs training. From left to right, the configurations of 36 units are shown at epoch 1,20,50,100, and at the end. The fine-grained green/cyan dots indicate the center of generated showers that pass the trigger condition. See the text for more detail.}
    \label{f:threeconfig36}
\end{figure*}

Results are summarized in \autoref{f:threeconfig36}, where the three initial configurations are shown on the left, and they are seen to evolve to triangle-symmetric configurations as the 12 triplets of units move coherently in search of the maximum utility. Despite the initial difference in the layouts, the final ones are very close to one another, confirming that the system successfully identified a stable maximum of the utility. The values of the utility of the three final configurations also converge to similar values, as shown in \autoref{t:utility_threeconfig36} below.  Overall, we observe that the utility improves by 40 to 60 percent, which is a quite significant achievement. 

\begin{table}[h!]
    \centering
    \begin{tabular}{c|c|c}
    Configuration & $U_{GF,in}$ & $U_{GF,fi}$ \\
    \hline
    Wide random ball     & $927 \pm 72$ & $1297 \pm 36$  \\
    Tightly packed ball  & $727 \pm 76$ & $1235 \pm 37$  \\
    Two annuli           & $714 \pm 69$ & $1235 \pm 66$  \\
%    Wide random ball & $926.8 \pm 71.9$ & $1297.0 \pm 36.2$  \\
%    Tightly packed ball     & $727.1 \pm 75.6$ & $1234.6 \pm 37.1$\\
%    Two annuli & $714.4 \pm 68.7$ & $1234.9\pm 66.5$\\
    \end{tabular}
    \caption{Values of the utility at the start of the optimization cycle and at the end, for the three configurations described in the text. }
    \label{t:utility_threeconfig36}
\end{table}

\begin{figure*}[h!]
\begin{center}
\includegraphics[width=0.95\linewidth]{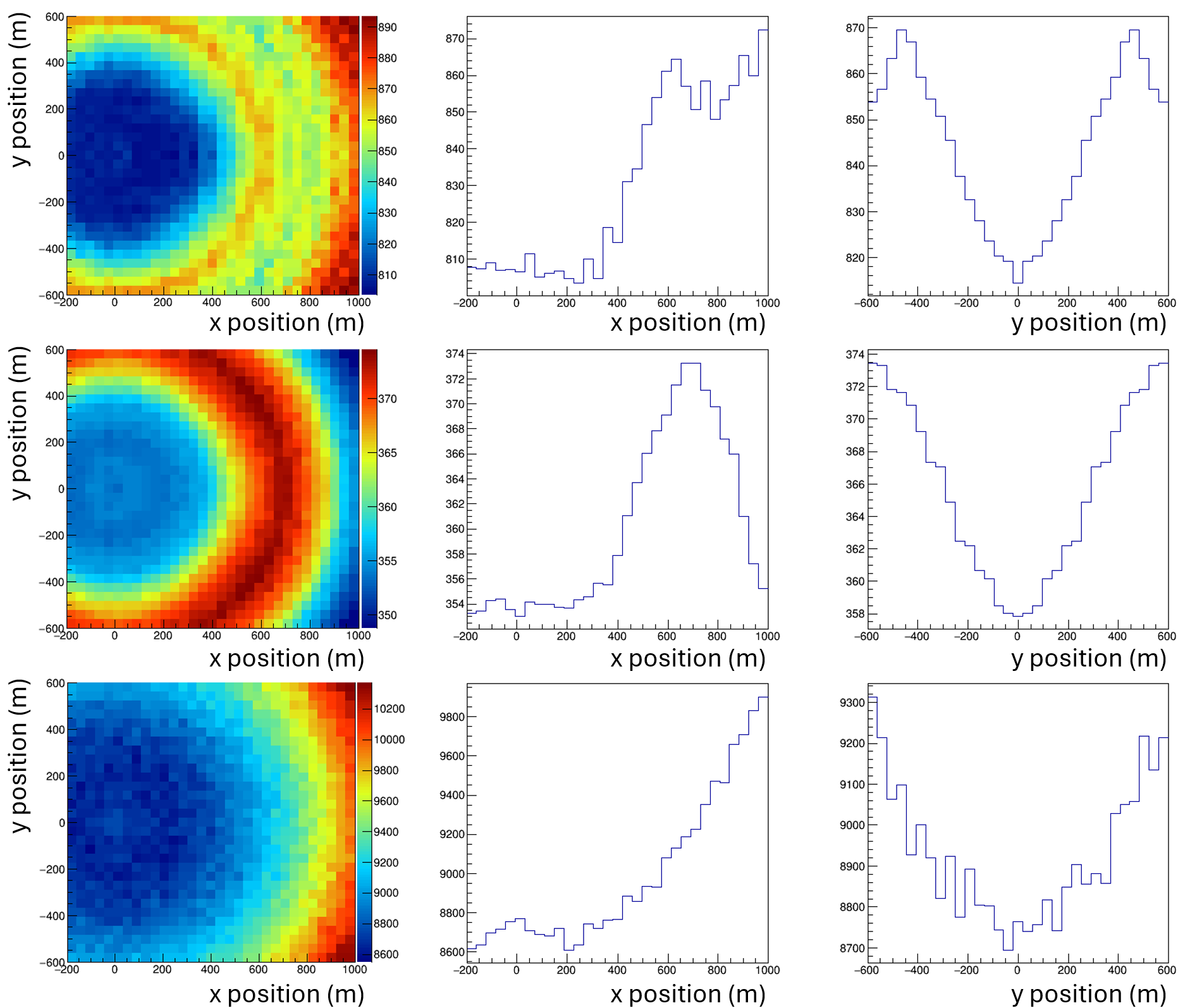}
%{Figures/UGF_5k_961.png}
%\includegraphics[width=0.95\linewidth]%%{Figures/Uscan_37_5k_010.png}
%{Figures/UIR_5k_961.png}
%\includegraphics[width=0.95\linewidth]{Figures/UPS_5k_961.png}
\caption{
Top: temperature map (left) of the value of the $U_{GF}$ utility term as a function of the position of a 19-unit aggregate detector complementing a circular array of 37 other macro-units arranged in a hexagonal pattern spanning a circular area of 400m radius. The center (right) histograms show the utility as a function of the $x$ (resp., $y$) coordinate of the unit, for $y=0$ (resp., $x=0$). Center, left: temperature map of the value of $U_{IR}$ utility term for the same array; the center and right histograms show the value of $U_{IR}$ as a function of $x$ and $y$ as above. Showers used in these simulations have an energy $E=1$ PeV. Bottom: the same graphs are shown for the $U_{PS}$ utility, given a point source of 2 PeV energy at a polar angle of 45 degrees.} 
\label{f:UtilityGradients}
\end{center}
\end{figure*}

\subsection{Size of the array}

The first thing that becomes apparent, once an optimization run is performed starting with a layout consisting of detectors uniformly spread within a small region on the ground, is that gradients of the utility --in particular, the $U_{GF}$ part in \autoref{eq:U1}-- point outwards (see \autoref{f:UtilityGradients}, top). In other words, the layout wants to expand to a larger area, in order to catch a higher flux of gamma showers including ones that fall away from the center, yield a signal only in part of the array, and are thus hard to reconstruct and discriminate from proton backgrounds. This phenomenon exists even if the $1/\sqrt{\rho}$ factor in $U_{GF}$ is omitted: a larger flux of secondaries from showers produced at \SI{1}{\kilo\meter} or more from the center of the array allows for a larger likelihood ratio of the gamma versus proton hypothesis of those showers, which ends up improving the precision of the flux estimate when a two-component fit to the test statistic is carried out. The optimization of the point source utility $U_{PS}$ produces a similar effect, whose strength however depends more directly on the energy of the generated gamma rays.
%The presence of the density factor further enhances the advantage of a radial expansion, as in hindsight should be easy to understand: a flux measurement will benefit from a larger sensitive area.

The $U_{IR}$ term in the utility also produces outward-pushing gradients (see \autoref{f:UtilityGradients}, bottom), but the gradient has a different shape as a function of the distance from the center of the array. 

\begin{figure*}[h!]
\begin{center}
\includegraphics[width=0.99\linewidth]{ 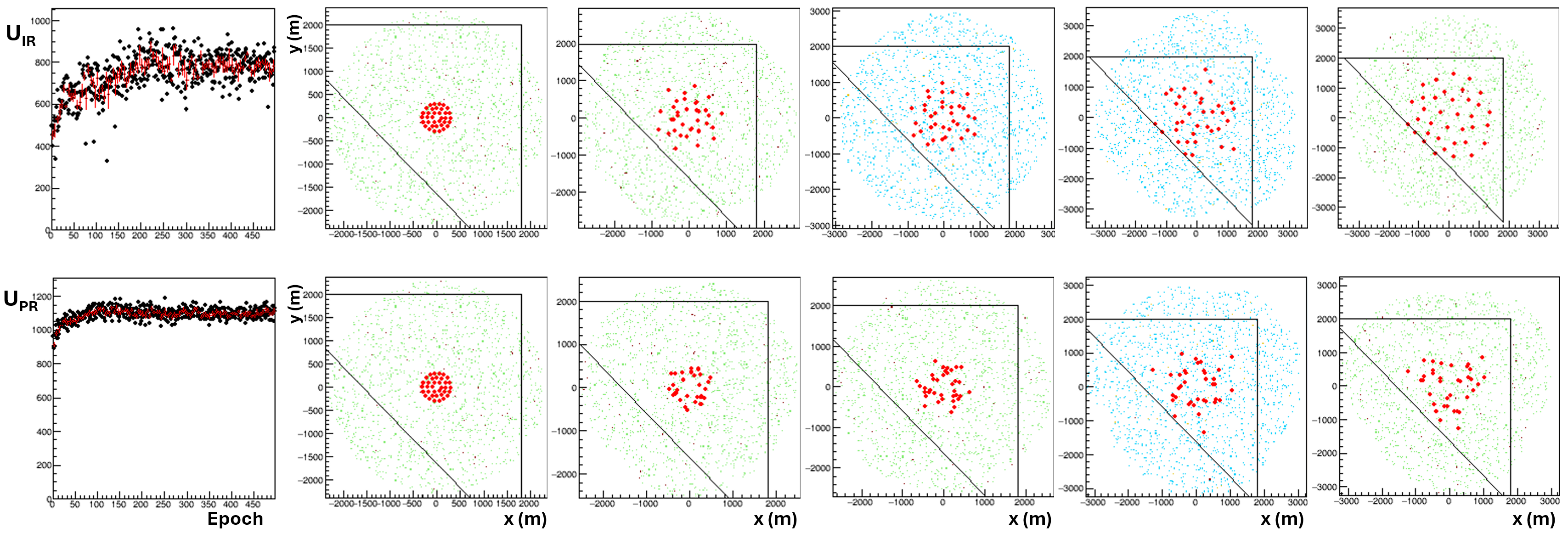} 
\caption{ Comparison of the optimization of $U_{IR}$ and $U_{PR}$ terms, for an array of 36 macro-tanks comprising 19 units each. The initial array is set in a packed circle, centered within the triangular constraints of the Pampa la Bola site. The top left (bottom left) panel show the $U_{IR}$ ($U_{PR}$) utility as a function of epoch. The five graphs to the right in each line show the progressive optimization of the $U_{IR}$ ($U_{PR}$) terms at epoch 1, 20, 50, 100, and 500, respectively. The background green and cyan points show the center of generated showers that pass the triggering criterion.}
\label{f:ir_vs_pr_36_500}
\end{center}
\end{figure*}

A comparison of the optimal layouts that can be obtained by maximizing exclusively the $U_{PR}$ or $U_{IR}$ terms is also instructive (see \autoref{f:ir_vs_pr_36_500}). While both utilities tend to expand the array, the $U_{IR}$ term is maximized when units are spread out evenly on the ground, while the $U_{PR}$ term maximization induces non-trivial configurations which may best capture, with at least 50 triggering units, the timing structure of the arrival front of secondary particles.

The bottom-line is that the $U_{GF}$ and $U_{IR}$ terms can be in apparent competition in some regions of the plane. This further highlights the importance of a careful definition of the relative merits of different desirable properties of the shower reconstruction by any given array. This kind of interplay between misaligned optimization criteria makes the appraisal of layouts a task unfit for solely speculative calculations. 
%It is also to be noted that the coefficient of the power spectrum of the generated gamma showers plays a very important role in determining both the size and the behavior of the $U_{IR}$ term. For this reason, we have set the power to zero in our studies, pending a discussion on the utility specification.

%\begin{figure}[h!]
%\caption { Progression of optimization jobs using 330 macro-tank aggregates initially arranged in a packed circle. Top: maximization of the standard utility (\ref{eq:U1}); center: maximization of the point source utility $U_{PS}$, with an energy spectrum centered at $E_\gamma=2 PeV$; bottom: maximization of the point source utility $U_{PS}$, with an energy spectrum centered at $E_\gamma=6 PeV$.}  
%\label{f:compilation}
%\end{figure}

\subsection{Radial distribution of tanks}

The second effect which can be observed almost immediately in optimization loops that start with a uniformly populated area on the ground (whatever its shape) is that, in conjunction with an expansion of the array, the program usually finds no rationale in populating with high density the center of any symmetrical configuration --in absolute contrast with the most striking commonality of the benchmark layouts of \autoref{sec:swgo}. While keeping a radially symmetric distribution, the utility gradients cause a shear of the center of the array, which pushes toward larger radii units placed at the center faster than units placed at the edge. This is a separate phenomenon from the one discussed above, although it originates from the same physical reason --namely, the existence of a dimensional scale in the problem (the transverse size of photon-originated showers). 

Why should a hole-in-the-middle, donut-like configuration be providing a smaller flux uncertainty than a flat disc distribution of units? The answer is simple: the center of the array is a point of measure zero; in other words, odds that a shower has its core in that place are null. Even the task of measuring gamma rays of very low energy (whose secondary particles distribute within a smaller spatial extension) is not better attended by a densely populated distribution of units, as there is a quadratic dependence of the number of showers as a function of radius. For that reason, a donut is to be preferred to a disc, and this appears a very robust conclusion, although the scale of the central void strongly depends on the target energy of the showers. Only if one had infinite integration time and were concerned solely with the energy resolution of the few showers one could collect in a small area on the ground, would the strategy of putting all the units in the ``same basket'' work. Any other experimental condition will instead prefer a more distributed layout which ``looks toward the boundary'', and tries to catch as much information as possible from the $R^2$-growing flux of showers. % (see Fig.\ref{f:330_initialblowup}).

%\begin{figure*}
%\begin{centering}
%    \includegraphics[width=0.8\linewidth]{330_initialblowup_1_100_300_ups2vs6.png}
%    \caption{Initial expansion of a tightly packed configuration of 330 detectors (19-unit macrotanks) optimizing the
%    point-source utility for $E_{PS}=2 PeV$ (left) or $E_{PS}=6 PeV$ (right). For each set of two graphs, the left panel shows the layout of macro-tanks on the ground (red points) and the distribution of shower centers (small coloured points); the histograms in the right panel instead show the initial (red) and current (blue) radial distribution of tanks around the array center. The first row of graphs refers to the initial configuration (a packed circle, with macro-units placed at $30m$ distance from one another), the second shows an improved configuration reached after 100 iterations, and the third shows the optimal configuration after 300 iterations. The
%    utility improves by 40\% - 50\% with this number of iterations, but longer runs would improve the layout and the utility further. 
%    The depletion of the center of the array in both runs is evident in the histograms as well as in the distribution of detectors on the plane.}
%    \label{f:330_initialblowup}
%\end{centering}
%\end{figure*}

\subsection{Effects of reconstruction requirements}

Since a more precise reconstruction of the five shower parameters increases the discriminating power of the likelihood ratio, and consequently reduces the uncertainty on the signal fraction in a batch, we may expect that optimal configurations be strongly dependent on how well the reconstruction can be carried out. A way to observe this phenomenon, with a view to gaining more understanding on the way optimal configurations are selected by the algorithm, is to check how the reduction of the complexity of the reconstruction task modifies the optimal layouts. Or, from the opposite standpoint, we may start with a simpler reconstruction task, by assuming that we know some of the shower parameters from external inputs, and progressively make it more complex, by gradually dropping those assumptions.

We may, {\it e.g.}, assume that angular and energy parameters are perfectly known, and only task the algorithm with measuring the shower cores. A further simplification is to assume that all showers have zero polar angle, {\it i.e.} that they come down orthogonally, leaving circular footprints. In this case, we observe that the system loses any interest in populating the ground with a two-dimensionally structured array, and produces optimal configurations of units arranged in a circumference, or a triangular line if a triangle degeneracy is enforced. The reason for this behavior is rather easily understood: once the energy and angles of the shower are given as known, the flux is a simple function of the distance of the core, so that no ``stereo'' information is needed to disentangle the effect of energy and distance on the intensity of the flux, and no timing information is required to figure out the polar and azimuthal angles. 

When angular parameters are also unknown, and showers are generated with the standard distribution of polar angles, the reconstruction problem becomes slightly more complex; timing information allows to determine the azimuthal direction of the shower and its tilt from the vertical only if some two-dimensional information on the actual shower footprint is collected by the array. We then observe that more distributed configurations are preferred, as expected. 

The integrated energy resolution term in the utility function, $U_{IR}$, plays a role when the energy of the showers is not assumed known. The optimal configurations can then be observed to distribute tanks on the ground in richer two-dimensional shapes, dictated by the need to acquire more information on the distance of the core together with the overall intensity. Of course, a modification of the relative value of the coefficients in the two terms of the utility function will significantly affect the optimal configurations.

In summary, we observe that some of the detailed characteristics of the reconstruction problem as posed have a strong impact in the outcome of the optimization task; others less so. The utility specification is however crucial. Rather than discouraging us from searching for the best configuration, this consideration and the richness of the problem should motivate us to study it further. We discuss {\it infra} the different behavior of optimal configurations when different utility specifications are imposed.

\subsection {Symmetries}

In the default mode of operation, triplets of units are locked together in an equilateral shape by imposing a 120-degree period on detector coordinates around the center. The optimization then converges to configurations in which a triangular symmetry is apparent; yet the array has complete freedom in arranging itself into non-trivial configurations within a 120-degree sector: the emergence of more regularity in the patterns arising after a few dozen loops of gradient descent updates signals that coherent gradients are winning over stochastic noise.

\begin{figure}[h!]
\includegraphics[width=0.99\linewidth]{ 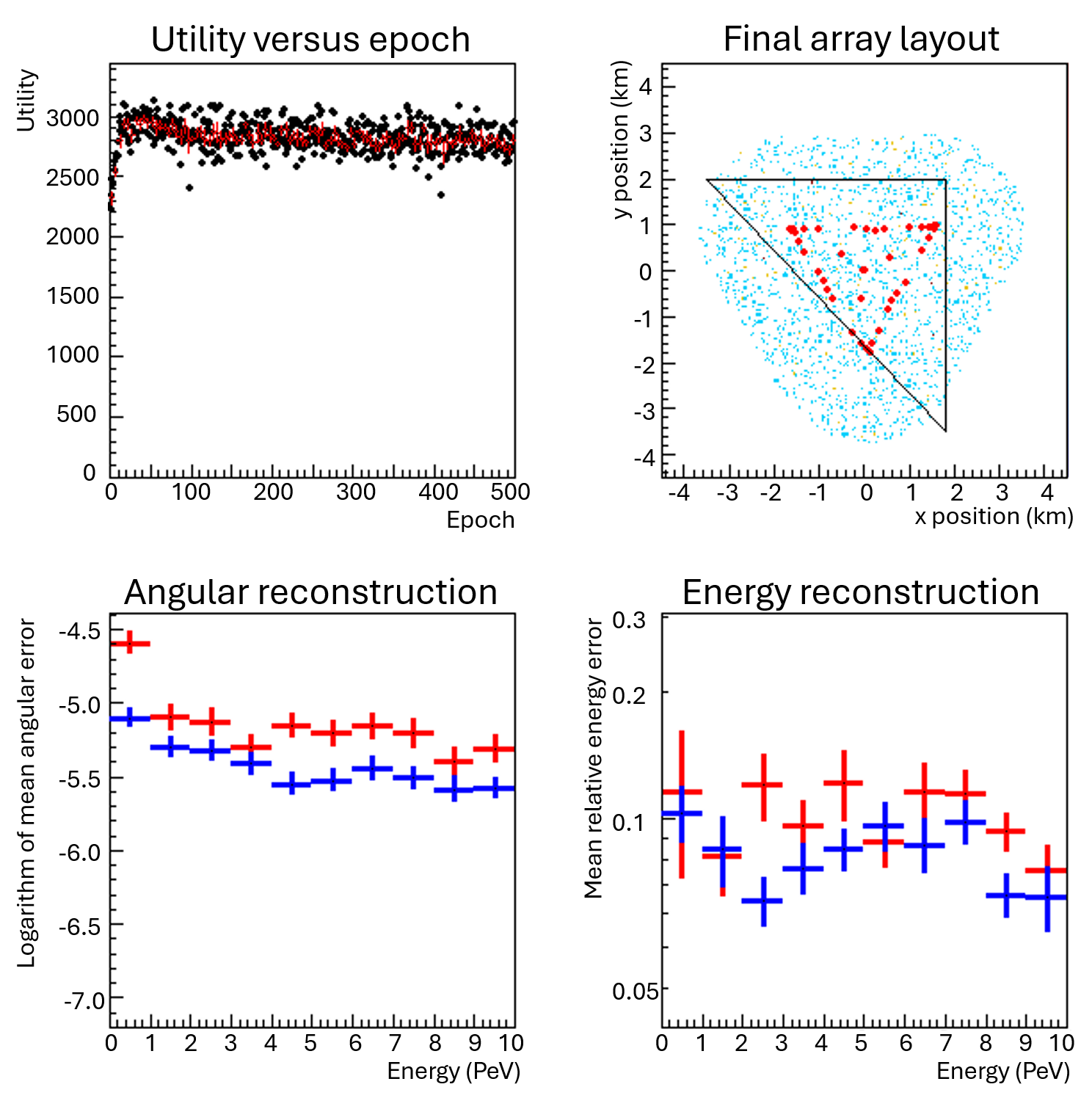}
\caption{Optimization of the $U_1$ utility function for 61 19-unit macro-tanks in the Pampa la Bola site. Top left: utility versus epoch; top right: final layout (macro-units are the red points), with in light cyan the center of triggering showers at the last epoch; bottom left: logarithm of the mean pointing error as a function of gamma energy, for the initial (red) and the optimized configuration (blue); bottom right: mean relative energy error as a function of gamma energy, for the initial (red) and the optimized configuration (blue). }
\label{f:u123_61_500}
\end{figure}

\begin{figure}[h!]
\includegraphics[width=0.99\linewidth]{ 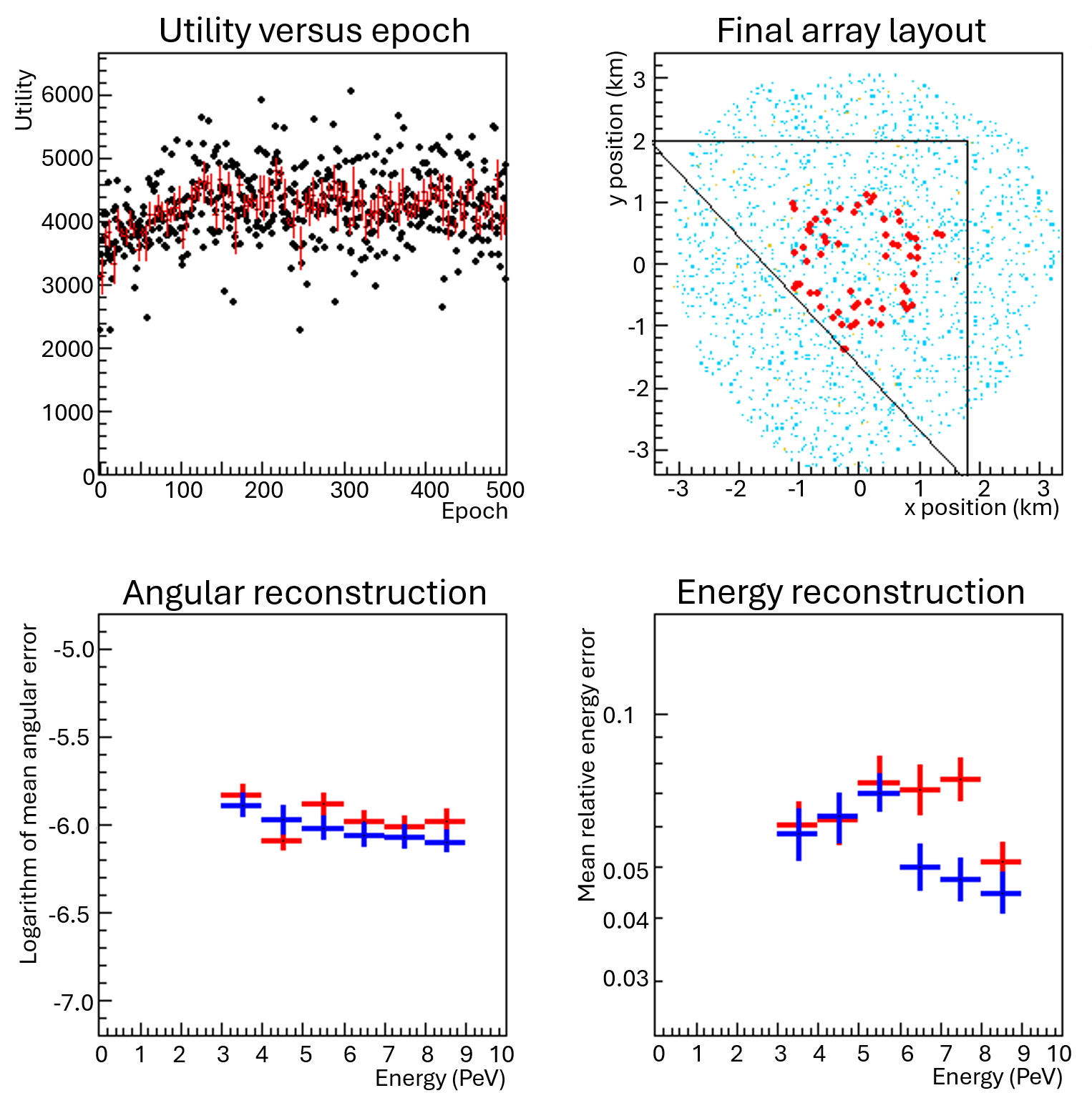}
\caption{Optimization of the $U_{PS}$ utility function (with a point-source energy of $E=6$ TeV) for 61 19-unit macro-tanks in the Pampa la Bola site. Top left: utility versus epoch, with the center of triggering showers at the last epoch shown as cyan dots; top right: final layout (macro-units are the red points); bottom left: logarithm of the mean pointing error as a function of gamma energy, for the initial (red) and the optimized configuration (blue); bottom right: mean relative energy error as a function of gamma energy, for the initial (red) and the optimized configuration (blue). }
\label{f:ups6_61_500}
\end{figure}

For example, a 500-epoch maximization of the standard utility starting from an initial tightly-packed circular array of 63 macro-units of 19 tanks, subjected to the physical constraints of the Pampa La Bola site and to a strong penalty factor hindering expansions of the array to area larger than $1.44^2 \pi$ km$^2$, produces a highly symmetrical configuration (see \autoref{f:u123_61_500})\footnote{The triangular shape of the final configuration reflects the effect of the area penalization, which was multiplied for this run by a larger coefficient for a test. This behavior is due to the prescription of generating showers within 2 $km$ from the closest detection unit: the triangular shape then ``sees'' a wider region impinged by showers than a circular shape fulfilling the same strong area constraint, and becomes favourable in the optimization.}. In this case, units are pushed toward the outer rim because the flux term in the utility wins over the terms that try to improve energy and pointing resolution; however, as shown by the two bottom panels, the resolution is still significantly improved in the final configuration with respect to the initial one. In contrast, the maximization of the $U_{PS}$ utility (with an energy of 6 PeV) for the same initial array subjected to the same constraints produces a richer structure, while still one exhibiting strong symmetry features (see \autoref{f:ups6_61_500}). 

The program allows one to experiment with a higher number of units locked together into squares, pentagons, hexagons, or even with full radial symmetry, when all units placed at the same initial radius will only move radially in synchrony. We however believe that the triangular symmetry configurations reached by runs with \texttt{CommonMode}=3 are the most instructive ones to study: the evolution from an initial uniform arrangement to shapes such as those of \autoref{f:u123_61_500} or \autoref{f:ups6_61_500} during gradient descent contains information on the stability of the intermediate solutions and hints at ways to explore more fully the configuration space.

The convergence to symmetrical configurations of varied complexity does not imply that the optimization task is completed. On the contrary, it in fact proves that the task of finding the {\it best} configuration in a landscape of similarly good-performing ones is a quite difficult one: the utility landscape is, in other words, riddled with a large number of local maxima. A more aggressive scheduling of the learning rate may be needed to break that impasse. However, this lies beyond the intended purpose of this document. In fact, that final step will be most proficuously attended to when the shortcuts and approximations we have taken at the present stage will have been sorted out.

\subsection {Comparison of different utility choices}

For a study of the impact of different optimality conditions, we run three simulations with 330 19-unit macro-tanks (thus equivalent to a 6270-unit array) initially arranged in a tightly packed circular shape, using the \texttt{commonmode=3} feature (120 degree symmetry) and the three following definitions of the utility function:\par

\begin{enumerate}
    \item the standard $U_1$ term of \autoref{eq:U}, with a flat energy spectrum of showers between 100 TeV and 10 PeV;
    \item a point source utility $U_{PS}$ with energy of 2 PeV;
    \item a point source utility $U_{PS}$ with energy of 6 PeV.
\end{enumerate}

%Some intermediate configurations reached by the optimization loops are shown in \autoref{f:330_comparison}. As expected, the lower energy of the 2 PeV point source case call for less spatial extension of the array than the other two cases, and a smaller voided region at the center of the array. For the rest, the three configurations exhibit a roughly cylindrical symmetry, and most of the information on their optimality condition resides in the radial distribution. 

\begin{figure*}[h!] 
\begin{center}
\includegraphics[width=0.87\linewidth]{ 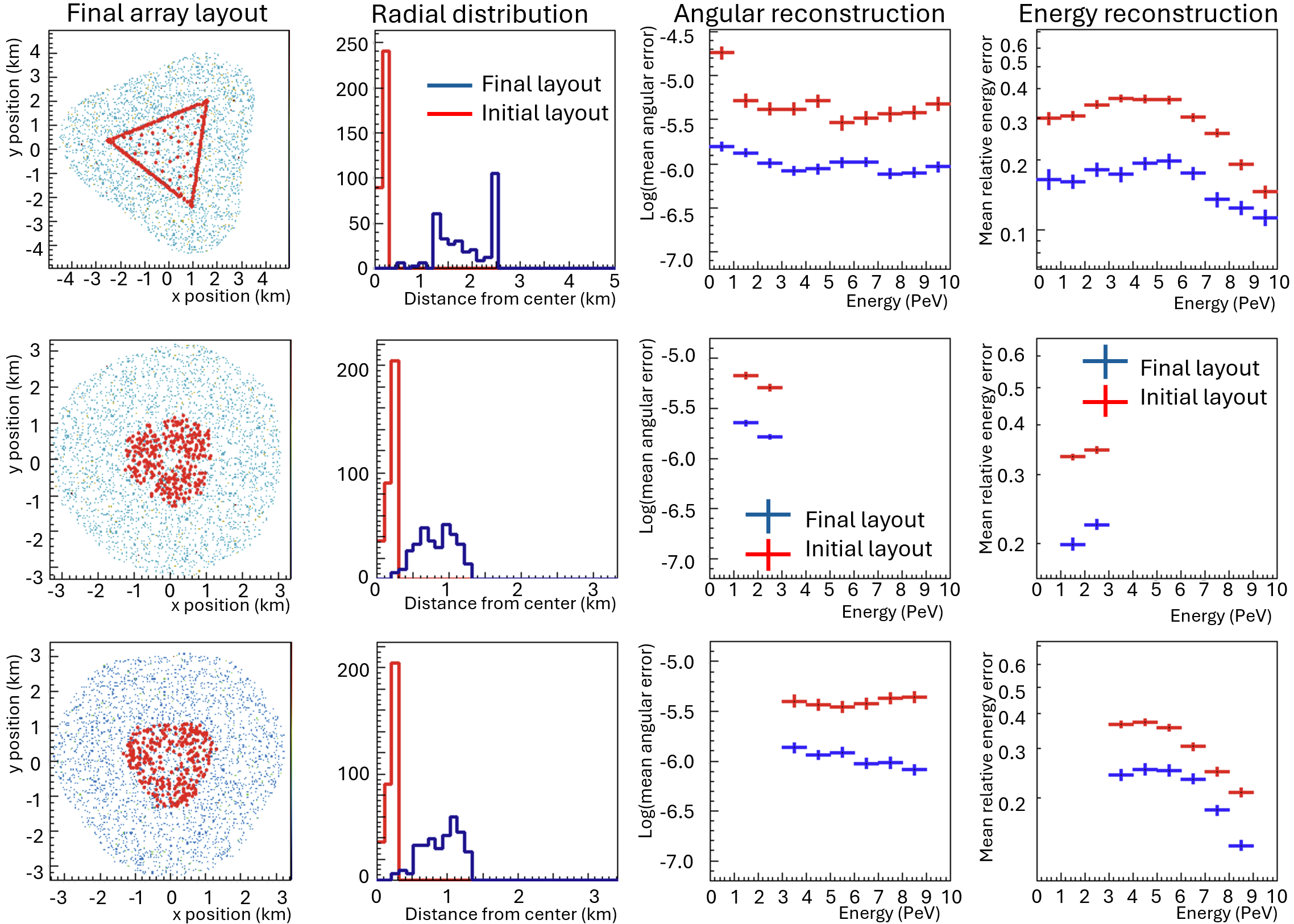}
\caption{Final layouts, radial distribution, and relative pointing and energy errors achieved by optimization runs maximizing the $U_1$ utility of Eq. \ref{eq:U1} (top row), the $U_{PS}$ utility with a point source energy of 2 PeV (middle row), and the $U_{PS}$ utility with a point source of 6 PeV (bottom row). The arrays, initially set in a tightly-packed configuration of 330 macro-units in a circle of $300$ m radius, converge to different layouts after 2000 iterations.}
    \label{f:330_3200_3}
\end{center}
\end{figure*}

By examining the final configurations reached by these optimization runs (\autoref{f:330_3200_3}) a few things become apparent. 
\begin{enumerate}
    \item As already noted earlier, in presence of an area constraint the $U_1$ maximization tends to push the majority of the units toward the outer boundary of the array to maximize the useful flux: due to the restriction in the generation of showers to less than 2~km from the closest unit, this configuration becomes an equilateral triangle. We consider this a feature rather than a bug, which can easily be modified as appropriate for the utility that is eventually chosen for realistic optimization runs. However, in the present case the outward-pushing gradients are seen to win over the area cost gradient, with the result that the final layout occupies an area about 1.8 times larger than the reference one. This highlights the difficulty of tuning the coefficients of the different utility terms in these runs. 
    \item Another feature that is apparent in the final configuration of the $U_1$ maximization is the approximate uniformity of the arrangement of macro-units within the boundary of the array. This was also noted earlier as a feature induced by the maximization of the integrated energy resolution term. 
    \item In all three runs, the already commented depletion of the center of the array (initially packed with detection units) is quite apparent.
    \item The maximization of the $U_{PS}$ utility produces similar layouts for the two source energies here considered (2 and 6 PeV). The final configuration has a larger hole in the center for the higher source energy, corresponding to the higher area spanned by the gamma showers at higher energy, which as already noted acts as a dimensional scale in the problem.
\end{enumerate}

\begin{table*}[h!]
\centering
\begin{tabular}{l|c|c|c|c|c|c}
& $U_{in}$ & $U_{fi}$ & $U_{fi}/U_{in}$ & $U_{GF,fi}/U_{GF,in}$ & $U_{IR,fi}/U_{IR,in}$ & $U_{PR,fi}/U_{PR,in}$ \\
\hline

$U_{1}$ & $2178 \pm 20$ & $3572 \pm 10$ & $ 1.64 \pm 0.02$  & $2.20 \pm 0.02$ & $1.06 \pm 0.02$ & $1.34 \pm 0.01$ \\
$U_{PS}$ (E = 2 PeV) & $1737 \pm 52$ & $2822 \pm 25$ & $1.62 \pm 0.04$ & $1.64 \pm 0.04$ & $1.07\pm 0.01$ & $1.41 \pm 0.01$ \\
$U_{PS}$ (E = 6 PeV) & $1732 \pm 34$ & $3071 \pm 17$ & $1.77 \pm 0.04$ & 1.50 $\pm 0.04 $ & $ 1.05 \pm 0.01$ & $1.40 \pm 0.01$ \\
\hline

% these results are from the logs available in the Outputs subdirectory, /lustre/cmswork/dorigo/swgo/MT/Outputs
% files are named RunDetails_Nb=32000_Nu=330_Ne=0-2000_Sh=3_Id=0.txt, =1.txt, =2.txt
% macro mean_sqm_utilities.C is used to get
%mean values for the 6 PeV case (Id=2), which only run for 1000 iterations at the time of writing.

%$U_{1}$ & $2178.19 \pm 20.18$ & $3586.96 \pm 20.70$ & $2.196 \pm 0.022$ & $1.060 \pm 0.018$ & $1.337 \pm 0.008$ \\
%$U_{PS}(E=2PeV)$ & $1736.76 \pm 52.34$ & $2812.48 \pm 72.99$ & $1.635 \pm 0.039$ & $1.070\pm 0.008$ & $1.407 \pm 0.008$ \\
%$U_{PS}(E=6PeV)$ & $1759.34 \pm 93.45$ & $2923.85 \pm 189.57$ & $1.462 \pm 0.087$ & $ 1.055 \pm 0.076$ & $1.402 \pm 0.048$ \\
\end{tabular}
\caption{Initial and final value of the utility for the three runs described in the text. The improvement in the utility from the initial packed circle configuration to the final ones shown in \autoref{f:330_3200_3} is shown for the separate factors making up the $U_1$ utility also for the runs maximizing the $U_{PS}$ utility, for comparison. }
\label{t:u_330_values}
\end{table*}

As reported in \autoref{t:u_330_values}, the increase in the utility provided by the final configurations ranges from {\bf 62\% to 75\%}; part of this increase is due to the fact that a more extended layout benefits from a higher flux of reconstructible showers, but the inhomogeneous, non-trivial configuration of the detectors on the ground plays a significant role. It is also noticeable how the energy resolution does not increase significantly during the optimization, while the position resolution term increases by $34\%-41\%$.

\begin{table}[h!]
\begin{centering}
\begin{tabular}{l|c}
& A7-like  \\
\hline
$U_1$ &    $2845 \pm 9$  \\
$U_{PS}$(E = 2 PeV)   & $2698 \pm 27$  \\
$U_{PS}$ (E = 6 PeV)  & $2739 \pm 19$  \\
\hline
\end{tabular}
\caption{Utilities computed with the same running parameters of the optimization runs discussed in the text, for an array of 330 19-unit macro-tanks arranged to mimic the layout of the A7 configuration. Results for the final utilities of the optimized results are reported for comparison.}
\label{t:330_A7}
\end{centering}
\end{table}

\begin{figure}[h!]
\begin{center}
\includegraphics[width=0.65\linewidth]{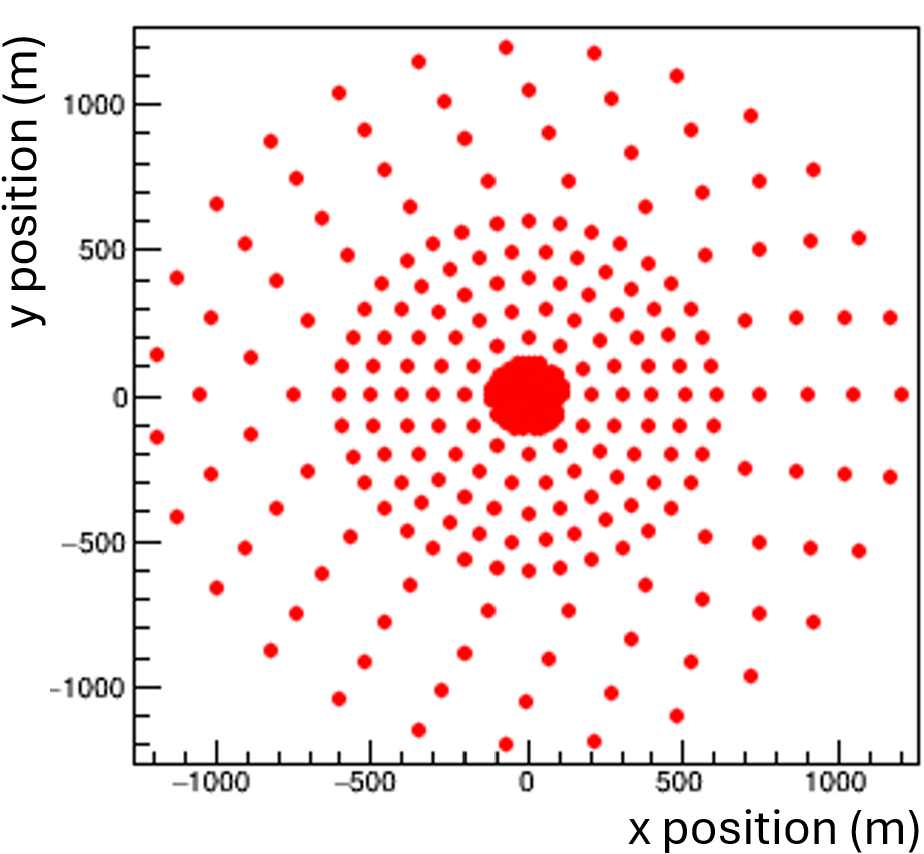}
\caption{"A7-like" layout of the 330 19-unit macro-tanks mimicking the fill factors of the A7 array described in the text.}
\label{f:A7like}
\end{center}
\end{figure}

It is important to note that the values of the utility functions reported in \autoref{t:u_330_values} cannot be directly compared to those obtained for the benchmark layouts in \autoref{s:benchmarks} (\autoref{t:layouts}), for two reasons. The first is that the number of events generated and accepted in each batch (using the trial procedure discussed in \autoref{s:methodology}) plays a role in the calculation of the utility; the second is that the macro-tank aggregates produce a smaller amount of information --in particular on the timing of secondary particle detection, but also on the distribution of secondaries-- than that produced by the corresponding number of single detection units. 
For a quick-and-dirty assessment of the results of \autoref{t:u_330_values} in relation to the 13 benchmarks, we may take the highest performer according to the three metrics in \autoref{t:layouts}, the A7 configuration, and construct with the layout of its 6571 units a 330-macro-tank layout that corresponds to its three fill factors and radii tabulated in  \autoref{t:layouts_sec2} (see \autoref{f:A7like}). We may then evaluate on that layout the three utilities with runs of 3200 showers as done for the results of \autoref{t:u_330_values}. Results are shown in \autoref{t:330_A7}: they indicate that the optimized layouts are better than the A7 configuration by $25.5\pm 0.5$ \%, $4.6 \pm 1.4$\%, and $12.1 \pm 1.0$\% as measured by the ratios of their $U_1$, $U_{PS}$ (2 PeV), and $U_{PS}$ (6 PeV) functions, respectively. As the focus of this work is mainly to demonstrate the optimization technique, which still includes many approximations and simplifications in order to allow its testing and development, we leave to future studies a more careful direct comparison and appraisal of specific configurations.

\section {Concluding remarks}
\label{sec:conclusion}

In this  concluding  section we offer a brief summary of the indications we have collected on the layout optimization problem from early tests of the developed software, and a short set of remarks on the future work that ought to be carried out in order to meaningfully exploit it. We are convinced of the importance of these studies, as we have observed very significant variations in the considered utility function as we move from simple-is-best layouts to more complex configurations.

\subsection {Discussion}

If we make no assumptions on the distribution of sources of gamma rays in the sky, we must consider that SWGO will be hit by showers from all angles, with a uniform distribution in celestial azimuth and right ascension that translates in a $p(\theta) \propto \sin(\theta)$ distribution of polar angles of incidence, and a uniform distribution in azimuthal angle on the ground. The optimal array should therefore, from geometrical considerations alone, have an axial symmetry. However, with a finite number of detection units, a fully symmetric distribution around a point on the plane is not achievable in practice, and one has to consider n-fold angular symmetries as a substitute. This matter seems irrelevant when we consider thousands of detection units, but it constitutes a constraint in an optimization search, and an opportunity to verify the consistency of the identifiable solutions by automated means. 

In the previous section we have shown how different N-fold symmetries tend to converge to similar arrangements on the ground, but that subtle effects --such as the one arising from the triggering conditions, or from physical constraints on the ground-- may influence it in non-trivial ways. We have also observed how the exact specification of the problem, and the details of the reconstruction procedure, have an impact in the determination of the solutions the algorithm converges to. With regards to the dependence on the reconstruction, which is at face value a troublesome feature of what we would like to dress up as an end-to-end solution, we note that we have so far considered a sub-optimal, but ``legal'' setup in our tests, which are preliminary and mainly meant to explore the behavior of the algorithm. A final search of the best solution can only be performed once a full definition of the utility function has been agreed upon; once detailed modeling of the chosen tank characteristics have been coded in the program; and once external constraints (such as total area available for the deployment of the tanks, total number and detailed specification of the tanks) have been considered and included. That is the task that awaits us in the near future.

\subsection{Future prospects}

The presented software, while already producing interesting results and offering meaningful suggestions for the placement of the units, requires further improvements before it can become a precise oracle of the final merits of different configurations. We list some of these improvements and shortly comment on them in the list below.

\begin{enumerate}
    \item Model improvements: as stated above, a number of improvements are possible in the parametric model, and they are liable to have an impact on the performance of reconstruction of the five shower parameters. The improvements include a more precise modeling of the polar angle distribution of primaries, a treatment of disuniformities in the composition of the shower front on the ground, and precise timing and angular profiles of the various components of secondaries.  
    \item Detection units response: depending on the precise choice for the geometry and specification of the tanks, we will be able to introduce in the model a realistic efficiency profile versus angle of incidence and energy of the secondaries, and counting and energy uncertainties at high fluence. The approach we have used so far (only reliant on the counting statistics of the two main particle species) needs to be revisited once the model is improved.
    \item Reconstruction method: the likelihood implemented in the current version of the software exploits a full knowledge of the model details, but is liable to be improved both in speed and in accuracy of estimates. Ultimately, we could entirely replace the likelihood reconstruction --and consequently the likelihood ratio test statistic for gamma/hadron separation-- with a deep neural network; indeed, we have already been developing one such tool, although we are not ready to present results based on it yet.
    \item Alternative optimization procedures: the procedure employed in this work --stochastic gradient descent-- is a well understood tool, but it is not necessarily the most effective one for the problem at hand; its shortcomings include the wide variation of the size of gradients that are being used for geometry updates, approximations in the derivation of the density functions of the test statistic employed to gauge the uncertainty on the signal fraction, and other implications, {\it e.g.} on the need for a fully parametric model. Genetic algorithms are a notable alternative which should be tested;
    another proposed method consists in starting with an overabundant number of detection units tightly arranged in a wide area, and then proceeding to estimate the quantities of interest by enforcing sparsity in the array by Lasso regression. We plan to consider these alternatives in the near future.
\item Computation of derivatives: Our present approach to compute
derivatives analytically is very flexible ({\it e.g.}, making it easy to operate shortcuts and workarounds for non-differentiable operations) and tangible (making it most obvious to developers which operations are performed on the CPU). An automatic differentiation approach is however more scalable: when we add more detail to our model, {\it e.g.} to better approximate the true performance of the detection units, automatic differentiation will become the only option to create and maintain an efficient optimization code.
\end{enumerate}

The success of the SWGO experiment will ultimately depend on a set of decisions --on allocation of resources, choice of site, choice of detection units, choice of layout of the units on the ground-- that ideally should collectively maximize the desired scientific outputs, once all possible confounding factors, external constraints, detector-related systematic uncertainties, and last but not least, monetary cost have been taken into account. The choice of layout of the detector units is thus only one of many ingredients, but it is also the one which has the loosest coupling to most of the others; it thus offers itself as a good candidate to be studied in isolation, pending the progressive integration of additional factors in the problem. We are convinced that the collaboration ought to work out a baseline definition of a global, experiment-wide utility function which included most if not all the scientific results that are defining its necessity as an addition to the present landscape of running and approved future astroparticle physics experiments. The better specified the global utility will be, the more useful will become all the studies aiming at optimized scientific output.

 We finally wish to clarify that at the time of writing there is no formal decision within the SWGO collaboration on the use of this software for the optimization of the array layout. At present the collaboration is pursuing a phased approach, starting with a proposed small array (SWGO-A) that targets low-energy gamma rays, whose effective detection and measurement require the highest fill factor within a small region on the ground, which cannot benefit significantly from an optimization such as the one the developed software can offer. In a later stage this may however become possible and advantageous, and we are working to make it possible. With minor adaptations, the developed software may also find future use in similar experiments that deploy a large number of units to measure cosmic rays on the ground; a couple of such use cases are presently under consideration.

\section*{Acknowledgments}
We gratefully acknowledge useful discussion within the SWGO collaboration. In particular, we acknowledge the work of BSc student T. Guercio (Univesity of Padova) for initial studies on the particle imprint at the ground and MSc student R. Lui (University of Padova) for the continuation of this work. We acknowledge the Italian National Center for Informatic Techologies (CNAF) where \texttt{CORSIKA} simulations were developed. Jan Kieseler is supported by the Alexander-von-Humboldt-Stiftung. Max Aehle and Nicolas Gauger gratefully acknowledge the funding of the German National High Performance Computing (NHR) association for the Center NHR South-West.

\section*{ Data availability statement }

 The code that supports the findings of this study is openly available in \href{https://github.com/tdorigo/SWGOLO}{https://github.com/tdorigo/SWGOLO}. The provided software generates the datasets required for all the optimization studies described in this work. Other datasets, including \texttt{CORSIKA} simulations that are the basis of the developed surrogate model, are available upon request to the contact authors. 

% @software{RepoName,
% author = "{GitHub User or Organization}",
% title = "{Repository Name}",
% url = {https://github.com/user/repo},
% version = {1.2.3}, % Replace with actual version if 
% applicable
% date = {2024-03-21} % Release or last update date
% }

%% The Appendices part is started with the command \appendix;
%% appendix sections are then done as normal sections
\appendix

\appendix
\label{sec:appendix}
\section{Details of the showers model}

As described in \S\ref{s:modelderiv} we parametrize the flux of electromagnetic and muon secondaries on the ground as a function of radius from the shower core using the following equations: \par

\begin{align}
    \frac{dN^{\mu,\gamma}(E_{\gamma} | \theta_{\gamma})}{dR} &= p_0^{\mu,\gamma}(E_{\gamma} | \theta_{\gamma}) \cdot \exp\left(-p_1^{\mu,\gamma}(E_{\gamma} | \theta_{\gamma}) R^{p_2^{\mu,\gamma}(E_{\gamma} |\theta{\gamma})}\right)
\\
    \frac{dN^{e, \gamma}(E_{\gamma} |\theta_{\gamma})}{dR} &= p_0^{e, \gamma}(E_{\gamma} |\theta_{\gamma}) \,  \exp\left(-p_1^{e,\gamma}(E_{\gamma} |\theta_{\gamma}) R^{p_2^{e, \gamma}(E_{\gamma} |\theta_{\gamma})}\right)
\\
    \frac{dN^{\mu,p}(E_{p} |\theta_{p})}{dR} &= p_0^{\mu,p}(E_p |\theta_p) \,  \exp\left(-p_1^{\mu,p}(E_p |\theta_p) R^{p_2^{\mu,p}(E_p |\theta_p)}\right)
\\
    \frac{dN^{e,p}(E_{p} |\theta_{p})}{dR} &= p_0^{e,p}(E_p |\theta_p) \,  \exp\left(-p_1^{e,p}(E_p |\theta_p) R^{p_2^{e,p}(E_{p} |\theta_p)}\right)
\end{align}

\noindent
We study the distribution of the four sets of parameters ${p_0, p_1, p_2}$ as a function of 20 values of the energy of the primary particle producing the EAS, for each of four values $\theta_0, \theta_1, \theta_2, \theta_3$ of the polar angle (centers of four bins in the $(\SI{0}{\degree}, \SI{65}{\degree})$ range), and interpolate them in order to obtain a prediction of the value of each parameter for any value of $E$ and for each of the four $\theta$ values. This is done by fitting the parameters $p_0$, $p_1$, and $p_2$ of each of the four theta values as a function of the logarithm of the primary particle energy: \par 
\begin{equation}
    p_i(E|\theta_j) = q_{i0}(\theta_j) \exp \left[ q_{i1}(\theta_j)  [\ln(E/E_0)]^{q_{i2}(\theta_j)} \right]
\end{equation}

\noindent
The value of parameters $q_{ij}$ are tabulated in \autoref{t:parsgamma} and \autoref{t:parsproton}  \footnote {The sets of parameters $p$ and $q$ are redundant, but their simultaneous presence simplifies the convergence of bulk fitting of the histograms in the case of the high-statistics distributions of electrons plus photons, $dN_{e,\gamma}/dR$.  For fits to muon fluxes we obtain sufficiently good convergence by fixing all $p_1$ parameters to 1.0 (and consequently the corresponding $q_{i1}$ parameters are set to $(1,0,0)$).}. The procedure to turn the sets of parameters into a continuous model is performed as follows: \par

\begin{enumerate}
    \item For each of the 20 studied values of shower energy we determine the parameters $\alpha, \beta, \gamma, \delta$ of the cubic function that reproduces the values of each of the $p_0, p_1, p_2$ parameters at the four available values of polar angle: {\it e.g.}, for the first parameter $p_0$ required to compute the flux of muons from proton primaries we write \par
\begin{equation}
     \alpha(E_i) + \beta(E_i) \theta_j + \gamma(E_i) \theta_j^2 + \delta(E_i) \theta_j^3 = p_0^{\mu,p}(\theta_j | E_i) 
\end{equation}
with $j=1,2,3,4$ spanning the four bins in polar angle, and
\begin{equation}
    \theta_j = \frac{j-0.5}{4} \left( \pi \frac{65}{180} \right)
\end{equation}
and with
\begin{equation}
    E_i = \exp \left[ (\ln 10 - \ln 0.1) \frac{i+0.5}{20} + \ln 0.1 \right] PeV.
\end{equation}

The above calculation is performed by computing: \par
\begin{equation}
\begin{pmatrix}
    \alpha(E_i) \\ \beta(E_i) \\ \gamma(E_i) \\ \delta(E_i)
\end{pmatrix} =  A^{-1} \begin{pmatrix} p_0^{\mu,p}(\theta_1)_{i}  \\ 
p_0^{\mu,p}(\theta_2)_{i}  \\
p_0^{\mu,p}(\theta_3)_{i}  \\
p_0^{\mu,p}(\theta_4)_{i}  \end{pmatrix} 
\end{equation}
where $A$ is the $4\times4$ matrix whose $k$-th row and $j$-th column contains the $(j-1)$-th power of $k$: \par
\[A = \begin{pmatrix}
   1 &1 &1 &1 \\
   1 &2 &4 &8 \\
   1 &3 &9 &27\\
   1 &4 &16 &64\\
\end{pmatrix}
\]

\item For each flux parameter $(p_i^{\{\mu,e\},\{\gamma,p\}}(E,\theta))$, the parameters defining the cubic functions are then computed for 100 values of $\theta$ and 100 values of $E$ evenly spaced in the relevant ranges of the two variables, and stored in a $100\times 100$ lookup table. This enables us to interpolate each flux parameter at any real value of energy and polar angle by linear interpolation from the known grid values.
\item Along with tables containing parameter values, separate $100\times 100$ tables are built to contain values of the first and second derivatives of parameters with respect to energy and polar angle. These are needed for calculation of derivatives of the fluxes (see \S\ref{s:modelderiv}).
\end{enumerate}

\noindent
Through the above procedure we obtain a continuous model of the flux of each secondary particle species for each primary particle of energy $E$ and polar angle $\theta$. 
%, while 
%Fig. \ref{f:lookuptables} shows the contents of the lookup tables storing the values of the flux parameters $p_i$.

%\begin{figure}[h!]
%\begin{minipage}{0.49\linewidth}
%\includegraphics[width=6.5cm]{}
%\end{minipage}
%\begin{minipage}{0.49\linewidth}
%\includegraphics[width=6.5cm]{}    
%\end{minipage}
%\begin{minipage}{0.49\linewidth}
%\includegraphics[width=6.5cm]{}
%\end{minipage}
%\begin{minipage}{0.49\linewidth}
%\includegraphics[width=6.5cm]{}
%\end{minipage}
%\caption{Temperature maps of the value of flux parameters obtained by the continuous model. Top left: muons from photon primaries; top right: muons from proton primaries. Bottom left: electrons+photons from photon primaries; bottom right: electrons+photons from proton primaries. In each panel the horizontal axis spans linearly the energy range [\SI{0.1}{\peta\eV}, \SI{10}{\peta\eV}], and the vertical axis spans linearly the polar angle range [\SI{0}{\degree}, \SI{65}{\degree}].}
%\label{f:lookuptables}
%\end{figure}

\begin{figure*}[h!t]
\centering
\includegraphics[width=0.9\linewidth]{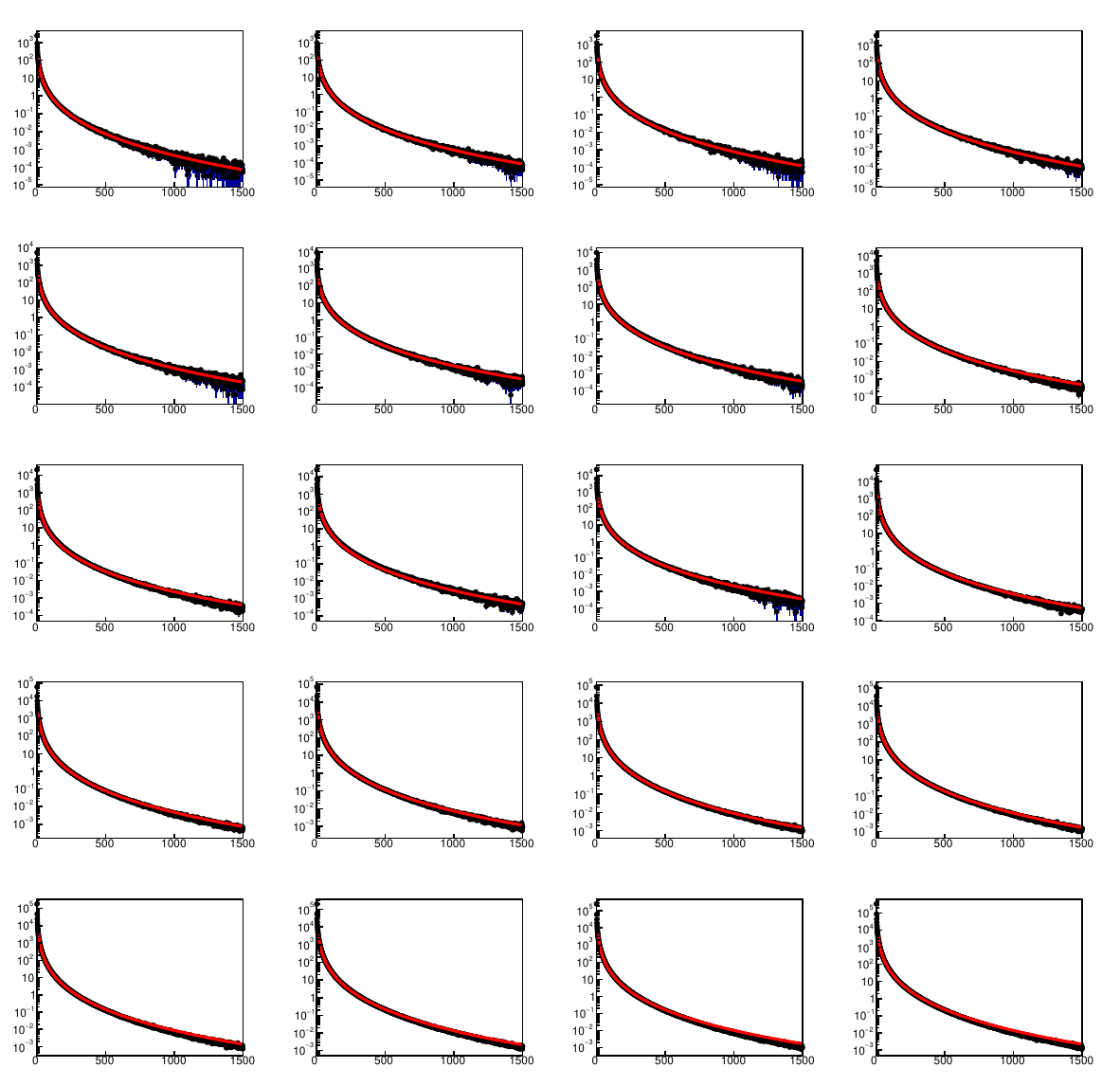}
\caption{Combined flux of electrons and photons (in m$^{-2}$, as a function of distance in meters from the core) from photon primaries of polar angle $\theta_{\gamma}$ in the (\SI{0}{\degree} , \SI{16.25}{\degree}) range, for energies of \SI{100}{\tera\eV} (top left) to \SI{10}{\peta\eV} (bottom right). The model is overlaid in red to the distributions.}
\label{fig:ele1}
\end{figure*}

\begin{figure*}[h!t]
\centering
\includegraphics[width=0.9\linewidth]{ 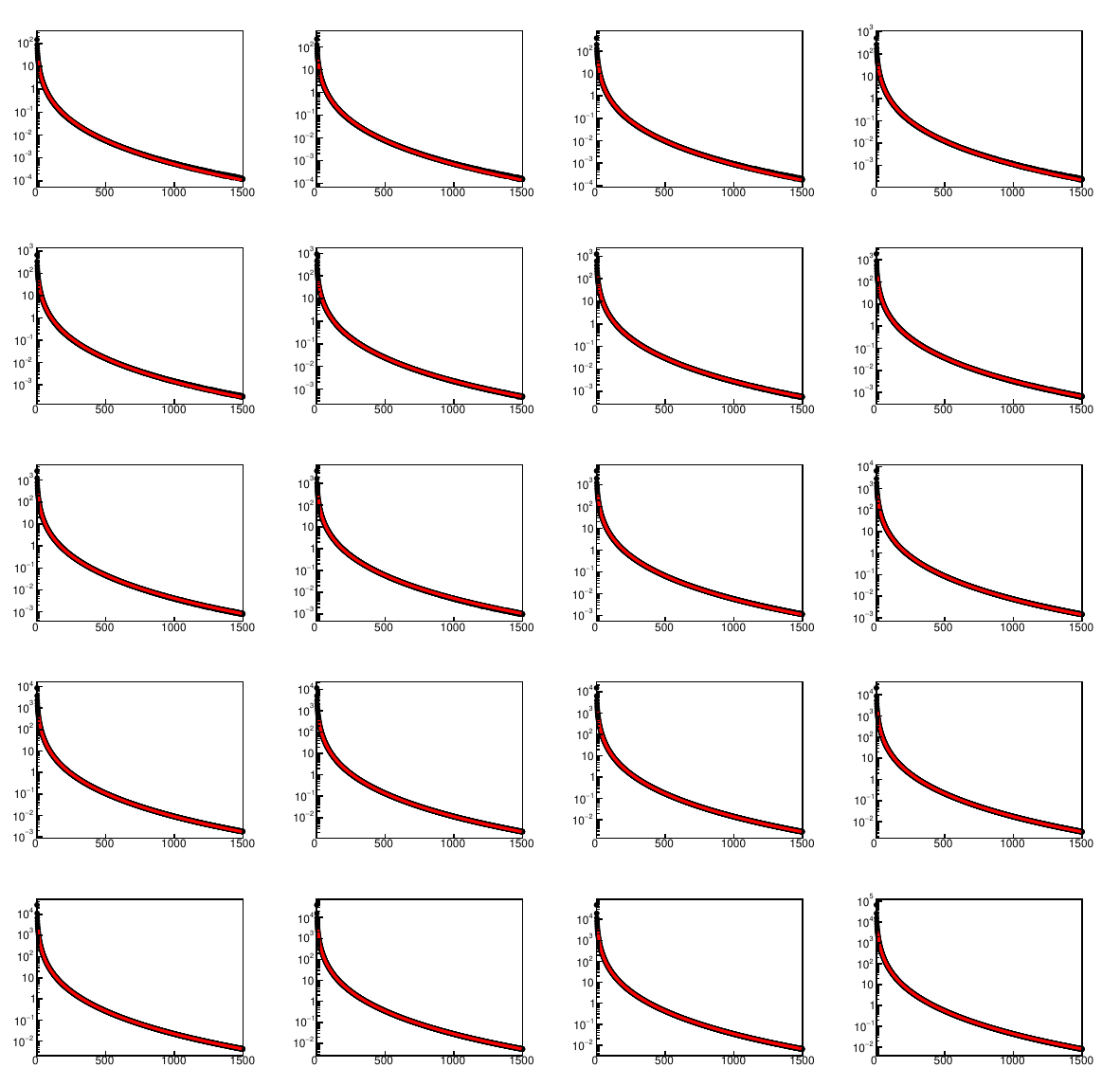}
\caption{\label{fig:ele2}Combined flux of electrons and photons  (in m$^{-2}$, as a function of distance in meters from the core) from proton primaries of polar angle $\theta_{p}$ in the (\SI{32.5}{\degree}, \SI{48.75}{\degree}) range, for energies of \SI{100}{\tera\eV} (top left) to \SI{10}{\peta\eV} (bottom right). The model is overlaid in red to the distributions.}
\end{figure*}

\begin{figure*}[h!t]
\centering
\includegraphics[width=0.9\linewidth]{ 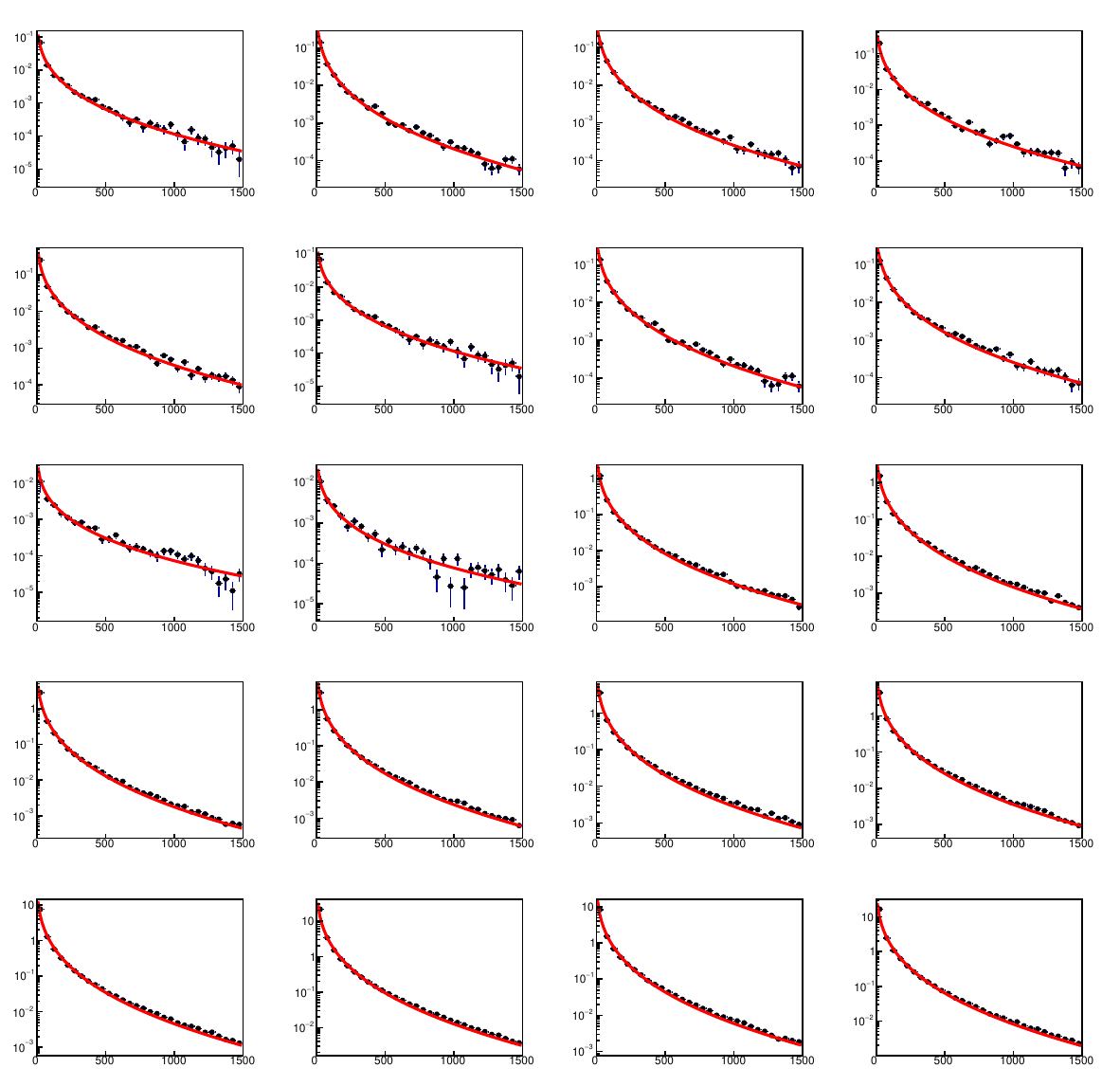}
\caption{Flux of muons (in m$^{-2}$, as a function of distance in meters from the core) from photon primaries of polar angle $\theta_{\gamma}$ in the (\SI{16.25}{\degree}, \SI{32.5}{\degree}) range, for energies of \SI{100}{\tera\eV} (top left) to \SI{10}{\peta\eV} (bottom right). The model is overlaid in red to the distributions.}
\label{f:muo1}
\end{figure*}

\begin{figure*}[h!t]
\centering
\includegraphics[width=0.9\linewidth]{ 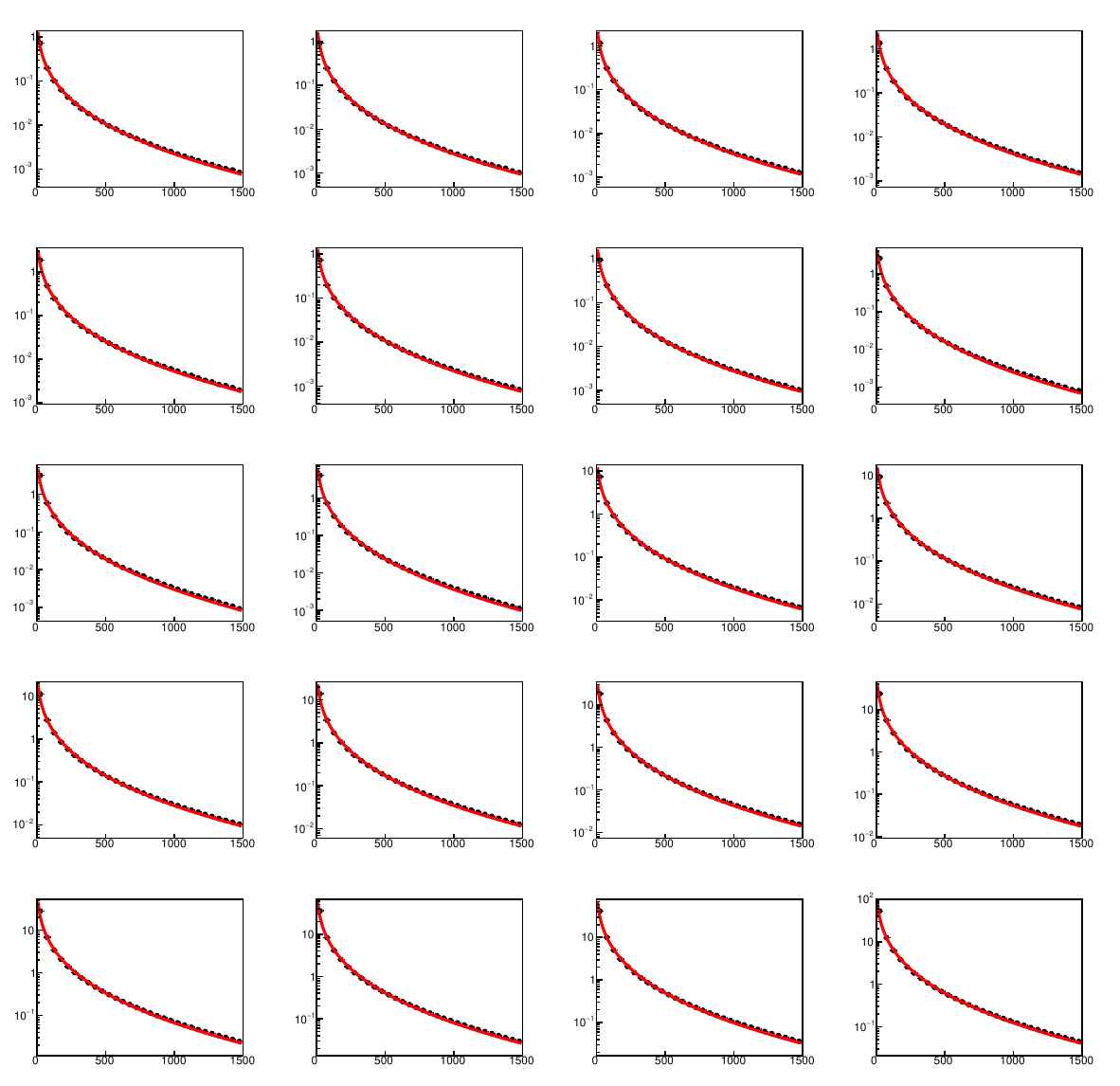}
\caption{Flux of muons (in m$^{-2}$, as a function of distance in meters from the core) from proton primaries of polar angle $\theta_{p}$ in the (\SI{48.75}{\degree}, \SI{65}{\degree}) range, for energies of \SI{100}{\tera\eV} (top left) to \SI{10}{\peta\eV} (bottom right). The model is overlaid in red to the distributions.}
\label{f:muo2}
\end{figure*}

Sample radial profiles of the particle densities along with extracted parametric fits are also shown in \autoref{fig:ele1} and \autoref{f:muo1} (\autoref{fig:ele2} and \autoref{f:muo2}) for photon (proton) primaries, respectively.

\begin{table}[h!]
\begin{center}
\begin{tabular}{l|c|c|c}
Parameter & $p_0$ (\si{\per\meter}) & $p_1$ & $p_2$ \\
\hline
$q_0(\theta_0)_{e,\gamma}$ & 36117.1 & 0.172385 & 1.21588 \\
$q_0(\theta_1)_{e,\gamma}$ & 24130.6 & 0.154078 & 1.23513 \\ 
$q_0(\theta_2)_{e,\gamma}$ & 1730.82 & 0.626387 & 0.82250 \\ 
$q_0(\theta_3)_{e,\gamma}$ & 83.7947 & 0.958974 & 0.71986 \\ 

$q_1(\theta_0)_{e,\gamma}$ & 3.29423 & 0.057859 & 1.73547e-3 \\
$q_1(\theta_1)_{e,\gamma}$ & 3.00016 & 0.069276 & 3.37280e-4 \\
$q_1(\theta_2)_{e,\gamma}$ & 2.70469 & 0.041916 & 3.68160e-4 \\
$q_1(\theta_3)_{e,\gamma}$ & 2.22895 & 0.116381 &-3.67932e-3 \\

$q_2(\theta_0)_{e,\gamma}$ & 0.252659 &-3.23064e-3  & 0.48205e-4 \\
$q_2(\theta_1)_{e,\gamma}$ & 0.266636 &-4.94512e-3  & 1.25588e-4 \\ 
$q_2(\theta_2)_{e,\gamma}$ & 0.249405 &-6.40193e-5  &-0.47208e-4 \\
$q_2(\theta_3)_{e,\gamma}$ & 0.262290 &-4.43699e-3  & 1.73817e-4 \\
\hline

$q_0(\theta_0)_{\mu,\gamma}$ & -358.593 & 0.151459 & 1.17910 \\
$q_0(\theta_1)_{\mu,\gamma}$ & -740.388 & 0.107846 & 1.31133 \\
$q_0(\theta_2)_{\mu,\gamma}$ & -113.901 & 0.064268 & 1.46762 \\
$q_0(\theta_3)_{\mu,\gamma}$ & -116.403 & 0.013792 & 1.90919 \\

$q_1(\theta_0)_{\mu,\gamma}$ & 1 & 0 & 0 \\
$q_1(\theta_1)_{\mu,\gamma}$ & 1 & 0 & 0 \\
$q_1(\theta_2)_{\mu,\gamma}$ & 1 & 0 & 0 \\
$q_1(\theta_3)_{\mu,\gamma}$ & 1 & 0 & 0 \\

$q_2(\theta_0)_{\mu,\gamma}$ & 0.325954 & -0.00141329 & 1.07998e-4 \\
$q_2(\theta_1)_{\mu,\gamma}$ & 0.321244 & -0.00156709 & 1.17792e-4 \\
$q_2(\theta_2)_{\mu,\gamma}$ & 0.309210 &  0.00104322 &-0.18057e-4 \\
$q_2(\theta_3)_{\mu,\gamma}$ & 0.289677 &  0.00248633 &-0.87118e-4 \\
\hline
\end{tabular}
\caption{\it Parameters of the energy dependence of parameters $p_i$ for the flux model of photon-originated secondaries. From top to bottom: e.m.\ particles from gamma showers; muons from gamma showers. The four $\theta$ values are the centers of four bins in the $(\SI{0}{\degree}, \SI{65}{\degree})$ polar angle range where primaries have been generated with \texttt{CORSIKA}, as discussed {\it supra}.}
\label{t:parsgamma}
\end{center}
\end{table}

\begin{table}[h!]
\begin{center}
\begin{tabular}{l|c|c|c}
Parameter & $p_0$ (\si{\per\meter}) & $p_1$ & $p_2$ \\
\hline
$q_0(\theta_0)_{e,p}$ & 9.61560 & 5.20551 & 0.364469 \\
$q_0(\theta_1)_{e,p}$ & 9.28613 & 4.84781 & 0.378406 \\
$q_0(\theta_2)_{e,p}$ & 8.30776 & 4.28549 & 0.394419 \\
$q_0(\theta_3)_{e,p}$ & 6.06557 & 3.95115 & 0.370725 \\

$q_1(\theta_0)_{e,p}$ & 3.00069 & 0.0715927 &-1.90026e-4 \\
$q_1(\theta_1)_{e,p}$ & 2.86508 & 0.0677157 &-4.22523e-4 \\
$q_1(\theta_2)_{e,p}$ & 2.65306 & 0.0549206 &-5.74074e-4 \\
$q_1(\theta_3)_{e,p}$ & 2.61631 & 0.0606328 &-1.55515e-3 \\

$q_2(\theta_0)_{e,p}$ & 0.245223 &-2.33057e-3 & 4.04411e-5 \\ 
$q_2(\theta_1)_{e,p}$ & 0.247950 &-2.24902e-3 & 4.35953e-5 \\
$q_2(\theta_2)_{e,p}$ & 0.250321 &-1.71418e-3 & 3.40669e-5 \\
$q_2(\theta_3)_{e,p}$ & 0.239306 &-1.15019e-3 & 2.81104e-5 \\

\hline 
$q_0(\theta_0)_{\mu,p}$ & 3.29812 & 1.49020 & 0.532621 \\
$q_0(\theta_2)_{\mu,p}$ & 3.20693 & 1.43384 & 0.542787 \\
$q_0(\theta_2)_{\mu,p}$ & 2.82742 & 1.26128 & 0.573445 \\ 
$q_0(\theta_3)_{\mu,p}$ & 1.82703 & 0.89117 & 0.652063 \\

$q_1(\theta_0)_{\mu,p}$ & 1 &0 &0 \\
$q_1(\theta_1)_{\mu,p}$ & 1 &0 &0 \\
$q_1(\theta_2)_{\mu,p}$ & 1 &0 &0 \\
$q_1(\theta_3)_{\mu,p}$ & 1 &0 &0 \\

$q_2(\theta_0)_{\mu,p}$ & 0.324353 & 3.37751e-4 & -1.30596e-6 \\
$q_2(\theta_1)_{\mu,p}$ & 0.322256 & 2.99934e-4 & -1.06650e-6 \\
$q_2(\theta_2)_{\mu,p}$ & 0.316132 & 2.93557e-4 & -1.65590e-6 \\
$q_2(\theta_3)_{\mu,p}$ & 0.303916 & 3.98738e-4 & -9.15743e-6 \\

\hline
\end{tabular}
\caption{\it Parameters of the energy dependence of parameters $p_i$ for the flux model of proton-originated secondaries. From top to bottom: e.m.\ particles from proton showers; muons from proton showers. The four $\theta_0 \div \theta_3$ values are the centers of four bins in the $(\SI{0}{\degree}, \SI{65}{\degree})$ polar angle range where primaries have been generated with \texttt{CORSIKA}, as discussed {\it supra}.}
\label{t:parsproton}
\end{center}
\end{table}

\section { Verification of the derivatives with \texttt{derivgrind}} 
\label{sec:derivgrind-checks}

Forming derivatives by hand, as we do above, is tedious and error-prone. For many pieces of our derivatives computation code, we have verified their correctness by comparing their results to derivatives computed with algorithmic differentiation (AD). AD is a set of techniques to evaluate derivatives of computer code with floating-point accuracy. Specifically, we use the forward mode of AD with a single AD
input $x$, which is based on the idea of keeping track of a ``dot value'' $\dot a = \tfrac{\partial a}{\partial x}$ for every real number $a$ handled by the computer code to be differentiated. Any global variables
have an initial dot value of $0$, and the AD input $x$ --{\it i.e.}, the
variable with respect to which we want to obtain algorithmic
derivatives-- receives a dot value of $\tfrac{\partial x}{\partial
x}=1$. Now, whenever a new real number is computed in the code to be
differentiated, typically by elementary floating-point arithmetic
operations ({\it e.g.} $c=a\cdot b$), its dot value can be computed
according to the appropriate differentiation rule ({\it e.g.} $\dot c =
\dot a \cdot b + a \cdot \dot b$). For any AD output --{\it i.e.}, any
variable that we want to differentiate with respect to the single AD
input $x$--, the derivative $\tfrac{\partial y}{\partial x}$ can then
be read from $\dot y$.

AD tools implement this procedure by analyzing computer code for real
arithmetic operations, and inserting such ``AD logic'', in a semi-
automatic fashion. Many AD tools operate directly on the source code
({\it e.g.}, \texttt{Tapenade}), use language features like operator overloading
({\it e.g.}, \texttt{CoDiPack}, \texttt{JAX}, and the internal tools of \texttt{PyTorch} and
\texttt{TensorFlow}), or hook into a compiler that builds the code to be
differentiated ({\it e.g.}, \texttt{Enzyme, Clad}). \texttt{derivgrind}~\citep{derivgrind} is a novel
AD tool that operates on the machine code of the compiled program. This
is advantageous for us as it minimizes the amount of modifications to
the source code necessary in order to apply the AD tool --to compute a
derivative $\tfrac{\partial y}{\partial x}$, we merely needed to
identify $x$ and $y$ in our C{\ttfamily ++} source code, and apply macros
\texttt{DG\_SET\_DOTVALUE} and \texttt{DG\_GET\_DOTVALUE} from a header \texttt{derivgrind.h}, respectively.

Parts of the optimization code are organized in subroutines that take a
specific integer argument ({\ttfamily mode}), evaluate a function if {\ttfamily
mode} is 0, and derivatives of the function with respect to relevant
variables if {\ttfamily mode} is larger than zero. For these routines,
the calculation with \texttt{derivgrind} is straightforward. %, as shown in
%\autoref{fig:derivgrind-example}.

\texttt{derivgrind} can handle third-party libraries like
ROOT out of the box. While \texttt{derivgrind} does not run as fast as source-code-based AD tools targeted to high performance setups, this is not a
concern for us as the derivative checks do not run in production.

In all of our checks, analytic derivatives match \texttt{derivgrind}'s
algorithmic derivatives. This indicates that we correctly derived and
implemented the analytic derivatives.

\begin{footnotesize}
\begin{lstlisting}[language=C++]
// Evaluation of the algorithmic derivative `df/dx1` of a function `f` with `derivgrind`. 
// The user can compare it with the analytic derivative.

#include <valgrind/derivgrind.h>

double f(double x1, double x2, int mode) {
  if(mode==0){
    // ... return f(x1,x2) ...
  } else if(mode==1) {
    // ... return df(x1,x2)/dx1
  } else if(mode==2) {
    // ... return df(x1,x2)/dx2
  }
}

int main(){
  // ... initializations ...
  double x1=3.1, x2=42.0; // some test values

  double one = 1.0;
  DG_SET_DOTVALUE( &x1, &one, sizeof(double) );
  // evaluate f, could also have multiple statements here
  double result = f(x1,x2,0);
  double df_dx1_algorithmic = 0;
  DG_GET_DOTVALUE( &result, &df_dx1_algorithmic, sizeof(double) );
  std::cout << "ALGORITHMIC DERIVATIVE: " << df_dx1_algorithmic <<
std::endl;

  double df_dx1_analytic = f(x1,x2,1);
  std::cout << "ANALYTIC DERIVATIVE: " << df_dx1_analytic << std::endl;
}
\end{lstlisting}    
\end{footnotesize}

%% \label{}

%% If you have bibdatabase file and want bibtex to generate the
%% bibitems, please use
%%

\bibliographystyle{elsarticle-harv} 
\bibliography{biblio}

%% else use the following coding to input the bibitems directly in the
%% TeX file.

\end{document}